\pdfoutput=1
\documentclass[a4paper,12pt]{article}

\usepackage{amsfonts}
\usepackage{mathrsfs}
\usepackage{amsmath}
\usepackage{amssymb}
\usepackage{framed} 
\usepackage{subfigure}
\usepackage[medium]{titlesec}
\usepackage{bm}
\usepackage{cite}

\usepackage[normalem]{ulem}
\usepackage{extarrows}
\usepackage{slashed}
\usepackage{isodateo}
\usepackage{graphicx}
\usepackage{xcolor}
\usepackage[bookmarksnumbered=true,bookmarksopen=true]{hyperref}
 \hypersetup{colorlinks,%
             linkcolor=[rgb]{0,0.3,0.6}, %
             citecolor=[rgb]{0,0.3,0.6}, %
             urlcolor=[rgb]{0,0.3,0.6}}
\usepackage[hmargin=.7in,vmargin=1.1in]{geometry}
\usepackage{indentfirst}
\usepackage{booktabs}

\usepackage{bbm}

\newcommand{\FR}[2]{\displaystyle\frac{\,{#1}\,}{#2}}
\newcommand{\fr}[2]{\mbox{$\frac{\,{#1}\,}{#2}$}}
\newcommand{\n}{\nonumber}

\graphicspath{{fig/}}

\newcommand{\red}{\color{red}}
\newcommand{\blue}{\color[rgb]{0,.3,1}}

\def\bge{\begin{equation}}
\def\ede{\end{equation}}
\def\bga{\begin{aligned}}
\def\eda{\end{aligned}}
\def\bgb{\begin{bmatrix}}
\def\edb{\end{bmatrix}}
\def\bgp{\begin{pmatrix}}
\def\edp{\end{pmatrix}}
\def\bgm{\begin{matrix}}
\def\edm{\end{matrix}}
\def\bgs{\begin{subequations}}
\def\eds{\end{subequations}}
\newcommand{\order}[1]{\mathcal{O}({#1})}
\def\di{{\mathrm{d}}}

\def\mb{\mathbf}

\def\pd{\partial}
\def\ld{{\mathscr{L}}}

\def\la{\langle}\def\ra{\rangle}

\def\to{\rightarrow}
\def\To{\Rightarrow}
\def\ii{\mathrm{i}}

\def\al{\alpha}
\def\be{\beta}
\def\ga{\gamma}
\def\de{\delta}

\def\ka{\kappa}
\def\lam{\lambda}

\def\si{\sigma}

\def\aa{\mathsf{a}}
\def\bb{\mathsf{b}}
\def\cc{\mathsf{c}}
\def\dd{\mathsf{d}}

\usepackage{mdframed}

\newmdenv[skipabove=0pt,%
          skipbelow=5pt,%
          leftmargin=0pt,%
          rightmargin=0pt,%
          innertopmargin=-5pt,%
          innerbottommargin=7pt,%
          innerleftmargin=2pt,%
          innerrightmargin=2pt,%
          splittopskip=0pt,%
          splitbottomskip=0pt,%
          linewidth=0pt,%
          nobreak=true]%
          {keyeqn2}

\newmdenv[backgroundcolor=gray!15,%
          skipabove=0pt,%
          skipbelow=5pt,%
          leftmargin=0pt,%
          rightmargin=0pt,%
          innertopmargin=-5pt,%
          innerbottommargin=7pt,%
          innerleftmargin=2pt,%
          innerrightmargin=2pt,%
          splittopskip=0pt,%
          splitbottomskip=0pt,%
          linewidth=0pt,%
          nobreak=true]%
          {keyeqn}

\newtheorem{theorem}{Theorem}
\newtheorem{lemma}{Lemma} 

\makeatletter
\@ifpackageloaded{hyperref}%
  {\newcommand{\mylabel}[2]
    {\protected@write\@auxout{}{\string\newlabel{#1}{{#2}{\thepage}%
      {\@currentlabelname}{\@currentHref}{}}}}}%
  {\newcommand{\mylabel}[2]
    {\protected@write\@auxout{}{\string\newlabel{#1}{{#2}{\thepage}}}}}
\makeatother

\usepackage{titlesec}          
\titleformat{\section}
{\normalfont\fontsize{15}{20}\bfseries}{\thesection}{1em}{}

\newcommand{\ob}[1]{\mkern 2mu \overline{\mkern -2mu #1 \mkern -2mu}\mkern 2mu}
\newcommand{\wt}[1]{\mkern 2mu \widetilde{\mkern -2mu #1 \mkern -2mu}\mkern 2mu}

\newcommand{\fnemail}[1]{\footnote{Email: \href{mailto:#1}{\nolinkurl{#1}}}}

\begin{document}

\title{\Large\textbf{Nonanalyticity and On-Shell Factorization of Inflation Correlators at All Loop Orders\\[2mm]}}

\author{Zhehan Qin\fnemail{qzh21@mails.tsinghua.edu.cn}~~~~~ and ~~~~~Zhong-Zhi Xianyu\fnemail{zxianyu@tsinghua.edu.cn}\\[5mm]
\normalsize{\emph{Department of Physics, Tsinghua University, Beijing 100084, China}}}

\date{}
\maketitle

\vspace{20mm}

\begin{abstract}
\vspace{10mm}

The dynamics of quantum fields during cosmic inflation can be probed via their late-time boundary correlators. The analytic structure of these boundary correlators contains rich physical information of bulk dynamics, and is also closely related to cosmological collider observables. In this work, we study a particular type of nonanalytic behavior, called nonlocal signals, for inflation correlators with massive exchanges at arbitrary loop orders. We propose a signal-detection algorithm to identify all possible sources of nonlocal signals in an arbitrary loop graph, and prove that the algorithm is exhaustive. We then present several versions of the on-shell factorization theorem for the leading nonlocal signal in graphs with arbitrary number of loops, and provide the explicit analytical expression for the leading nonlocal signal. We also generalize the nonlocal-signal cutting rule to arbitrary loop graphs. Finally, we provide many explicit examples to demonstrate the use of our results, including an n-loop melon graph and a variety of 2-loop graphs.

\end{abstract}

\newpage
\tableofcontents

\newpage
\section{Introduction}
\label{sec_intro}

The inflation patch of the $(3+1)$-dimensional de Sitter spacetime features an exponential expansion of 3-dimensional flat space towards a spacelike future boundary. The exponential expansion can on the one hand trigger active particle productions through quantum fluctuation of the vacuum state, and on the other hand redshift all those quantum fluctuations to superhorizon scales. If there is an observer who can access a finite portion of the future boundary, then this observer could access the quantum field theory processes in the bulk by measuring the equal-time correlation functions of field operators. These equal-time correlation functions at the future boundary of the inflation patch are what we call \emph{inflation correlators}. 

It is widely believed that our own universe has undergone a period of inflation at very high energy scales which lasted for at least several tens of $e$-folds \cite{Achucarro:2022qrl}. A happy consequence of this scenario is that we are the observers who can measure inflation correlators from large-scale nonuniformity of the universe. This type of measurements has been emphasized in recent years as promising tools to study not only the primordial evolution of the universe, but also the particle physics at presumably the highest energy scale we can ever reach, a program known as the cosmological collider (CC) physics \cite{Chen:2009we,Chen:2009zp,Baumann:2011nk,Chen:2012ge,Pi:2012gf,Noumi:2012vr,Gong:2013sma,Arkani-Hamed:2015bza,Chen:2015lza,Chen:2016nrs,Chen:2016uwp,Chen:2016hrz,Lee:2016vti,An:2017hlx,An:2017rwo,Iyer:2017qzw,Kumar:2017ecc,Chen:2017ryl,Tong:2018tqf,Chen:2018sce,Chen:2018xck,Chen:2018cgg,Chua:2018dqh,Wu:2018lmx,Saito:2018omt,Li:2019ves,Lu:2019tjj,Liu:2019fag,Hook:2019zxa,Hook:2019vcn,Kumar:2018jxz,Kumar:2019ebj,Alexander:2019vtb,Wang:2019gbi,Wang:2019gok,Wang:2020uic,Li:2020xwr,Wang:2020ioa,Fan:2020xgh,Aoki:2020zbj,Bodas:2020yho,Maru:2021ezc,Lu:2021gso,Sou:2021juh,Lu:2021wxu,Pinol:2021aun,Cui:2021iie,Tong:2022cdz,Reece:2022soh,Qin:2022lva,Chen:2022vzh,Niu:2022quw,Niu:2022fki,Aoki:2023tjm,Chen:2023txq,Tong:2023krn,Jazayeri:2023xcj,Jazayeri:2023kji,Meerburg:2016zdz,MoradinezhadDizgah:2017szk,MoradinezhadDizgah:2018ssw,Kogai:2020vzz}.  

Similar to ordinary scattering amplitudes, inflation correlators are functions of external momenta of field operators. As it happened repeatedly in the history, it turns out very useful to extend those external momenta to \emph{complex values} and to study the analytical properties of inflation correlators on the complex planes. Indeed, the study of analytic properties of scattering amplitudes has led to many fruitful results in the past decades \cite{Eden:1966dnq,Kruczenski:2022lot,Bern:2022jnl}. 

The analytic property of dS amplitudes was much less explored than their flat-space counterparts, but this direction has attracted lots of attentions recently. New results include various bootstrap methods using a variety of basic physical principles as input \cite{Arkani-Hamed:2018kmz,Baumann:2019oyu,Baumann:2020dch,Pajer:2020wnj,Hillman:2021bnk,Baumann:2021fxj,Hogervorst:2021uvp,Pimentel:2022fsc,Jazayeri:2022kjy,Wang:2022eop,Baumann:2022jpr}, 
the study of unitarity, causality, symmetries, and their implications to analytical structures \cite{Goodhew:2020hob,Goodhew:2021oqg,Melville:2021lst,Meltzer:2021zin,DiPietro:2021sjt,Tong:2021wai,Salcedo:2022aal,Agui-Salcedo:2023wlq}, 
the Mellin-space approach to dS amplitudes \cite{Sleight:2019hfp,Sleight:2019mgd,Sleight:2020obc,Sleight:2021plv,Jazayeri:2021fvk,Premkumar:2021mlz},
the study of analytical structure and explicit results using techniques of partial Mellin-Barnes (MB) representation \cite{Qin:2022lva,Qin:2022fbv,Qin:2023ejc,Qin:2023bjk}, spectral decomposition \cite{Xianyu:2022jwk,Loparco:2023rug},
and new results for spinning fields \cite{Maldacena:2011nz,Baumann:2017jvh,Bonifacio:2022vwa,Lee:2022fgr} and parity violations \cite{Cabass:2022rhr,Cabass:2022oap,Lee:2023jby}.
There are other approaches to explore the analytical structure of dS amplitudes, including the cosmological polytopes \cite{Arkani-Hamed:2017fdk,Arkani-Hamed:2018bjr}, the scattering equation \cite{Gomez:2021qfd,Gomez:2021ujt}. 
These researches are mostly concerned with the wavefunction coefficients, while correlators are gaining more interest recently since they are physical observables. Wavefunction coefficients and correlators are related to each other, and thus we collectively call them dS amplitudes. However, we will focus on inflation correlators in this work.

Through recent studies, analytical structures of tree-level dS amplitudes have been understood quite well. In general, a tree-graph contribution to a $B$-point amplitude $\mathcal{T}(\mb k_1,\cdots,\mb k_B)$ is regular in all physically accessible configurations (henceforth physical region), but possesses singularities in the unphysical region. Generic singularities include the \emph{total-energy pole}, which is a divergence of the amplitude when the sum of magnitudes of all external momenta goes to zero: $k_1+\cdots+ k_B\to 0$.  The magnitude of a momentum $k_i\equiv|\mb k_i|$ $(i=1,\cdots,B)$ is called an energy in the literature, hence the name total-energy pole. Likewise, a tree graph generally possesses poles when the sum of all energies at an interaction vertex goes to zero, which is called a partial-energy pole. Technically, the total-energy and partial-energy limits are singular due to the divergence of the (SK) time integrals in the early-time limit. In the early times, the spacetime curvature and the future boundary become irrelevant, and one expects that the residues of these poles are given by flat-space quantities. Indeed, for example, the residue of the total-energy pole is exactly given by the corresponding scattering amplitude in flat space. However, the total-energy and partial-energy singularities are in unphysical region and thus inaccessible if one holds all individual energies finite.

\begin{figure}
\centering 
\includegraphics[width=0.4\textwidth]{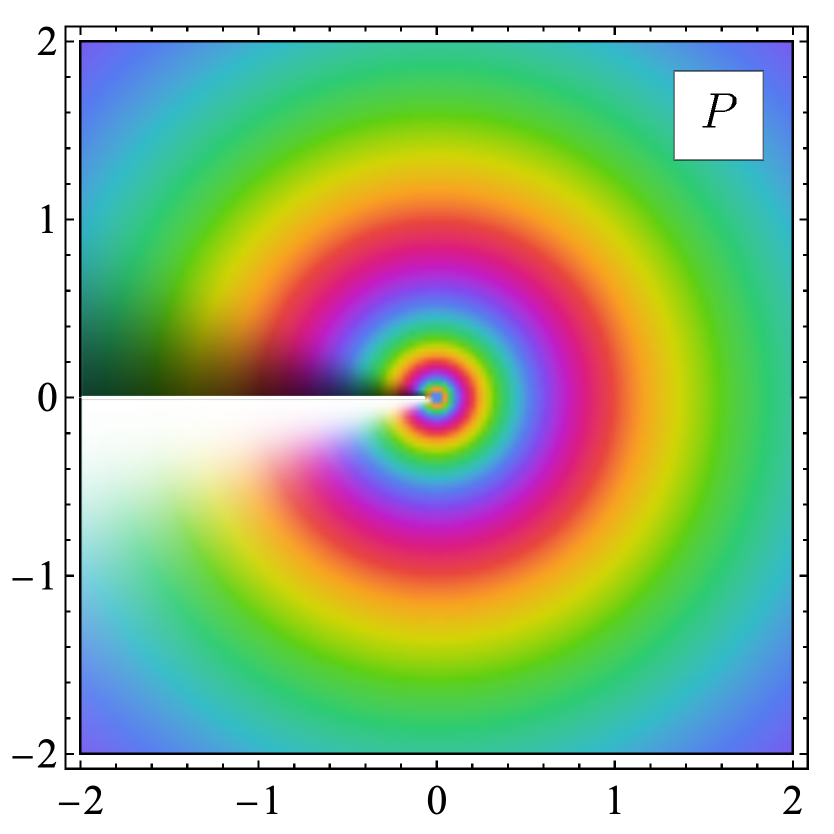}
\caption{The function $P^{\ii\omega}$ on the complex plane with $\omega=4$, taken from \cite{Qin:2023bjk}. The color shows the phase of the function and the brightness shows the magnitude.} 
  \label{fig_complexpower}
\end{figure}

Besides the total-energy and partial-energy singularities, there is another class of singularities which are characteristic of inflation correlators with massive exchanges. These are branch points sitting at the boundary of physical region, when a partial sum of external momenta goes to zero. For a $B$-point amplitude $\mathcal{T}(\mb k_1,\cdots,\mb k_B)$, this limit can be specified by $\mb P_i\equiv \mb k_1+\cdots+\mb k_i\to \mb 0$ $(i<B)$. We note that these soft limits are physically accessible even when individual $k_i$ remains finite. The branch points at such limits are generated by complex powers $P_i^{\pm\ii\omega}$ where  $\omega$ is a real parameter related to the mass of intermediate particles. As shown in Fig.\ \ref{fig_complexpower}, the complex power function $P^{\ii\omega}$ is nonanalytic at $P=0$: It produces logarithmic oscillations in $P$ in the physical region $P>0$, and generates a branch cut when $P<0$. The logarithmic oscillation is exactly the CC signals extensively studied in recent years. The detailed shape of these oscillations contains rich information about particle physics at the inflation scale. Thus, such complex-power branch points are phenomenologically important and deserve more studies. 

Following our earlier works along this line, we shall call such complex-power singularities in partial sums of momenta \emph{nonlocal signal}, to highlight their relation to CC observables. There is a similar type of complex-power branch points at the boundary of the physical region in the complex plane of partial sum of energies $k_1+\cdots+k_i$ instead of partial momenta. These complex energy powers are called \emph{local signal} which are also CC observables. The physical origin of nonlocal and local signals are quite different \cite{Tong:2021wai}. In this work, we focus on nonlocal signals, and a study of local signals will be presented elsewhere.\footnote{There are other situations where (non-oscillatory) branch-point singularities arise in dS amplitudes. For instance, if the time integral at an interaction vertex diverges in the infrared (late-time) limit, the resulting amplitude could contain (poly-)logarithmic terms that contain branch points \cite{Arkani-Hamed:2015bza,Baumann:2021fxj,Wang:2022eop}. In this work, we exclude this possibility by requiring that all couplings are infrared safe. A related situation is the generation of anomalous dimension for a bulk field when one resums the chain of infrared-divergent bubble graphs \cite{Premkumar:2021mlz}.}

It has been realized in the early studies that the form of a nonlocal or local signal is typically simpler than the whole amplitude \cite{Chen:2015lza,Arkani-Hamed:2015bza}. From the viewpoint of SK integral, the signals are generated from factorized time integrals; One does not need to worry about time-orderings in the multi-layer time integrals so far as the (nonlocal or local) signal is the only concern. From the bootstrap viewpoint, the signal part corresponds to the solution to the homogeneous (sourceless) bootstrap equations \cite{Arkani-Hamed:2015bza,Arkani-Hamed:2018kmz,Qin:2022fbv}. All these facts show that the nonlocal signal in an amplitude should be much simpler to understand and compute than the full results. 

Indeed, in \cite{Tong:2021wai} it was shown that one can formulate a cutting rule for signals in a tree graph, in the sense that one only needs to compute factorized time integrals. We proved this cutting rule in a more rigorous way in \cite{Qin:2022fbv} with the partial Mellin-Barnes representation. Later in \cite{Qin:2023bjk}, we generalized this result to arbitrary 1-particle-irreducible 1-loop graphs, and showed that the nonlocal signals in 1-loop graphs are also free from time orderings. More importantly, we provided an explicit analytic expression for the leading nonlocal signal (leading in the partial sum $\mb P_i$ as $\mb P_i\to \mb 0$), and showed that the leading nonlocal signal factorizes into two subgraphs together with a ``bubble signal'' at the 1-loop order. This factorization is very similar to the on-shell factorization of tree-level scattering amplitude in flat spacetime, as we shall elaborate in Sec.\ \ref{sec_nlsignal}. Thus, we will borrow the term ``on-shell factorization'' to describe the factorization of nonlocal signals in inflation correlators. We choose to use this term also to avoid potential confusion with the factorization of correlators at their partial-energy poles \cite{Arkani-Hamed:2018kmz,Baumann:2019oyu,Baumann:2020dch,Goodhew:2020hob}. 

In this work, we generalize our earlier study of 1-loop nonlocal signals to graphs with an arbitrary number of loops, and thus provide a more complete understanding of leading nonlocal signals within the scope of perturbative diagrammatic expansion. The main results of this work are the following:
\begin{enumerate}
  \item We properly define what a nonlocal signal is in an arbitrary massive inflation correlator. Specifically, a nonlocal signal is defined with respect to a particular nonlocal soft limit, where a particular partial sum of external momenta goes to zero, while all other independent partial sums remain finite. The nonlocal signal is then defined as the complex power dependence in the soft partial sum. 
  
  \item Intuitively, a nonlocal signal arises when some of the bulk propagators in a graph become soft simultaneously. We formulate this intuition into a signal-detection algorithm which enables us to locate all sources of nonlocal signal in a given nonlocal soft limit. This is the first main result of our work, summarized in Theorem \ref{thm_signal} in Sec.\ \ref{sec_detect}. Essentially, the algorithm says that we take all possible nonlocal cuts of the diagram with respect to a fixed bipartition. Then, the kinematic region where all cut lines become soft gives a candidate of nonlocal signals. With the partial Mellin-Barnes representation, we prove rigorously that our signal-detection algorithm is exhaustive. Technically, we show that the nonlocal signal must be from the ``degenerate singularities'' of the loop integral, and we show that the ``degenerate regions'' and nonlocal cuts have a one-to-one relationship. 
  
  \item Given the complication of arbitrary loop inflation correlators in the absence of full dS isometry, we are unable to rigorously formulate a factorization theorem in the most general case. Instead, we can prove the factorization of leading nonlocal signal in graphs with an arbitrary number of loops in two broad cases: a) The graph can contain arbitrary dS-symmetry-breaking propagators and interactions, but the graph has no internal vertices, i.e., all vertices are connected to at least one external line; 2) The graph can have arbitrary loop topology but is dS covariant. In either of these two cases, we can show that the leading nonlocal signal in a given nonlocal soft limit is the sum of finite terms, each of which corresponds to a ``minimal cut'' of the diagram. With a given minimal cut, the nonlocal signal factorizes into three pieces: a left subgraph, a right subgraph, and a $(D-1)$-loop ``melon signal,'' where $D$ is the number of lines in the minimal cut. This is the second main result of this paper, summarized in Theorem \ref{thm_bulkfree} in Sec.\ \ref{sec_bulkfree} and Theorems \ref{thm_UMC} and \ref{thm_generalgraph} in Sec.\ \ref{sec_arbitrarygraph}.
 
\end{enumerate}

The rest of the paper is structured as follows. Given the technical complication of this work, we add a non-technical discussion about the nonlocal signal in Sec.\ \ref{sec_nlsignal}, where we start from flat-space scattering amplitudes, explain the origin of singularities in QFT amplitudes, and show why the factorization of a nonlocal signal is similar to the on-shell factorization of tree amplitudes in flat space. Then, in Sec.\ \ref{sec_tree}, we provide a discussion of nonlocal signals from cutting a tree line in arbitrary inflation correlators, as a warm-up example for more complicated cases of cutting loops. 

Then, in Sec.\ \ref{sec_detect} we set out to study nonlocal signals in arbitrary loop graphs, defining the nonlocal signal properly (Sec.\ \ref{sec_def}), formulating the signal-detection algorithm (Sec.\ \ref{sec_algorithm}), and proving the algorithm with the partial Mellin-Barnes representation (Sec.\ \ref{sec_algorithm_proof}). We also collect a few useful lemmas about cutting a graph in Sec.\ \ref{sec_lemma}.

In Sec.\ \ref{sec_bulkfree}, we formulate and prove the on-shell factorization theorem (Theorem \ref{thm_bulkfree}) of leading nonlocal signal in an arbitrary graph without internal vertices, which we call a bulk-free graph. 

Then, in Sec.\ \ref{sec_arbitrarygraph}, we discuss the complications of graphs with internal vertices (Sec. \ref{sec_complication}), and then formulate the on-shell factorization theorem of nonlocal signals for graphs with unique minimal cut (Sec.\ \ref{sec_uniquecut}) and graphs with multiple minimal cuts (Sec.\ \ref{sec_multiplecuts}). 

We then provide in Sec.\ \ref{sec_example} a number of explicit examples, including the tree-level mixing graph, the $L$-loop melon graphs, and many two-loop examples. These examples serve as demonstrations of nonlocal signals in many different situations in multiloop graphs. The conclusion and outlook are in Sec.\ \ref{sec_conclusion}. We collect frequently used mathematical functions and notations in App.\ \ref{app_notation}, and useful integrals in App.\ \ref{app_int}.

\paragraph{Notations and conventions.}
Most of the notations and conventions of this work are in line with the ones adopted in our previous works, especially \cite{Qin:2023bjk}. We use mostly plus signature for the spacetime metric: $\di s^2 = a^2(\tau)(-\di\tau^2+\di\mb x^2)$, where $a(\tau)=-1/(H\tau)$ is the scale factor, $\tau\in (-\infty,0)$ is the conformal time, and $H$ is the inflation Hubble parameter. We take $H=1$ throughout this work for simplicity. 
We use bold letters such as $\mb k$ to denote 3-momenta and the corresponding italic letter $k \equiv |\mb k|$ to denote its magnitude.
When encountering the sum of variables with different indices, we shall use a shorthand notation $k_{12}\equiv k_1+k_2$, $E_{12s}\equiv E_1+E_2+E_s$, etc.
We shall use Mellin variables such as $s_i$ and $\bar s_i$ extensively. We often use a pair of barred and unbarred variables to denote the two (independent) Mellin variables on the same bulk line. We shall also use $\bar s_i$ and $s_{\bar i}$ interchangeably. Following our notations, we shall denote the sum of Mellin variables with the shorthand $s_{ij} = s_i+s_j$, $s_{i\bar i j\bar j} = s_i + \bar s_i + s_j + \bar s_j$, etc.

\section{Nonlocal Signals: Physical Understanding}
\label{sec_nlsignal}

\subsection{From scattering amplitudes to inflation correlators}
\label{sec_scatamp}

\paragraph{Flat-space scattering amplitudes and correlators.}
The occurrence of a singularity in amplitudes typically has a physical meaning. It is instructive to review a familiar example from scattering amplitudes in flat spacetime. Consider a simple model of a massless scalar $\varphi$ and a massive scalar $\si$ of mass $m$, interacting through a simple cubic vertex $\ld\supset -\fr12\lam\si\varphi^2$. Then, the scattering $\varphi(k_1)\varphi(k_2)\to\varphi(k_3)\varphi(k_4)$ at the leading order is mediated by a single $\si$ exchange at the tree level, with three independent channels. Take the $s$-channel exchange as an example, the scattering amplitude simply reads:
\bge
\label{eq_Tflat}
  \mathcal{T}_s(\varphi\varphi\to\varphi\varphi)=\FR{\lam^2}{-s+m^2},
\ede
where $s=-(k_1^\mu+k_2^\mu)^2$ is the squared 4-momentum of the $s$-channel. Clearly, the amplitude possesses a simple pole at $s=m^2$, where the intermediate particle is on shell. In addition, the amplitude factorizes into three pieces at this pole:
\bge
  \lim_{s\to m^2}\mathcal{T}_s(\varphi\varphi\to\varphi\varphi)=\lam\times\FR{1}{-s+m^2}\times \lam.
\ede
It looks funnily trivial for this simple process, but this factorization is actually a very general result which holds nonperturbatively \cite{Weinberg:1995mt}. The crucial ingredient of this factorization is an intermediate state going on shell. To appreciate the physical meaning of this pole, we switch to a manifestly on-shell language, writing $s=(k_s^0)^2-\mb k_s^2=(E_1+E_2)^2-\mb k_s^2$ and using the on-shell energy $E_s^2\equiv \mb k_s^2+m^2$. (We emphasize that $E_s\neq k_s^0$.) Then:
\begin{align}
  \FR{\lam^2}{-s+m^2} 
  =- \FR{\lam^2}{2E_s}\Big(\FR{1}{E_1+E_2-E_s}-\FR{1}{E_1+E_2+E_s}\Big).
\end{align}
This is nothing but what we would get by solving the Lipman-Schwinger equation to the first nontrivial order in the old-fashioned perturbation theory. In this formalism, all the states stay manifestly on shell, at the expense that the energy is not conserved at each interaction vertex, i.e., $E_1+E_2\neq E_s$, etc. (Indeed, the energy difference in the denominator is familiar in the standard perturbation theory in quantum mechanics.) The uncertainty principle tells us that the energy needs not to be conserved so long as the process happened within a short period of time $\Delta t\sim (E_1+E_2-E_s)^{-1}$. Therefore, when the intermediate particles ``go on shell'' in the covariant language, it really means that the energy conservation is satisfied at each interaction vertex so that the interaction can happen indefinitely long, which becomes the source of the divergence. Also, since the intermediate on-shell particle can travel a long distance, the left and right subgraphs are naturally separated from each other. This gives us an intuitive understanding of the on-shell pole and the factorization of the graph.
 
Although the old-fashioned perturbation theory provides a clear physical intuition for the scattering process, it is rarely used in practical calculations due to the manifest breaking of Lorentz covariance. However, the noncovariant approach based on the Schwinger-Keldysh (SK) formalism (also called the in-in formalism) \cite{Weinberg:2005vy,Schwinger:1960qe,Feynman:1963fq,Keldysh:1964ud,Chen:2017ryl}, proves enormously useful in cosmology with flat spatial slices, where the Lorentz symmetry is typically absent but the 3-dimensional translation and rotation symmetries are present. In this noncovariant approach, we use the time and 3-momentum as independent variables, so that every interaction vertex is associated with a 3-momentum-conserving $\de$-function as well as an integral over the time variable. There are four types of frequently used propagators, the Feynman $D_{++}$, the anti-Feynman $D_{--}$, and the two Wightman functions $D_{\pm\mp}$, which for a scalar of mass $m$ are respectively given by:
\begin{align}
\label{eq_DflatInh}
  D_{\pm\mp}(k;t_1,t_2)=&~\FR{1}{2\sqrt{\mb k^2+m^2}}e^{\pm\ii \sqrt{\mb k^2+m^2} (t_1-t_2)}, \\
\label{eq_DflatHom}
  D_{\pm\pm}(k;t_1,t_2)=&~D_{\mp\pm}(k;t_1,t_2)\theta(t_1-t_2)+D_{\pm\mp}(k;t_1,t_2)\theta(t_2-t_1).
\end{align}
As a warm-up example, we use this formalism to recompute the scattering amplitude $\mathcal{T}(\varphi\varphi\to \varphi\varphi)$. Note that the scattering amplitude is an in-out amplitude connecting states at $t=-\infty$ and $t=+\infty$, and the four external legs should be amputated, while the internal leg should take the Feynman propagator $D_{++}$. Therefore:
\begin{align}
 &(2\pi)\de(E_{1234})\ii\mathcal{T}_s(\varphi\varphi\to\varphi\varphi)\n\\
 =& -\FR{\lam^2}{2E_s}\int_{-\infty}^{+\infty}\di t_1\di t_2\,e^{\ii E_{12}t_1+\ii E_{34}t_2}\Big[e^{-\ii E_s(t_1-t_2)}\theta(t_1-t_2)+e^{+\ii E_s(t_1-t_2)}\theta(t_2-t_1)\Big]\n\\
 =&~ \ii (2\pi)\de(E_{1234}) \FR{\lam^2}{E_s^2-E_{12}^2} .
\end{align}
Here and below we use the shorthand $E_{ij\cdots}=E_i+E_j+\cdots$. This is exactly the familiar result (\ref{eq_Tflat}).

In cosmology, observables usually correspond to equal-time correlation functions rather than scattering amplitudes. The analytical structure for the correlation functions is already different from the corresponding scattering amplitudes in flat space. Still take the above $\varphi\varphi\to\varphi\varphi$ process as an example. Now we want to compute the equal-time correlation function $\la\varphi_{\mb k_1}\varphi_{\mb k_2}\varphi_{\mb k_3}\varphi_{\mb k_4}\ra'$ instead of the scattering amplitude. For this purpose, we should take the final time at $t=0$, and include all possible four propagators according to the SK formalism \cite{Chen:2017ryl}. Also, the external legs are not amputated. The $s$-channel exchange then gives the following contribution to the correlator:
\begin{align}
  \la\varphi_{E_1}\varphi_{E_2}\varphi_{E_3}\varphi_{E_4}\ra_s'=\FR{-\lam^2}{16E_1\cdots E_4}\sum_{\aa,\bb=\pm}\aa\bb\int_{-\infty}^0\di t_1\di t_2 e^{+\ii\aa E_{12}t_1+\ii\bb E_{34}t_2}D_{\aa\bb}(k_s;t_1,t_2).
\end{align}
Here the prime in $\la\cdots\ra'$ means that we have removed the 3-momentum-conserving $\de$-function from the correlator.
Using the explicit expressions for the propagators given in (\ref{eq_DflatInh}) and (\ref{eq_DflatHom}), we can evaluate the above integral directly, and the result is:
\bge
  \la\varphi_{E_1}\varphi_{E_2}\varphi_{E_3}\varphi_{E_4}\ra_s'=\FR{\lam^2}{8E_1\cdots E_4E_s}\FR{E_{1234s}}{E_{1234}E_{12s}E_{34s}}. 
\ede
An immediate observation is that, unlike the scattering amplitude, the correlation function is regular in the interior of physically reachable region. (In this case, the interior of the physical region is specified by the joint condition $E_i>0$ ($i=1,2,3,4$), $E_s>m$, and $E_{12}, E_{34}>\sqrt{E_s^2-m^2}$.) In particular, there is no physical analog of ``on-shell poles'' since we are not using the 4-momentum language, and all quantities are manifestly on shell. However, if we are allowed to go to unphysical region by analytical continuation of some energy or momentum variables, we can find new singularities. For instance, we will hit poles when we send either of the three factors in the denominator to zero: $E_{1234}$, $E_{12s}$, or $E_{34s}$. The first is called a total-energy pole and the last two are called the partial-energy poles in the literature. In addition, if we want to view $k_s=\sqrt{E_s^2-m^2}$ as an independent variable and consider the complex $k_s$-plane, we will also find a pair of branch points at $k_s=\pm \ii m$. Or, if we stick to energy variables $E_i$ $(i=1,\cdots,4,s)$, we only find poles in the complex energy plane, without getting any branch cut \cite{Salcedo:2022aal}. There are certain freedom and ambiguities in choosing variables to do analytic continuation and it seems that there is no unique optimal choice. One can make use of freedom to choose appropriate variables depending on the problem they want to solve. In any case, the lesson here is that, unlike scattering amplitudes, a graph contribution to the correlator is fully regular in the physical region, and singularities appear only in unphysical regions, or at most at the boundary of the physical region. (For instance, the total-energy pole can be physically approached by simultaneously sending $E_1,\cdots,E_4\to 0$.) This result also extends to loop orders: New singularities such as branch cuts can appear in the complex energy plane at loop orders, but they are all in unphysical region.

\paragraph{Inflation correlators.}
Now we proceed to inflation correlators. Here we are interested in the correlation function of final time slice, marked by the conformal time $\tau_f\simeq 0$. Due to the exponential expansion, it is more convenient to use the comoving momentum $\mb k$ as the kinematic variable. So, the correlator we are interested in has the form $\la\varphi_{\mb k_1}\cdots\varphi_{\mb k_4}\ra_{\tau=0}'$. By a slight abuse of terminology, the magnitude $k_i\equiv|\mb k_i|$ is sometimes called the energy in the literature, although it is not directly related to the energy of the mode. Indeed, there is no simple time-independent relation between the energy and the comoving momentum in an inflating background. 

Even with this kinematic difference, the analytic structure of inflation correlators, when viewed as functions of complex ``energies'' $k_i$, has similarities with the flat-space correlation functions. In particular, inflation correlators are also regular in the interior of the physical region. Singularities emerge only in unphysical regions, or at most at the boundary of the physical region. The difference is that the curved spacetime background distorts the mode function, which makes the expression of correlation functions more complicated, but also a lot more interesting. As a result, the tree-level correlators may also possess branch cuts in unphysical regions, which is a central topic of this work.

Again, let us take the 4-point $s$-channel exchange graph as the example, but in dS. (As a reminder, we always take $H=1$.) Here we take the four external modes to be conformal scalar field $\varphi_c$ with mass $m_c^2=2$. (It is the conformal scalar rather than the massless scalar in dS that resembles a flat-space massless scalar field.) For the $s$-channel exchanged particle, we take it to be a massive scalar $\si$ in the principal series, namely, with mass $m>3/2$. Still consider the direct cubic coupling $\sqrt{-g}\varphi_c^2\si$, we can then compute the correlation function in the standard way. The complete expression is rather long, and here we only write it in the following schematic way:
\begin{align}
\label{eq_4ptTree}
  &\la\varphi_{\mb k_1}\varphi_{\mb k_2}\varphi_{\mb k_3}\varphi_{\mb k_4}\ra_s'=\FR{\tau_f^4}{16k_1k_2k_3k_4k_s}\n\\
  &\times\bigg\{\sum_{\cc=\pm}\bigg[
  \underbrace{\mathcal{F}^\text{(NL)}_\cc\Big(\FR{k_s}{k_{12}},\FR{k_s}{k_{34}}\Big)\Big(\FR{k_s^2}{k_{12}k_{34}}\Big)^{+\ii\cc\wt\nu}}_\text{nonlocal signal}
  +\underbrace{\mathcal{F}^\mathrm{(L)}_\cc\Big(\FR{k_s}{k_{12}},\FR{k_s}{k_{34}}\Big)\Big(\FR{k_{34}}{k_{12}}\Big)^{+\ii\cc\wt\nu}}_\text{local signal}\bigg] 
   +\underbrace{\mathcal{F}^\text{(BG)} \Big(\FR{k_s}{k_{12}},\FR{k_s}{k_{34}}\Big)}_{\text{background}}\bigg\}.
\end{align}
Here we have defined $\wt\nu\equiv \sqrt{m^2-9/4}$, and $|\tau_f|\ll 1$ is a late-time cutoff inserted to properly take account of the late-time fall off of a conformal scalar. All the three functions $\mathcal{F}^\text{(NL)}, \mathcal{F}^\mathrm{(L)}, \mathcal{F}^\text{(BG)}$ are analytic in $k_s$ at $k_s=0$ although each of them contain singularities in other places. Important is that the nonanalytic behavior in $k_s$ at $k_s=0$ is fully from the first term $\propto k_s^{2\ii\cc\wt\nu}$, which is called the nonlocal signal. In addition, there is a piece analytic in $k_s$ but nonanalytic in $k_{34}/k_{12}$ which is called the local signal. Finally, there is a piece analytic in all energy variables, and is called the background. 

Away from the special point $k_s=0$, the full correlator $\la\varphi_{\mb k_1}\varphi_{\mb k_2}\varphi_{\mb k_3}\varphi_{\mb k_4}\ra_s'$ is actually an analytic function on the whole complex $k_s$ plane except isolated poles and branch cuts. As mentioned above, these singularities are all outside the physical region. The expression (\ref{eq_4ptTree}) can be viewed as an expansion of the full correlator around $k_s=0$. 

At this point we want to make a comment on the 3-point limit of (\ref{eq_4ptTree}). If we now send $k_4\to 0$ to get a 3-point correlator, we see that this amounts to sending $k_{34}\to k_s$ due to the momentum conservation. As a result, the nonlocal and the local signals become indistinguishable. Since we are only considering nonlocal signals in this work, we will stay away from such degenerate configurations.   

\begin{figure}[ht]
\centering 
\subfigure[Resonance pole]{\label{fig_poleA}\includegraphics[width=0.4\textwidth]{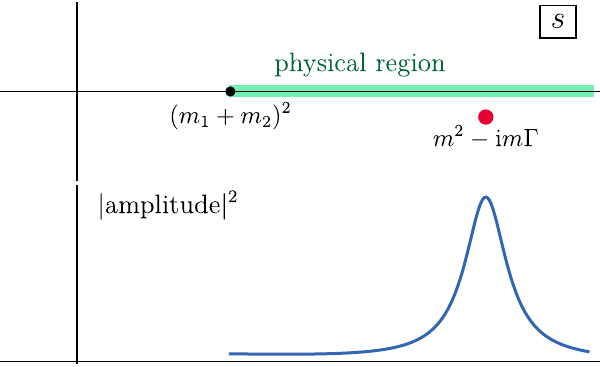}} 
\hspace{6mm}
\subfigure[Unphysical pole]{\label{fig_poleB}\includegraphics[width=0.4\textwidth]{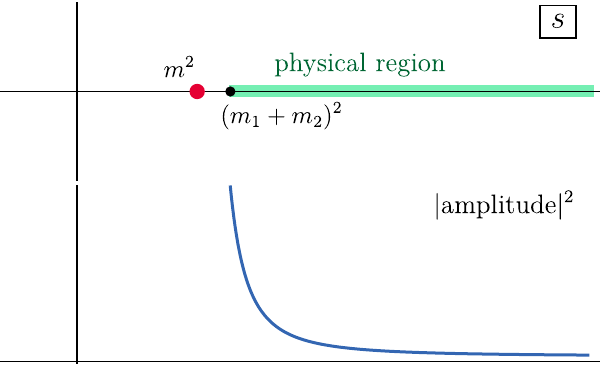}}  \\[4mm]
\subfigure[Branch cut]{\label{fig_cutC}\includegraphics[width=0.4\textwidth]{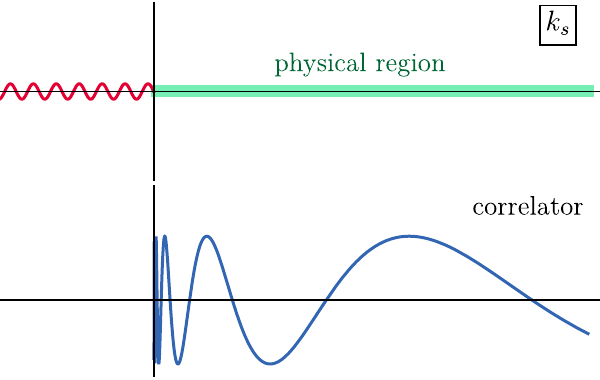}}  \caption{Singularities of amplitudes and their effects in physical observables. The upper two figures show that in flat spacetime, a kinematically reachable pole results in a resonance pick in the cross section as a function of $s$ (left), while an unreachable pole implies a tail of the pick (right).
The lower figure shows that in dS, a branch cut starting from the origin point of $k_s$ in the unphysical region is associated with a logarithmic oscillatory signal (nonlocal signal) in the physical region.} 
  \label{fig_poleandcut}
\end{figure}

Readers may wonder why we care about the singularities of correlation functions if they all appear outside of the physical region. In fact, even in scattering amplitude, the pole should not really appear in the physical region. The usual on-shell mass pole in a tree-level scattering amplitude, if kinematically reachable, implies that the intermediate particle cannot be stable. Its finite lifetime $T$ then has the effect of shifting the position of the on-shell pole off the real axis, by an amount of $T^{-1}$, which is the well-known Breit-Wigner construction. In this case, the pole off the real axis produces a peak in the scattering amplitude as a function of $s$. As is well known, searching for such peaks is a standard way to find new particles in collider experiments; See Fig.\ \ref{fig_poleA}. Even when the on-shell pole is not kinematically reachable (for example, when $m<m_1+m_2$ as shown in Fig.\ \ref{fig_poleB}), the tail of the pole may still be seen in the scattering amplitude near the threshold, so long as the pole is not too far away from the physical region. This is the case, for example, for the pion pole in the nucleon-nucleon scattering \cite{Weinberg:1995mt}. 

In the case of inflation correlators, the appearance of a complex power at $k_s=0$ has even more interesting consequences: it gives rise to an oscillatory dependence on the \emph{logarithm} of a momentum ratio when this ratio approaches zero; See Fig.\ \ref{fig_cutC}. Thus it becomes a true signal to be searched for from cosmological data, and this is the ultimate reason that we call it a signal. The main aim of this work is to define, to locate, and to characterize such signals in inflation correlators with an arbitrary number of external points and an arbitrary number of loops. However, before embarking on a full and technical discussion, let us first explain the physics behind these nonlocal signals.

\paragraph{Interpretation of nonlocal signals.}
There are two useful ways to understand the appearance of the nonlocal signal. Physically, it can be understood as a resonance between the soft massive mode and the adjacent hard modes. Mathematically, it can be understood as a manifestation of the conformal two-point correlator of boundary operators with complex scaling dimensions.
\begin{enumerate}
  \item \emph{Resonance effect.} The fast expansion of the spacetime background can trigger spontaneous pair production of massive scalar particles. Technically, the scalar mode function of a fixed comoving momentum develops a negative-frequency mode when its physical wavelength $(k\tau)^{-1}$ is redshifted to a value comparable to its Compton wavelength $m^{-1}$. As a result, we find on-shell excitations and a nonzero occupation number for low momentum modes with $|k\tau|\lesssim m$. This production can also be seen from the propagator of the massive particle in the late-time limit \eqref{eq_DNL}: 
  \begin{align} 
  \label{eq_Dlate}
  D_{\aa\bb}(k_s;\tau_1,\tau_2)\simeq \FR{(\tau_1\tau_2)^{3/2}}{4\pi}\bigg[\Gamma^2(-\ii\wt\nu)\Big(\FR{k_s^2\tau_1\tau_2}4\Big)^{\ii\wt\nu}+\Gamma^2(\ii\wt\nu)\Big(\FR{k_s^2\tau_1\tau_2}4\Big)^{-\ii\wt\nu}\bigg]+\cdots.
\end{align} 
Here we have retained the terms nonanalytic in $k_s$ at the leading order. If we now sandwich this late-time propagator with the external modes and perform the two time integrals of the two interaction vertices, we get, schematically:
\begin{align}
\label{eq_resonance}
    \int\di\tau_1  \di\tau_2 \, e^{+\ii k_{12}\tau_1+\ii k_{34}\tau_2}(k_s^2\tau_1\tau_2)^{\pm\ii \wt\nu}\sim \Big(\FR{k_s^2}{k_{12}k_{34}}\Big)^{\pm\ii\wt\nu}.
\end{align}
We see that the integral receives most of its contribution from the saddle point $|k_{12}\tau_1|\sim \wt\nu$ and $|k_{34}\tau_2|\sim \wt\nu$, where the nonrelativistic massive mode resonates with the external modes at the two endpoints. This resonance condition tells us that, by sending $k_s/k_{12}$  and $k_s/k_{34}\to 0$, we are effectively probing the $\tau\to 0$ limit of the soft mode. This turns out to be a useful physical intuition for our later analysis. Indeed, the resulting momentum dependence on the right hand side of (\ref{eq_resonance}) is exactly the nonlocal signal shown in (\ref{eq_4ptTree}). This is also called a clock signal in the literature, as the massive mode can be viewed as a clock clicking with fixed physical frequenecy, which records the expansion history $a(t)$ through the resonance \cite{Chen:2015lza}.

  \item \emph{Conformal two-point correlator.} The nonlocal signal can be understood from a late-time boundary point of view in terms of CFT correlators \cite{Antoniadis:2011ib,Mata:2012bx,Qin:2023bjk}. Very often, the massive propagator is covariant under the bulk isometries, which is SO(4,1) in the case of dS. This is mapped to the conformal symmetry at the future boundary. Thus, when we pull a bulk 2-point function to the boundary, it will naturally satisfy all the boundary conformal Ward identities, and thus are classified by the scaling dimension. For instance, a principal scalar with mass $m>3/2$ in the bulk will give rise to a pair of operators on the boundary with scaling dimension $\Delta_\pm=\fr32\pm\ii\wt\nu$. Thus, its 2-point function with two $\Delta_\pm$ operators in the position space has the form $1/|\mb x|^{2\Delta_\pm}$. In the momentum space, this translates to $\int\di^3\mb x\,e^{-\ii\mb k\cdot\mb x}|\mb x|^{-2\Delta_\pm}\sim k^{2\Delta_\pm-3}=k^{\pm2\ii\wt\nu}$. Now, by going to a squeezed limit, we are effectively pushing the massive bulk propagator to the boundary, and therefore we do expect to see the $k^{\pm2\ii\wt\nu}$ behavior in the correlation function, which is exactly the nonlocal signal. Incidentally, there is also a local part of the two-point function $\propto \de(\mb x)$, which arises from the two-point correlator of $\Delta_\pm$ with its shadow counterpart $\Delta_\mp$ \cite{Sun:2021thf}. This local part leads to the local signal in (\ref{eq_4ptTree}). Let us emphasize that we are not assuming any CFT dual of the bulk theory. All we need is the kinematic information of the boundary correlators which is constrained by the conformal symmetry.

\end{enumerate}

\subsection{Nonlocal signals in tree graphs}
\label{sec_tree}

The above observations for the nonlocal signal in a 4-point correlator is readily generalized to arbitrary tree graphs. As we have shown, the essence of a nonlocal signal is that an intermediate massive propagator becomes soft relative to all other propagators it connects to, so that a late-time expansion of this soft propagator can be performed, yielding a nonanalytic term in the soft momentum. In \cite{Qin:2022fbv}, it was shown using the partial MB representation that this late-time limit of the internal soft propagator is the only source of the nonlocal signal. We shall generalize these arguments to arbitrary loop graphs in the subsequent sections. Now, focusing on tree graphs, the momentum of any internal line is fully controlled by the external momenta, and therefore, we can adjust the external momenta to make any internal line soft. This makes the computation of nonlocal signals rather straightforward. 

In fact, if we adopt the late-time expansion of the soft internal propagator, we can directly work out an explicit expression for its nonlocal signal. So, let us consider an arbitrary tree graph, containing an internal massive scalar line $D_{\aa\bb}(P;\tau_1,\tau_2)$ with soft momentum $P$ and two time variables $\tau_{1,2}$. We assume there are $I_L$ ($I_R$) additional tree internal lines, denoted by $D_{\aa\aa_i}$, $i=1,\cdots,I_L$ ($D_{\bb\bb_j}$, $j=1,\cdots,I_R$), connected to the vertex at $\tau_1$ ($\tau_2$). There may also be a number of bulk-to-boundary propagators connected to the vertex at $\tau_1$ ($\tau_2$), with total incoming energy being $k_L$ ($k_R$). Then, the relevant part of the SK integral for this tree graph can be written as:
\begin{align}
\label{eq_treeInt}
  \mathcal{G}_\text{tree}=&\sum_{\aa,\bb=\pm}(\ii\aa)(\ii\bb)\int_{-\infty}^0\di\tau_1\di\tau_2(-\tau_1)^{p_1}(-\tau_2)^{p_2}D_{\aa\bb}(P;\tau_1,\tau_2)e^{+\ii \aa k_L\tau_1+\ii \bb k_R\tau_2}\n\\
  &\times\prod_{i=1}^{I_L}\Big[D_{\aa\aa_i}(p_i;\tau_1,\tau_{Li})\Big]\prod_{j=1}^{I_R}\Big[D_{\bb\bb_j}(q_j;\tau_1,\tau_{Lj})\Big]\cdots.
\end{align}
Terms not explicitly shown and denoted by ``$\cdots$'' are irrelevant to our current argument.
At this point we have to introduce the explicit form of the bulk massive scalar propagator. The two Wightman functions $D_{\mp\pm}(k;\tau_1,\tau_2)$ are given by \eqref{eq_Dmp} and \eqref{eq_Dpm}:
\begin{align}
\label{eq_Dmp}
 D_{-+}^{(\wt\nu)}(k;\tau_1,\tau_2)
 =&~\FR{\pi}{4}e^{-\pi\wt\nu}(\tau_1\tau_2)^{3/2}\mathrm{H}_{\ii\wt\nu}^{(1)}(-k\tau_1)\mathrm{H}_{-\ii\wt\nu}^{(2)}(-k\tau_2),\\
\label{eq_Dpm}
 D_{+-}^{(\wt\nu)}(k;\tau_1,\tau_2)
 =&~\Big[D_{-+}^{(\wt\nu)}(k;\tau_1,\tau_2)\Big]^*,
\end{align}
where $\text{H}_\nu^{(j)}(z)$ is the Hankel function of $j$'th kind with $j=1,2$,
and the two (anti-)time-ordered propagators are constructed out of the two Wightman functions and given by \eqref{eq_Dpmpm}:
\begin{align}
\label{eq_Dpmpm}
  D_{\pm\pm}^{(\wt\nu)}(k;\tau_1,\tau_2)=&~D_{\mp\pm}^{(\wt\nu)}(k;\tau_1,\tau_2)\theta(\tau_1-\tau_2)+D_{\pm\mp}^{(\wt\nu)}(k;\tau_1,\tau_2)\theta(\tau_2-\tau_1).
\end{align}
These are complicated expressions. However, as mentioned above, so far as the nonlocal signal is the only concern, it is valid to expand the soft propagator $D_{\aa\bb}(P;\tau_1,\tau_2)$ in the late-time limit:
\begin{align}
\label{eq_DNL}
  \lim_{\tau_1,\tau_2\to 0}D_{\aa\bb}^{(\wt\nu)}(P;\tau_1,\tau_2)\simeq \FR{(\tau_1\tau_2)^{3/2}}{4\pi}\bigg[\Gamma^2(-\ii\wt\nu)\Big(\FR{P^2\tau_1\tau_2}4\Big)^{\ii\wt\nu}+\Gamma^2(\ii\wt\nu)\Big(\FR{P^2\tau_1\tau_2}4\Big)^{-\ii\wt\nu}\bigg]+\cdots.
\end{align}
Here we have ignored the terms that are analytic in $k$ as $k\to 0$, since we are only interested in the nonanalytic part in the final result. Importantly, the nonanalytic part of the propagator in the soft limit $k\to 0$ is manifestly real. As a result, this part is independent of the two SK indices, and thus independent of the time ordering. This is in nice agreement with the aforementioned physical argument: The two endpoints of a soft propagator are in space-like separation, where the time ordering is irrelevant. Technically, it means that the two time integrals over $\tau_1$ and $\tau_2$ are factorized. This is the essence of the cutting rule for inflation correlators. 

\begin{figure}
\centering 
\includegraphics[width=\textwidth]{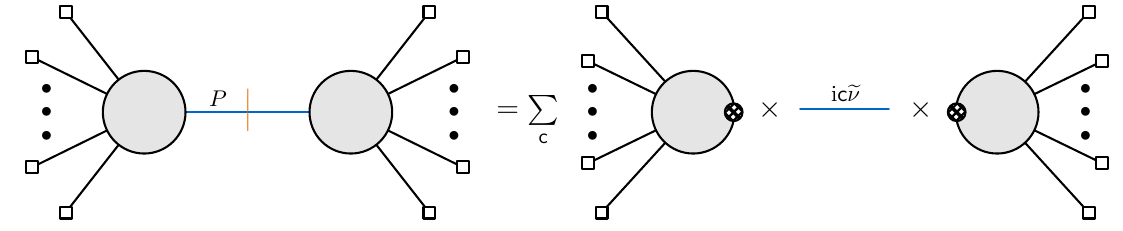}  
\caption{The factorization of the nonlocal signal from a tree propagator in a general inflation correlator in the soft limit $P\to 0$. The orange cut on the left graph means to take the nonanalytic part of the soft (blue) propagator. On the right hand side, the three graphs correspond to the left subgraph, the tree signal, and the right subgraph. The two hatched vertices on the left/right subgraphs denote the complex-power-of-time insertions, i.e., the factors in blue in (\ref{eq_treeSignal}).} 
  \label{fig_treecut}
\end{figure}

Now, substituting (\ref{eq_DNL}) into the SK integral (\ref{eq_treeInt}), we get the nonanalytic part of the tree graph in the limit $P\to 0$:
\begin{align}
\label{eq_treeSignal}
  &\lim_{P\to 0}\big[\mathcal{G}_\text{tree}\big]_\mathrm{NL}
  =\sum_{\cc=\pm}\bigg\{\sum_{\aa=\pm} \ii\aa\int_{-\infty}^0\di\tau_1{\blue(-\tau_1)^{p_1+3/2+\ii\cc\wt\nu}}e^{\ii\aa k_L\tau_1}\prod_{i=1}^{I_L} D_{\aa\aa_i}(p_i;\tau_1,\tau_{Li}) \cdots\bigg\}\n\\
   &\times\bigg\{\sum_{\bb=\pm} \ii\bb\int_{-\infty}^0\di\tau_2{\blue(-\tau_2)^{p_2+3/2+\ii\cc\wt\nu}}e^{\ii\bb k_R\tau_2}\prod_{j=1}^{I_R} D_{\bb\bb_j}(q_j;\tau_2,\tau_{Ri}) \cdots\bigg\} \FR{\Gamma^2(-\ii\cc\wt\nu)}{4\pi}\Big(\FR{P}2\Big)^{2\ii\cc\wt\nu}.
\end{align}
This shows that, in the soft limit $P\to 0$, the tree diagram possesses a nonlocal signal $\propto P^{\pm2\ii\wt\nu}$. The signal factorizes into three pieces: a left subgraph and a right subgraph, given respectively by the expressions in the first and the second curly brackets in (\ref{eq_treeSignal}), as well as a nonanalytic ``signal,'' given by $\fr1{4\pi}\Gamma^2(\mp\ii\wt\nu)(P/2)^{\pm2\ii\wt\nu}$. This can be viewed as an on-shell factorization theorem for inflation correlator at the tree level, which is very similar to the on-shell factorization of scattering amplitudes discussed previously. We illustrate the factorization of nonlocal signal in Fig.\ \ref{fig_treecut}.

The above result can be pushed to all orders in $k_s$. In principle, we can simply do the late-time expansion of the soft propagator to all orders in $k_s$ and keep the nonanalytic part as in (\ref{eq_DNL}). In practice, we find it more straightforward to work with the partial Mellin-Barnes representation. In this formalism, the leading signal (\ref{eq_treeSignal}) corresponds to the residue of the leading IR poles in the Mellin variables, while the higher order terms can be found by including residues from more IR poles. (Don't worry about the jargon; More explanations will be given in subsequent sections.) The result is:
\begin{align}
  \big[\mathcal{G}_\text{tree}\big]_\mathrm{NL}=\sum_{\cc=\pm}\sum_{n_1,n_2=0}^\infty \mathcal{G}^\text{(L)}_{n_1}\mathcal{G}^\text{(R)}_{n_2}\FR{(-1)^{n_{12}}}{n_1!n_2!}\Gamma\Big[n_1-\ii\cc\wt\nu,n_2-\ii\cc\wt\nu\Big]\Big(\FR{P}{2}\Big)^{2\ii\cc\wt\nu+2n_{12}}.
\end{align}
Here the left and right subgraphs $\mathcal{G}_{n_1}^\text{(L)}$ and $\mathcal{G}_{n_2}^\text{(R)}$ are obtained from the leading order results in (\ref{eq_treeSignal}), by replacing the factor $(-\tau_1)^{p_1+3/2+\ii\cc\wt\nu}\to(-\tau_1)^{p_1+3/2+\ii\cc\wt\nu+2n_1}$ and $(-\tau_2)^{p_2+3/2+\ii\cc\wt\nu}\to (-\tau_2)^{p_2+3/2+\ii\cc\wt\nu+2n_2}$, respectively.

The tree-level result in (\ref{eq_treeSignal}) can also be easily generalized to arbitrary loop graphs so long as we are cutting a tree line, because the momentum of a tree line can always be controlled by the external momenta. One may find it slightly uncomfortable if the soft internal line is connected to loops at the two sides, as in this case the adjacent bulk propagators in the loop do not have definite momenta, and the resonance argument in the last subsection does not immediately apply to this case. However, so long as one still accepts that the nonlocal signal comes entirely from the late-time limit of the soft tree propagator, one can still derive the on-shell factorization of the nonlocal signal following exactly the same steps to get (\ref{eq_treeSignal}). As mentioned above, the validity of this late-time expansion can be more rigorously justified using the partial MB representation. This generalization to arbitrary loop graphs when cutting a tree line is implicitly included in Fig.\ \ref{fig_treecut}: The two blobs can contain arbitrary loop structures. 
 
It is when we try to extract nonlocal signals in loop propagators in a general loop graph that we find a real need to go beyond the simple late-time expansion, and to develop more advanced techniques. As we showed in previous works \cite{Qin:2022lva,Qin:2022fbv,Qin:2023bjk}, the partial MB representation is suitable for this task. So, the the subsequent sections, we will use this method to develop a general algorithm and a theorem to detect and also to compute leading nonlocal signals in arbitrary loop graphs. Obviously, many technical details are involved. To help the readers keep track of what is going on, we will devote the subsection below to a brief discussion of the basic physical picture of nonlocal signals in loop diagrams.

\subsection{Nonlocal signals in loop graphs}

Intuitively, we can imagine that a loop inflation correlator also possesses nonlocal signals when some of its loop propagators become soft. In such situations, a late-time expansion of the soft loop propagator would be possible, and one can again generate oscillatory signals through the interference between the late-time massive modes and the effectively massless modes at an interaction vertex. However, here we immediately face a complication: The momentum of a loop propagator is not uniquely fixed by the choice of external momenta. Even if we can set up a momentum configuration such that the momentum transfer from one part of the graph to the other part is soft, the momentum of each individual loop line is not guaranteed to be soft. Given that the loop momentum can in principle be arbitrarily high, it is not immediately justified that we can make a late-time expansion to any internal loop propagator. 

More careful analysis with partial MB representation for 1-loop graphs reveals that, for the purpose of generating a nonlocal signal from a loop process, it is \emph{not} essential to make any loop propagator soft. Instead, what is important is that the external momentum configuration should be suitably chosen, such that the sum of momenta of at least two loop propagators is forced to approach zero \cite{Qin:2023bjk}.

This observation can be generalized to graphs with an arbitrary number of loops. The result is similar: To generate a nonlocal signal, one should take a soft limit of the external momenta configuration, such that the sum of momenta in several loop propagators is forced to approach zero. In this case, we can always make use of the freedom of redefining loop momentum variables to make all these loop lines soft simultaneously. Consequently, there is always a parameter space where all these loop lines go on shell in the sense described around \eqref{eq_Dlate}. It is trivial to see that, when all these loop lines become soft with their momentum approaching zero, we can simply remove them without spoiling the momentum conservation in the whole graph. That is, we can take a \emph{cut} of graph. Of course, there are additional technical requirements for this cut to generate a nonlocal signal which will be detailed later, but this simple line of reasoning forms the intuitive basis behind our signal-detection algorithm to be introduced in Sec.\ \ref{sec_detect}. 

\begin{figure}
\centering 
\includegraphics[width=\textwidth]{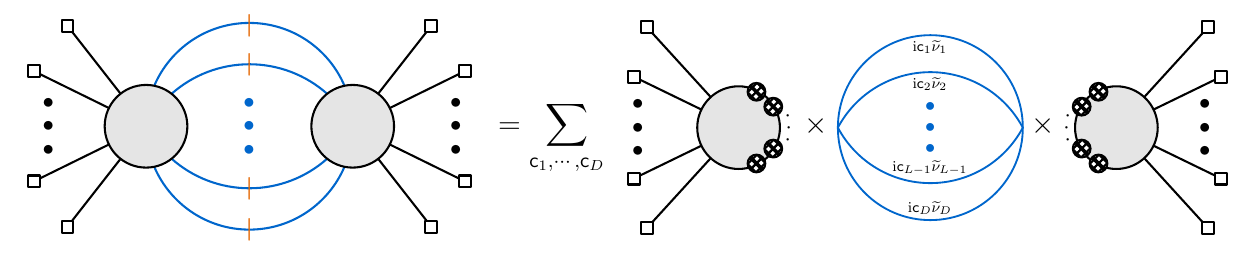}  
\caption{An illustration of the factorization theorem \eqref{eq_gfacUMC} of the nonlocal signal in an arbitrary loop level inflation correlator in the soft limit $P\to 0$. The orange cut on the left graph means to take the nonanalytic part of the soft (blue) propagator. On the right hand side, the three graphs correspond to the left subgraph \eqref{eq_TL}, the melon signal \eqref{eq_frakM}, and the right subgraph \eqref{eq_TR}. The hatched vertices on the left/right subgraphs denote the complex-power-of-time insertions.}
  \label{fig_nloopcut}
\end{figure}

In short, our signal-detection algorithm says that we take all possible nonlocal cuts of a  graph with respect to a given bipartition. Then, each cut is a candidate of the nonlocal signal in the sense that the parameter region with all cut lines getting soft could (though not always) generate a complex-power contribution to the graph. It still remains to find an explicit expression for the nonlocal signal from a given cut. We explore this topic in Sec.\ \ref{sec_bulkfree} and Sec.\ \ref{sec_arbitrarygraph}, where we present and prove various versions of on-shell factorization theorem for general loop graphs, where the signal generated from a $(D-1)$-loop cut can be analytically computed, and is called a ``melon signal,'' as shown in Fig.\ \ref{fig_nloopcut}. This is exactly the loop version of tree-level factorization shown in (\ref{eq_treeSignal}) and in Fig.\ \ref{fig_treecut}.  

It may come as a surprise that the phenomenon of on-shell factorization appears at loop levels for inflation correlators. In flat-space scattering amplitude, putting loop lines on shell yields branch cuts rather than poles. The optical theorem then relates the discontinuity of this branch cut to a total cross section, where we need to sum over all possible final states. This summation over states shows that the factorization no longer holds when cutting loop lines. The situation for the inflation correlator is in fact similar: We also need to sum over all possible ``intermediate states,'' which correspond to infinite towers of modes with definite scaling dimensions. For a melon graph of $D$ lines, this scaling dimension can be organized as $3(D-1)/2+\ii\sum\limits_{\ell}\cc_\ell\wt\nu_\ell+n$. However, an important point here is that, in the squeezed limit, there is only a finite number of terms with $n=0$ are dominant, which we call the leading signals. Then, each term in the leading signal is factorized.

\section{Detecting Nonlocal Signals in Arbitrary Loop Graphs}
\label{sec_detect}

\subsection{Defining nonlocal signal}
\label{sec_def}

Above we have explained the occurrence of nonlocal signals in a given inflation correlator in a heuristic way. With this physical picture in mind, now we are going to develop the formalism to systematically investigate the nonlocal signals. The first task is to precisely define what we mean by a nonlocal signal in an arbitrary inflation correlator. This is the main topic of the current subsection.

\paragraph{Basic setup.} We first introduce the basic setup. We shall begin with a $B$-point inflation correlator, by which we mean a correlation function of $B$ boundary operators $\varphi^{(i)}$ $(i=1,\cdots,B)$. To make the nonlocal signal distinct from local signals, we require $B\geq 4$. (See comments below (\ref{eq_4ptTree}).) The superscript $(i)$ denotes the species of the boundary operator. Thus these $B$ operators can be either identical or distinct. For CC applications, these boundary operators correspond to bulk fields which have nonvanishing boundary limit, such as a (nearly) massless scalar field, including the inflaton fluctuation, or the massless spin-2 graviton. For theoretical investigations, one also frequently considers the case where $\varphi$ is a conformal scalar, namely a scalar of effective mass $m^2=2$ in dS with 3 spatial dimensions. A conformal scalar dies away at the future boundary $\tau\to 0$ as $\varphi\propto |\tau|$, and therefore it does not immediately correspond to a realistic CC scenario. However, cases with conformal-scalar external modes are often simplest from a technical viewpoint, and conformal-scalar correlators are interesting theoretical objects on their own right.

We shall assume that the translation and rotation of the 3-dimensional spatial slices are kept by all classical backgrounds in the problem, and therefore it is beneficial to work in the 3-momentum space. Let us work in the Fourier space and assume that each of the $B$ operators $\varphi_{\mb k_i}^{(i)}$ carries a fixed 3-momentum $\mb k_i$. Clearly, the total momentum should be conserved, so the $B$-point correlator should be proportional to an overall $\de$-function of total momentum conservation. So, we can write the $B$-point correlator as:
\bge
  \Big\la \varphi_{\mb k_1}^{(1)}\cdots \varphi_{\mb k_B}^{(B)}\Big\ra=(2\pi)^3\de^{(3)}\Big(\sum_{i=1}^B\mb k_i\Big) \mathcal{T}\Big(\{\mb k\}\Big)+\text{disconnected contributions}.
\ede
We assume that the cluster decomposition holds for this correlator: The function $\mathcal{T}(\{\mb k\})$ does not contain further momentum-conserving $\de$-function factors. All such factors are classified as ``disconnected contributions.'' 

In this work, we shall always assume that the bulk theory is a weakly coupled quantum field theory, such that the weak coupling expansion is valid at least as an asymptotic series. In such cases, we can approximate the amplitude $\mathcal{T}(\{\mb k\})$ as a sum of connected graphs truncated at a finite order in the number of loops:
\bge
  \mathcal{T}(\{\mb k\})\simeq \sum\mathcal{G}(\{\mb k\}).
\ede
Here $\mathcal{G}(\{\mb k\})$ is the amplitude of an individual graph. Below we do not distinguish between the graph and the associated amplitude, and we shall directly call $\mathcal{G}(\{\mb k\})$ a graph without causing any confusion. Given the perturbativity of the problem, it is legitimate to work with individual graph $\mathcal{G}(\{\mb k\})$ instead of the total amplitude $\mathcal{T}(\{\mb k\})$. The analysis below will exclusively focus on individual graphs. The readers should keep in mind that the full result of the amplitude should be the sum of all diagrams to a desired order in the perturbation theory.

In this work, we shall mostly focus on the case where the internal lines of a graph $\mathcal{G}(\{\mb k\})$ propagate scalars $\si$ of arbitrary mass $m>3/2$. The couplings among these scalars and with the boundary operator $\varphi$ are assumed to be local, in the sense that the spatial or temporal derivatives only appear as polynomials of finite degree.

\paragraph{Nonlocal bipartition and nonlocal soft limit.} The nonanalyticity we shall consider in this work arises from a particular soft limit where the sum of a subset of external momenta goes to zero. To characterize this type of soft limits more precisely, we introduce the concept of \emph{nonlocal bipartition} of the graph and the corresponding \emph{nonlocal soft limit}. 

Let $\mathcal{G}(\{\mb k\})$ be a graph of $B\geq 4$ external points and $L$ independent loops as prescribed above. Here $\{\mb k\}$ denotes the collection of all 3-momenta of the $B$ external legs. Then, a \emph{nonlocal bipartition} of the graph $\mathcal{G}$ means a separation of the set $\{\mb k\}$ into two disjoined subgroups, $\{\mb k^\mathrm{(L)}\}$ and $\{\mb k^\mathrm{(R)}\}$. That is, we require that $\{\mb k^\mathrm{(L)}\}\cup\{\mb k^\mathrm{(R)}\}=\{\mb k\}$ and $\{\mb k^\mathrm{(L)}\}\cap\{\mb k^\mathrm{(R)}\}=\varnothing$. Let $\{\mb k^\mathrm{(L)}\}$ and $\{\mb k^\mathrm{(R)}\}$ have $B_L$ and $B_R$ elements, respectively. Then, as a part of the definition for the nonlocal bipartition, we demand $B_L\geq 2$ and $B_R\geq 2$. Clearly, $B=B_L+B_R$, and therefore, a nonlocal bipartition is possible only when $B\geq 4$. As discussed in the previous section, this assumption is necessary in order to identify nonlocal signals from local signals.

We then introduce the concept of \emph{partial sum} of external momenta. By a partial sum, we mean a linear combination:
\bge
\label{eq_partialSum}
  \wt{\mb P}_i=\sum_{j=1}^B \be_{ij} \mb k_j,
\ede
where the coefficients $\be_{ij}$ take values from $\{0,+1\}$, with the requirement that at least one $\be_{ij}=0$ for any given $i$. We shall denote the set of all partial sums of the external momenta by $\{\wt{\mb P}\}$.

Now, we are ready to introduce the concept \emph{nonlocal soft limit}. By a nonlocal soft limit associated with the nonlocal bipartition $\{\mb k^\mathrm{(L)}\}\cup\{\mb k^\mathrm{(R)}\}$, we mean the limit specified by the following two conditions:
\begin{enumerate}
  \item The partial sum $\mb P\equiv\sum\mb k^\mathrm{(L)}\to\mb 0$ (which then automatically implies $\sum\mb k^\mathrm{(R)}=-\mb P\to\mb 0$ due to the total momentum conservation);
  \item All external momenta $\mb k_j$ and all partial sums $\wt{\mb P}_i$ except $\pm\mb P$ remain finite in the physical region.
\end{enumerate}
Thus, by a nonlocal soft limit, we are considering a particular partial sum $\mb P$ being much softer than any external momenta and any other independent partial sums of them. That is, we only consider one soft configuration at a time, and this turns out to be a useful strategy to isolate nonlocal signals of the graph and analyze them one by one. Below whenever we mention a nonlocal soft limit $P\to 0$, we always mean that the above two conditions are met. 

\paragraph{Nonlocal signal.} Now we introduce our definition of the nonlocal signal.
As we shall see below, schematically, in a given nonlocal soft limit $P\to 0$, the graph $\mathcal{G}$ in general has the following behavior: 
\begin{align}
\label{eq_NLSignalDef}
  \lim_{P\to 0}\mathcal{G}(\{\mb k\}) = \sum_{\ell=1}^{C}\sum_{\bb=\pm}\mathcal{A}_{\bb|\ell}(\{\mb k\})P^{\al_\ell+\ii\bb\omega_\ell}+\text{terms analytic in $P$}.
\end{align}
Here $\mathcal{A}_{\bb|\ell}(\{\mb k\})$ is a generally complex amplitude analytic and finite at $P=0$. So, it allows a Taylor expansion in $P$ with a nonvanishing $\order{P^0}$ term. $\al_\ell$ and $\omega_\ell$ are both real and positive numbers. The complex exponent $\al_\ell\pm\ii\omega_\ell$ then implies the existence of a branch cut on the negative real axis of $P$, and also a logarithmic oscillatory signal on the real positive axis. This kind of nonanalyticity is the central topic of this work. As is made clear, different terms in the $\ell$-summation are distinguished by different values of $\al_\ell\pm\ii\omega_\ell$, and we shall call each pair of terms with $\bb=\pm$ in the $\ell$-summation a \emph{nonlocal signal}. We will sometimes just call it a signal when no confusion could arise. This is clearly a generalization of the nonlocal signal defined for a tree diagram in the last section.

It can be shown that the total number $C$ of nonlocal signals in a given nonlocal soft limit of a given graph is finite. Let us organize these $C$ signals such that $\al_1\leq\al_2\leq\cdots\leq\al_C$. Clearly, the nonlocal signal with the smallest $\al_\ell$ gives the leading contribution. We shall call such signal(s) the \emph{leading signal}. It is possible that the leading signals are not unique. For instance, we may have $\al_1=\al_2<\al_3<\cdots$. A major task of this work is to work out as explicitly as possible the form of the leading signal in a given nonlocal soft limit of an arbitrary loop graph. Below, we will first propose an algorithm to ``locate'' the origin of nonlocal signals. Then, in subsequent sections, we shall refine our result by giving more explicit expressions for the nonlocal signals for various special cases.

\subsection{Signal detection algorithm}
\label{sec_algorithm}

Given the complexity of an arbitrary loop diagram, not all nonlocal soft limits contain nonlocal signals. Thus, it would be useful first to have a practical algorithm to detect and locate all nonlocal signals in a given soft limit. We shall now present such an algorithm. At the end of this subsection, we shall introduce a theorem of signal detection, showing that our algorithm is exhaustive.

\paragraph{Nonlocal cut and its degree.} The key concept of our algorithm is the nonlocal cut. Given a bipartition $\{\mb k^\mathrm{(L)}\}\cup\{\mb k^\mathrm{(R)}\}$ of the graph $\mathcal{G}(\{\mb k\})$, we define a \emph{nonlocal cut} of the graph with respect to this bipartition to be the following operation on the graph: We remove some of the \emph{bulk} propagators from the graph such that the resulting cut graph satisfies the two conditions: 
\begin{enumerate}
  \item The left set $\{\mb k^\mathrm{(L)}\}$ is totally disconnected from the right set $\{\mb k^\mathrm{(R)}\}$;
  \item The left set $\{\mb k^\mathrm{(L)}\}$ itself is fully connected, so is the right set $\{\mb k^\mathrm{(R)}\}$.
\end{enumerate}
A removed line in this operation is called a \emph{cut line}, and the number of lines removed in the operation is called the \emph{degree} of the cut.

\paragraph{Irreducible cut and minimal cut.}
It is useful in the following analysis to introduce the concept of the irreducible cut and the minimal cut. 

\begin{figure}
\centering 
\includegraphics[width=\textwidth]{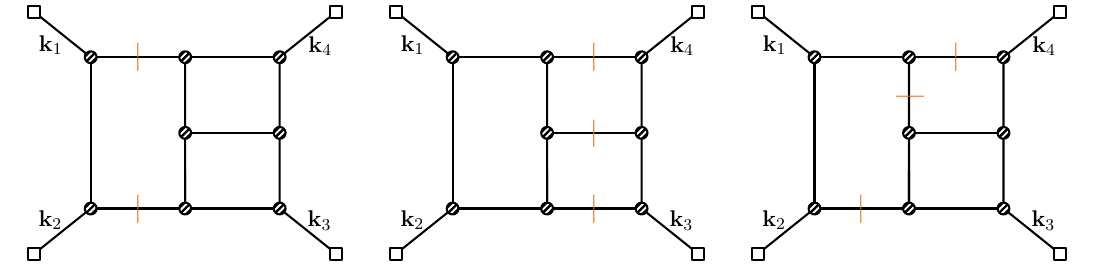}  
\caption{Three (out of many) valid nonlocal cuts associated with the bipartition $\{\mb k_1,\mb k_2\}\cup\{\mb k_3,\mb k_4\}$. All are irreducible but only the left one is minimal.} 
  \label{fig_tribox}
\end{figure}

By an \emph{irreducible cut}, we mean a nonlocal cut which ceases to be a nonlocal cut if we put any one cut line back into the cut graph. Clearly, a nonlocal cut is either irreducible or reducible, and any nonlocal cut can be made irreducible by putting some cut line back. However, we note that the resulting irreducible might not be unique. There may be many distinct ways to reduce a nonlocal cut, and they may end up with distinct irreducible cuts. For instance, in Fig.\ \ref{fig_tribox}, we can get a nonlocal cut by removing all horizontal internal lines. This nonlocal cut can be reduced to either of the two distinct irreducible cuts in the left and middle diagrams of Fig.\ \ref{fig_tribox}.

 By a \emph{minimal cut}, we mean a nonlocal cut of the smallest degree for a given nonlocal bipartition. Clearly, a minimal cut is irreducible, but an irreducible cut does not have to be minimal. See Fig.\ \ref{fig_tribox}. Also, we note that the minimal cuts may not be unique for a given bipartition. For example, in the double triangle diagram in Fig.\ \ref{fig_2triangle}, we can cut the two skew lines in either of the left and right triangle loops. Each of them is a minimal cut. However, if there is a unique irreducible cut for a given nonlocal bipartition, then it is automatically a minimal cut, and it is also the unique minimal cut for the given bipartition.

\paragraph{Signal detection algorithm.}
Now we are ready to present one main result of this work, namely an algorithm to detect all possible nonlocal signals in an arbitrary graph. We state it as the following theorem.

\begin{theorem}[Signal detection algorithm]\label{thm_signal}
Let $\mathcal{G}(\{\mb k\})$ be an arbitrary $B$-point ($B\geq 4$) and $L$-loop $(L\geq 0)$ graph as specified in the last subsection, let $\{\mb k^\mathrm{(L)}\}\cup \{\mb k^\mathrm{(R)}\}$ be a nonlocal bipartition of $\mathcal{G}(\{\mb k\})$,  and let $\mb P\equiv\sum\mb k^\mathrm{(L)}$. It is further assumed that the graph $\mathcal{G}(\{\mb k\})$ is properly regularized in the ultraviolet (UV) region, and all the bulk propagators are massive with arbitrary (possibly dS-boost-breaking) dispersions. Then, the following two statements hold:
\begin{enumerate}

\item The graph is an analytic function of $P$ in the interior of the physical region. 

\item In the limit $P\to 0$, every nonlocal cut with respect to the nonlocal bipartition $\{\mb k^\mathrm{(L)}\}\cup \{\mb k^\mathrm{(R)}\}$ gives rise to a \emph{candidate} of nonlocal signal when all cut lines become soft simultaneously.
\end{enumerate}
\end{theorem}
The first part of this theorem ensures that no nonanalytic behavior appears in the interior of the physical region. So, to find a nonlocal signal, we do need to consider a soft limit. Then, in the second part, the theorem tells us that the nonlocal signal when $P\to 0$ can be exhausted by enumerating all possible nonlocal cuts for a given nonlocal bipartition of the graph. When all the cut lines of a nonlocal cut become soft simultaneously, they could (though not always) contribute a nonlocal together. Whether such soft configurations really lead to nonlocal signals also depends on other details of the graph such as interaction types. Examples of nonlocal cuts with or without nonlocal signals will be given in Sec.\ \ref{sec_example}. So, the existence of nonlocal cuts is a necessary but not sufficient condition for nonlocal signals. Also, the above theorem only tells us where nonlocal signals could possibly appear; It does not give us an explicit expression for these signals. All these remaining questions will be addressed in the following two sections.

\subsection{Proof of the signal detection theorem} 
\label{sec_algorithm_proof}

Now we are going to prove Theorem \ref{thm_signal}. The proof is somewhat lengthy. So, we provide an outline of the proof here, to give readers a quick idea of what is going on:
\begin{enumerate}
  \item We first specify the diagram we are going to analyze, and then introduce the key tool, namely the partial MB representation. In one sentence, the partial MB representation means that we use the MB representation for all the bulk propagators, but leave all bulk-to-boundary propagators unchanged. The partial MB representation allows us to isolate a relatively simple loop momentum integral, which encodes all the dependences on the momentum partial sums. Therefore, to analyze the analyticity of the graph as functions of momentum partial sums, we can focus on the loop integral only.
  
  \item We then present a \emph{lemma of degenerate singularity}, which asserts that all possible nonanalytic behavior of the loop integral as a function of momentum partial sums must be from a region where two or more internal propagators carry linearly dependent momenta and these momenta all become soft simultaneously. We call such regions in the loop integral space \emph{degenerate region}, and the resultant nonanalytic behavior the \emph{degenerate singularity}. Phrased in another way, a necessary condition for the existence of a nonlocal signal is the existence of a degenerate region.
    
  \item We then show that each degenerate region of the loop integral space leads to a nonlocal cut and vice versa. Thus, the existence of a nonlocal cut is a necessary condition of a nonlocal signal. Therefore, by enumerating all possible nonlocal cuts with a given bipartition, we are guaranteed to get all possible nonlocal signals in the corresponding soft limit. 
\end{enumerate}

Below we carry out these steps in detail.

\paragraph{Specification of the graph.} We assume that the graph $\mathcal{G}(\{\mb k\})$ has $B$ external legs, $I$ internal legs, $V$ vertices, and $L$ independent loops. For definiteness, we take the external legs to be the bulk-to-boundary propagators of a conformal scalar with $m^2=2$:
\bge
\label{eq_CSProp}
  C_\aa(k;\tau)=\FR{\tau\tau_f}{2k}e^{\aa\ii k\tau},
\ede
On the other hand, we take all of the $I$ internal lines to be bulk propagators of massive scalar in principal series, i.e., $m_i>3/2$ $(i=1,\cdots, I)$. For these so-called principal scalars we define the mass parameter $\wt\nu\equiv\sqrt{m^2-9/4}$ which is a positive real number. Then, the SK bulk propagators of a principal scalar with mass parameter $\wt\nu$ are given as in (\ref{eq_Dmp})-(\ref{eq_Dpmpm}). Below, we shall also call the two Wightman functions $D_{\mp\pm}$ in (\ref{eq_Dmp}) and (\ref{eq_Dpm}) the \emph{homogeneous propagators}, since they are solutions to the homogeneous scalar equation of motion. On the other hand, the Feynman and anti-Feynman propagators $D_{\pm\pm}$ in (\ref{eq_Dpmpm}) are also called \emph{inhomogeneous propagators}, as they are solutions to scalar field equation with $\de$-function source.

The interaction vertices can be taken to be arbitrary local couplings for the validity of the theorem, which can have arbitrary polynomial dependences on the space and time derivatives, and also general polynomial dependences on the time variable of the vertex, so long as the integral is well behaved in the late-time limit.\footnote{However, there is a subtlety when spatial derivatives act on a bulk mode. In Fourier space, the action of a spatial derivative becomes the momentum of the propagator and thus will contribute to extra (polynomial) factors in the Mellin-space loop integral \eqref{eq_MellinLoopInt}. Furthermore, if the spatial derivatives act on a cut line, it will make the signal decrease faster, which should be carefully treated when picking out the dominant signal.} If the theory possesses the dilation symmetry, then the explicit time dependence of an interaction vertex is uniquely fixed. Purely for notational simplicity, we shall take all interactions to be direct couplings with arbitrary power dependences on $\tau_i$, namely $(-\tau_i)^{p_i}$ for the $i$'th vertex. Again, it is understood that $p_i$ is always chosen such that the integral is well behaved in the late-time limit. 

With all the propagators and interactions specified, we can write down the SK integral for the graph $\mathcal{G}(\{\mb k\})$ as:
\begin{align}
\label{eq_Tk}
  \mathcal{G}(\{\mb k\})
  =&\sum_{\aa_1,\cdots,\aa_V=\pm}\int_{-\infty}^0\prod_{i=1}^V\Big[ \ii\aa_i\di \tau_i(-\tau_i)^{p_i}\Big]\prod_{j=1}^{B}\Big[C_{\aa_j}(k_j;\tau_j)\Big]\n\\
  &\times\int\prod_{k=1}^L\bigg[\FR{\di^3\mb q_k}{(2\pi)^3}\bigg]\prod_{\ell=1}^I\Big[ D_{\aa_{\ell 1}\aa_{\ell2}}(p_\ell;\tau_{\ell1},\tau_{\ell2})\Big].
\end{align}
Here $\aa_i=\pm$ $(i=1,\cdots,V)$ is the SK index of the $i$'th vertex. We take the $L$ independent loop momentum variables to be $\mb q_k$ $(k=1,\cdots,L)$. The momentum flowing in the $\ell$'th bulk propagator is denoted as $\mb p_\ell$. Also, in the bulk propagator $D_{\aa_{\ell 1}\aa_{\ell2}}(p_\ell;\tau_{\ell1},\tau_{\ell2})$, the two time variables $\tau_{\ell 1},\tau_{\ell 2}$ should be identified with the time variables of the two vertices to which the propagator is attached, respectively, so are the two SK indices. The same identifications are also made for the time variables and SK indices of the bulk-to-boundary propagators $C_{\aa_j}(k_j;\tau_j)$.

\paragraph{Partial Mellin-Barnes representation.}
The basic tool of the proof is the partial MB representation introduced in \cite{Qin:2022lva,Qin:2022fbv}. In short, the partial MB representation suggests that we use the MB representation for all the bulk propagators. In particular, for the two homogeneous propagators, we have
\begin{align}
\label{eq_DScalarMB1}
    D_{\pm\mp}^{(\wt\nu)}(k;\tau_1,\tau_2) =&~ \FR{1}{4\pi}
    \int_{-\ii\infty}^{+\ii\infty}
    \FR{\di s}{2\pi\ii}\FR{\di \bar s}{2\pi\ii}\,
    e^{\mp\ii\pi(s-\bar s)}\Big(\FR{k}2\Big)^{-2(s+\bar s)}
    (-\tau_1)^{-2s+3/2}(-\tau_2)^{-2\bar s+3/2}\n\\
    &\times \Gamma\Big[s-\FR{\ii\wt\nu}2,s+\FR{\ii\wt\nu}2,\bar s-\FR{\ii\wt\nu}2,\bar s+\FR{\ii\wt\nu}2\Big],
\end{align} 
and the partial MB representation for the two inhomogeneous propagators can be obtained by substituting (\ref{eq_DScalarMB1}) into (\ref{eq_Dpmpm}). 

After adopting the partial MB representation for all bulk propagators, the graph (\ref{eq_Tk}) takes the following form:
\begin{align}
\label{eq_TkMellin}
  \mathcal{G}(\{\mb k\})=\sum_{\aa_1,\cdots,\aa_V=\pm}\int_{-\ii\infty}^{+\ii\infty}\prod_{i=1}^I\bigg[\FR{\di s_i}{2\pi\ii}\FR{\di\bar s_i}{2\pi\ii}\mathbb{D}(s_i,\bar s_i)\bigg]\mathbb{T}\Big(\{k\};\{s,\bar s\}\Big)\mathbb{L}\Big(\{\mb k\};\{s,\bar s\}\Big).
\end{align}
Here we have used $(s_i,\bar s_i)$ as a pair of Mellin variables for the $i$'th internal propagator, and we have suppressed the SK-index dependences in the integrand for clarity. The integral then breaks into several pieces. The Mellin-space time integral $\mathbb{T} (\{\mb k\};\{s,\bar s\} )$ take the following form:
\begin{align}
\label{eq_TIntMellin}
  \mathbb{T}\Big(\{\mb k\};\{s,\bar s\}\Big)
  =&\int_{-\infty}^0\prod_{i=1}^V\Big[ \di \tau_i\,(\ii\aa_i)(-\tau_i)^{P_i-2S_i}\Big]\prod_{j=1}^{B}\Big[C_{\aa_j}(k_j;\tau_j)\Big]\mathcal{N}(\tau_1,\cdots,\tau_V)\n\\
  =&\int_{-\infty}^0\prod_{i=1}^V \di \tau_i\,(\ii\aa_i)(-\tau_i)^{P_i-2S_i}e^{\ii\aa_i E_i\tau_i}  \mathcal{N}(\tau_1,\cdots,\tau_V),
\end{align}
where the function $\mathcal{N}(\tau_1,\cdots,\tau_V)$, called the \emph{nesting function}, denotes all possible combinations of $\theta$ functions that order the time variables. The factor $e^{\ii \aa_i E_i\tau_i}$  comes from the bulk-to-boundary propagators $C_{\aa_j}(k_j;\tau_j)$, where $E_i$ denote the sum of the \emph{magnitudes} of all 3-momenta of the bulk-to-boundary propagators attached to the $i$'th vertex. In the power function $(-\tau_i)^{P_i-2S_i}$, $P_i$ is a fixed real number which receives three contributions:
\begin{align}
  P_i=&~p_i+(\text{number of bulk-to-boundary propagators attached to the $i$'th propagator})\n\\
  &~+\FR32\times(\text{number of bulk propagators attached to the $i$'th propagator}),
\end{align}
which can be understood by looking at (\ref{eq_Tk}), (\ref{eq_CSProp}), and (\ref{eq_DScalarMB1}). On the other hand, the term $S_i$ in the exponent of $(-\tau_i)^{P_i-2S_i}$ denotes the sum of all Mellin variables for the bulk propagators attached at the $i$'th vertex.

We note in passing that, in the time integral $\mathbb{T}$, all dependences on external momenta are through the ``partial energies'' $E_i$ ($i=1,\cdots, V$), namely the partial sum of the magnitudes of external momenta. The graph $\mathcal{G}(\{\mb k\})$ does have nonanalytic behavior on partial energies which, in the boundary of physical regions, corresponds to the so-called ``local signals.'' We will study the local signals elsewhere. 

Next, we have a factor of loop momentum integral $\mathbb{L}(\{\mb k\};\{s,\bar s\})$:
\begin{align}
\label{eq_MellinLoopInt}
  \mathbb{L}\Big(\{\mb k\};\{s,\bar s\}\Big)=\int\prod_{\ell=1}^L\bigg[\FR{\di^3\mb q_\ell}{(2\pi)^3}\bigg]\prod_{i=1}^I\big|\mb p_i\big|^{-2s_{i\bar i}},
\end{align}
where $\mb p_i$ is the momentum of the $i$'th internal propagator. We emphasize that the Mellin-space loop integral $\mathbb{L}(\{\mb k\};\{s,\bar s\})$ is completely independent of SK indices.

Finally, we collect all other time- and momentum-independent factors into $\mathbb{D}(s_i,\bar s_i)$, including the exponential factors and $\Gamma$ factors from the MB representation of the bulk propagators as in (\ref{eq_DScalarMB1}).

The break of the graph $\mathcal{G}(\{\mb k\})$ into different pieces in (\ref{eq_TkMellin}) is what makes the partial MB representation useful for our analysis. For example, in this work, we are interested in the analyticity of $\mathcal{G}(\{\mb k\})$ as a function of partial sums of external momenta. Now, (\ref{eq_TkMellin}) tells us that only the loop integral $\mathbb{L}(\{\mb k\};\{s,\bar s\})$ could possibly depend on the partial sums of external momenta. Therefore, we only need to make a careful analysis of $\mathbb{L}(\{\mb k\};\{s,\bar s\})$ for our purpose, which is what we shall do as the next step.
  
\paragraph{Parametrization of internal momenta.}
We can choose loop momentum variables $\mb q_\ell$ $(\ell=1,\cdots,L)$ in such a way that the momentum $\mb p_i$ of the $i$'th loop propagator $(i=1,\cdots,I)$ takes the following form:
\begin{align}
\label{eq_pi}
  \mb p_i=
  \begin{cases} 
  \mb q_i, &(i=1,\cdots, L) \\[1mm]
  \sum\limits_{\ell=1}^{L}\al_{i\ell}\mb q_\ell+\sum\limits_{j=1}^{B}\be_{ij}\mb k_j, &(i=L+1,\cdots, I)
  \end{cases}
\end{align}
where the coefficients $\al_{i\ell}$ take values from $\{0,\pm 1\}$. The representation in the second sum $\sum\be_{ij}\mb k_j$ is not unique due to the total momentum conservation $\sum\mb k=\mb 0$. This freedom allows us to choose a representation in which $\be_{ij}$ takes value from $\{0,+1\}$.\footnote{\label{fn_beta}Proof: We can remove the first $L$ internal propagators carrying the loop momenta $q_i$ and set $q_i=0$ for simplicity $(i=1,\cdots,L)$. The resultant graph is a tree graph, whose internal propagators all have fixed momenta determined by the external momenta $\mb k$. Then, consider any internal line with momentum $\mb p_i$ in this tree diagram, there are two possibilities. Either $\mb p_i=\mb 0$, meaning $\be_{ij}=0$ for all $j$, or, $\mb p_i\neq\mb 0$ and is determined by external momenta $\{\mb k\}$. In the latter case, removing this internal line with momentum $\mb p_i$ would give a bipartition of the graph $\{\mb k\}\to \{\mb k^\mathrm{(L)}\}\cup \{\mb k^\mathrm{(R)}\}$, and therefore, we have $\mb p_i=\sum\mb k^\mathrm{(L)}$. Thus we see that, for this internal line, $\be_{ij}=+1$ for all $\mb k_j\in\{\mb k^\mathrm{(L)}\}$ and $\be_{ij}=0$ for all $\mb k_j\in\{\mb k^\mathrm{(R)}\}$.} Then, the combination $\sum\be_{ij}\mb k_j$ is simply a partial sum of external momenta as defined in (\ref{eq_partialSum}).

Now, we want to understand the analyticity of $\mathcal{G}(\{\mb k\})$ as a function of a particular partial sum $P$ at either finite $P$ in the physical region or at $P=0$, with all other partial sums $\wt{\mb P}_i\neq \pm \mb P$ held fixed and finite. Therefore, let us write
\bge
  \mb P=\ob{\mb P}+\de\mb P,
\ede
where $\ob{\mb P}$ is either finite or zero. and we want to understand what happens when we send $\de \mb P\to \mb 0$. 

A first observation is that the momentum $\mb p_i$ of the $i$'th internal propagator, as given in (\ref{eq_pi}), can be Taylor expanded around $\de \mb P= \mb 0$, with all $\order{\de P^2}$ and higher order terms vanish. Let us denote the zeroth order term in this expansion as $\ob{\mb p}_i$, then, we can write:
\begin{align}
\label{eq_piExpand}
  \mb p_i=\ob{\mb p}_i+\ga_i\de \mb P,
\end{align}
where $\ga_i$ is a momentum-independent number. We note that all loop momentum dependences in the above expansion are in the zeroth order term $\ob{\mb p}_i$.

It turns out that all the nonanalyticity of the loop integral $\mathbb{L}(\{\mb k\};\{s,\bar s\})$ as $\mb P\to\mb 0$ comes from a particular region in the loop momentum space, where two or more \emph{linearly dependent} internal momenta become soft simultaneously. We call a singularity from such a region the \emph{degenerate singularity.} Below, we use a lemma to make this point more explicit. 

\begin{figure}
\centering
\includegraphics[width=0.4\textwidth]{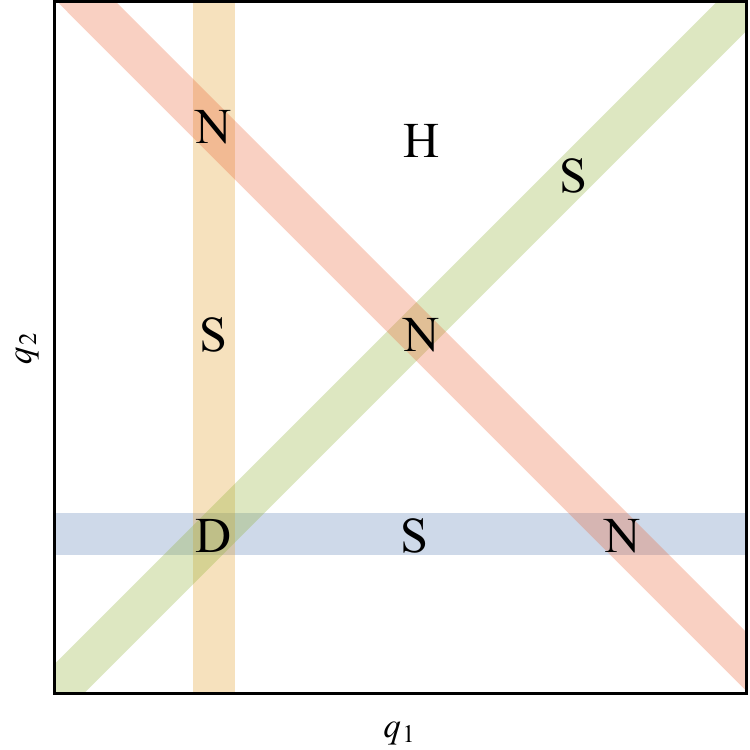}
\caption{An illustration of our separation of regions. In this plot, each of the two axes of $q_1$ and $q_2$ represents a three-dimensional space. Each of the four shaded strips represents a soft zone. The regions marked with H, S, N, and D represent the hard region, the singly soft region, the nondegenerate region, and the degenerate region, respectively. } 
  \label{fig_region}
\end{figure}

\paragraph{Degenerate singularity.}
Now suppose $\ob{\mb P}$ is fixed at a physical value either at or away from $\mb 0$. By construction of the nonlocal soft limit, all partial sums of external momenta $ \wt{\mb P}_i$ except $\ob{\mb P}$ (and, trivially, $-\ob{\mb P}=\sum \mb k^\mathrm{(R)}$) are finite. This fact allows us to introduce a comoving cutoff $\Lambda$ in the momentum integral, which satisfies
\bge
  \de P\ll\Lambda\ll\min\{\wt{\mb P}\}.
\ede
Now, the loop integral is performed over a $3L$-dimensional space $\mathscr{Q}$ spanned by $L$ independent loop momenta. To analyze the analyticity of the integral, we separate this loop momentum space into different regions. Specifically, for the $i$'th internal propagator with momentum $\mb p_i$ given in (\ref{eq_pi}), we  define a \emph{soft zone} $\mathscr{U}_i$, which is a codimension-3 sheet with $\order{\Lambda}$ thickness:
\bge
  \mathscr{U}_i\equiv\Big\{ \big(\mb q_1 ,\cdots ,\mb q_{L} \big)~;~\ob{p}_i<\Lambda\Big\}.~~~~(i=L+1,\cdots, I)
\ede
Here $\ob{p}_i\equiv|\ob{\mb p}_i|$ and $\ob{\mb p}_i$ is defined in (\ref{eq_piExpand}).
Then, the whole loop momentum space $\mathscr{Q}$ can be separated into the following four types of regions:
\begin{enumerate}
  \item \emph{Hard region}, which is the region outside of all soft zones, namely $\mathscr{Q}-\bigcup\limits_{i=1}^{L}\mathscr{U}_i$. In this region, the momenta of all loop propagators are harder than $\Lambda$.
  \item \emph{Singly soft region}, which is the interior of a given soft zone $\mathscr{U}_i$ but with all zone-intersections removed, namely $\mathscr{U}_i-\bigcup\limits_{j\neq i}\mathscr{U}_j$. In this region, there is a single loop propagator with momentum $\mb p_i$ softer than $\Lambda$, while the momenta of all other loop propagators are harder than $\Lambda$. 
  \item \emph{Nondegenerate region}. A nondegenerate region could appear when $\bigcap\limits_{i\in I}\mathscr{U}_i\neq \varnothing$ for a label set $I\subset\{1,\cdots,L\}$ that contains at least two elements. An intersection $\bigcap\limits_{i\in I}\mathscr{U}_i$ is called nondegenerate, if all momenta $\mb p_i$ with $i\in I$ are linearly independent. In this region, the propagators with momenta $\mb p_i$ $(i\in I)$ are softer than $\Lambda$, while all other loop propagators have momenta harder than $\Lambda$. 
  \item \emph{Degenerate region}. A degenerate region means a nonempty intersection $\bigcap\limits_{i\in I}\mathscr{U}_i$ where all momenta $\mb p_i$ with $i\in I$ are linearly dependent. The relative softness or hardness of momenta in this region is similar to the nondegenerate region, but the degeneracy of $\mb p_i$ with $i\in I$ makes a huge difference in the analytic property of the loop momentum integral, as will be detailed below. 
\end{enumerate}
We illustrate the separation into these four types of regions in Fig.\ \ref{fig_region}. The usefulness of this separation comes from the following ``lemma of degenerate singularity'':\\[1mm]

\begin{lemma}[Degenerate singularity]\label{lem_deg}
 Given the bipartition $\{\mb k^\mathrm{(L)}\}\cup\{\mb k^\mathrm{(R)}\}$ and the nonlocal soft limit $\mb P\equiv \sum\mb k^\mathrm{(L)}\to\mb 0$, the only possible singularities of the Mellin-space loop integral $\mathbb{L}(\{\mb k\};\{s,\bar s\})$ in (\ref{eq_MellinLoopInt}) as $P\to 0$ come from the integral over the degenerate region. The integral, restricted in either the hard region, the singly soft region, or the nondegenerate region, is analytic in $P$ as $P\to 0$. 
\end{lemma} 
\emph{Proof:}
\begin{enumerate}
\item We first show that the loop integral over the hard region $\mathscr{Q}-\bigcup\limits_{i=1}^{L}\mathscr{U}_i$ is analytic as $\de P\to 0$. In this region, we have $\ob{p}_i>\Lambda>\de P$ for all $\ob{\mb p}_i$, and thus we can Taylor expand the $i$'th propagator into convergent power series of $\de P$ for all $i=1,\cdots,I$:
\bge
  \big|\mb p_i\big|^{-2s_{i\bar i}}=\Big|\ob{\mb p}_i+\ga_i\de \mb P\Big|^{-2s_{i\bar i}}=\big|\ob{\mb p}_i\big|^{-2s_{i\bar i}}+\order{\de P}.
\ede
So the momentum integral over the region outside of all soft cones is clearly analytic in $\de P$ as $\de P \to 0$. 

\item We then show the loop integral over the singly soft region $\mathscr{U}_i-\bigcup\limits_{j\neq i}\mathscr{U}_j$ for all $i=1,\cdots, I$ is analytic as $\de P\to 0$. In this region, we can choose $\ob{\mb p}_i$ together with other $N_L-1$ momenta to form a new basis of $\mathscr{Q}$ and thus they can be used as a new set of loop momentum variables. Then,  in the interior of $\mathscr{U}_i-\bigcup\limits_{j\neq i}\mathscr{U}_j$, we have $p_j\simeq \ob p_j>\Lambda$ for any $j\neq i$. Then, we can Taylor expand $|\mb p_j|^{-2s_{j\bar j}}$ around $\de P=0$ and $\ob p_i=0$ for all $j\neq i$, and we can keep the leading term only. The leading term is simply independent of $\de P$ and $\ob p_i$. So we conclude that, within $\mathscr{U}_i-\bigcup\limits_{j\neq i}\mathscr{U}_j$, at the leading order in $\de P$ and $\ob{p}_i$, the whole loop integral is proportional to
\bge
\label{eq_singlySoftInt}
  \int^\Lambda\FR{\di^3\ob{\mb p}_i}{(2\pi)^3}\Big|\ob{\mb p}_i+\ga_i\de \mb P \Big|^{-2s_{i\bar i}}.
\ede
This integral is analytic at $\de P=0$. To see this, we rewrite this integral as:
\bge
\label{eq_singlySoftInt2}
  \int_0^\Lambda\FR{\di^3\ob{\mb p}_i}{(2\pi)^3}\Big|\ob{\mb p}_i+\ga_i\de \mb P \Big|^{-2s_{i\bar i}}
  =\int_0^\infty\FR{\di^3\ob{\mb p}_i}{(2\pi)^3}\Big|\ob{\mb p}_i+\ga_i\de \mb P \Big|^{-2s_{i\bar i}}-\int_\Lambda^{\infty}\FR{\di^3\ob{\mb p}_i}{(2\pi)^3}\Big|\ob{\mb p}_i+\ga_i\de \mb P \Big|^{-2s_{i\bar i}}.
\ede
The first integral on the right hand side of (\ref{eq_singlySoftInt2}) can be carried out directly, which is independent of $\de P$:
\bge
  \int_0^\infty\FR{\di^3\ob{\mb p}_i}{(2\pi)^3}\Big|\ob{\mb p}_i+\ga_i\de \mb P \Big|^{-2s_{i\bar i}}=\FR1\pi \de\big[\ii (3-2s_{i\bar i})\big]. 
\ede
On the other hand, in the second integral on the right hand side of (\ref{eq_singlySoftInt2}), we can expand the integrand around $\de P=0$ since $\ob p_i>\Lambda>\de P$ in this region. Thus this integral is analytic in $\de P$, so is the original integral (\ref{eq_singlySoftInt}). So we conclude that the loop momentum integral over the singly soft region is also analytic at $\de P=0$.

\item Next we show that the loop integral over any nondegenerate region $\bigcap\limits_{i\in I}\mathscr{U}_i$ is analytic at $\de P= 0$. In this case, we can choose $\ob{\mb p}_i$ $i\in I$, together with some other independent momenta to form a new set of loop integral variables. Then, in parallel with the above argument, the integral in the region $\bigcap\limits_{i\in I}\mathscr{U}_i\neq \varnothing$ can be rewritten as factorized integrals:
\bge 
  \prod_{i\in I}\int^\Lambda\FR{\di^3\ob{\mb p}_i}{(2\pi)^3}\Big|\ob{\mb p}_i+\ga_i\de \mb P\Big|^{-2s_{i\bar i}},
\ede
which is again analytic at $\de P=0$ by the same reasoning as above.  

\item So, any possible singular behavior of the loop integral as $\de P\to 0$ must arise from a degenerate region $\bigcap\limits_{i\in I}\mathscr{U}_i$. In this case, we can no longer take all $\ob{\mb p}_i$ with $i\in I$ as independent variables. Suppose such a degenerate region is an intersection of $D$ soft zones, and the order of degeneracy is 1. (That is, there are $D-1$ vectors $\ob{\mb p}_i$ out of $D$ which are linearly independent. Higher degrees of degeneracy can be handled similarly.) Then we can write, say:
\bge
\ob{\mb p}_D=\sum_{i=1}^{D-1}\ka_i \ob{\mb p}_i.
\ede
Then the integral within this region will have the following form:
\begin{align}
  \mathbb{L}_\text{degen} \propto\int^\Lambda\prod_{i=1}^{D-1}\bigg[\FR{\di^3\ob{\mb p}_i}{(2\pi)^3}\Big|\ob{\mb p}_i+\ga_i\de \mb P\Big|^{-2s_{i\bar i}}\bigg]\bigg|\sum_{i=1}^{D-1}\ka_i\ob{\mb p}_i+\ga_D\de\mb P\bigg|^{-s_{D\ob D}}
\end{align}
We can certainly shift and rescale the $D-1$ integral variables $\ob{\mb p}_i$ such that the above integral becomes:
\begin{align}
  \mathbb{L}_\text{degen}\propto \int \prod_{i=1}^{D-1}\bigg[\FR{\di^3\ob{\mb p}_i}{(2\pi)^3}\big|\ob{\mb p}_i \big|^{-2s_{i\bar i}}\bigg]\bigg|\sum_{i=1}^{D-1} \ob{\mb p}_i+\wt\ga_D\de\mb P\bigg|^{-s_{D\ob D}}+\text{analytic terms in $\de P$}.
\end{align}
Here we have removed the upper limit $\Lambda$ of the integral at the expense of introducing an (irrelevant) analytic term in $\de P$. The integral is nothing but the melon integral of degree $D$ which we shall encounter frequently in this work. We work out this integral explicitly in App.\ \ref{app_int}. The result is:
\begin{align}
\mathbb{L}_\text{degen}\propto \FR{(-\wt\ga_D\de P)^{3(D-1)-2s_{1\bar1\cdots D\ob{D}}}}{(4\pi)^{3(D-1)/2}}\Gamma\bgb s_{1\bar1\cdots D\ob{D}}-\fr{3(D-1)}2 \\ \fr{3D}2-s_{1\bar1\cdots D\ob{D}}\edb\prod_{\ell=1}^{D}\Gamma\bgb\fr32-s_{\ell\bar\ell}\\ s_{\ell\bar\ell}\edb+\cdots,
\end{align}
where the omitted terms are analytic in $\de P$. Therefore, we see that the loop momentum integral in the degenerate region produces a term $\propto \de P^{-2s_{1\bar 1\cdots D\ob D}}$, and this would contribute to a nonlocal signal whenever the exponent $3(D-1)-2s_{1\bar 1\cdots D\ob D}\notin \mathbb{N}$.

\end{enumerate}
Thus we conclude that a necessary condition for the appearance of singular behavior in $\de P\to 0$ is the existence of a degenerate region in the loop momentum space. This completes our proof of the lemma of degenerate singularity.

\paragraph{Equivalence between nonlocal cut and degenerate region.} The lemma of degenerate singularity tells us that, when $\mb P\to \mb 0$, nonanalyticity could (though not necessarily) arise from each degenerate region. Now, we are going to show that the existences of a nonlocal cut and a degenerate region imply each other. Thus, the existence of a nonlocal cut is a necessary condition for the existence of a nonlocal signal. By enumerating all nonlocal signals in a given nonlocal soft limit, we are guaranteed to get all possible nonlocal signals in this limit.

Naturally, our proof comes in two steps, one with ``$\To$'' and the other with ``$\Leftarrow$.''

\noindent\textbf{Nonlocal cut $\bm\To$ degenerate region:} First, we show that each nonlocal cut of the graph implies the existence of a degenerate region in the loop momentum space. In fact, a nonlocal cut of degree $D$ means that we can set the momenta of all cut lines to zero under the constraint of the nonlocal soft limit. (That is, all partial sums except $\mb P$ remain finite.) This implies the existence of an intersection of at least $D$ soft zones, and we only need to show that this intersection is degenerate. Suppose it were otherwise and the soft intersection is nondegenerate. Then it implies that we can choose the $D$ momenta in these $D$ lines as independent loop momentum variables. This means that removing these $D$ lines should result in a connected graph. However, being a nonlocal cut means that removing these $D$ lines should result in a disconnected graph. This contradiction shows that the intersection of the $D$ soft zones must be degenerate. 
 
\noindent\textbf{Degenerate region $\bm\To$ nonlocal cut:} Second, we show that a degenerate region from the intersection of $D$ soft zones gives a nonlocal cut. In fact, the existence of such a degenerate region means that we can set the momenta of the corresponding $D$ lines to zero up to $\order{\Lambda}$ without affecting the momenta of all other lines. So we can as well remove these $D$ lines without affecting other internal lines. Then, we only need to show that removing these $D$ lines is a valid nonlocal cut. This comes in two steps:
\begin{enumerate}
  \item We need to show that, after the removal of these $D$ lines, the left set $\{\mb k^\mathrm{(L)}\}$ is totally disconnected from the right set $\{\mb k^\mathrm{(R)}\}$. For simplicity, assuming the degeneracy of this region is order 1. (Again, degeneracies of higher order can be handled similarly.) Then we can select the momenta $\ob{\mb p}_i$ of $D-1$ lines as independent loop momentum variables, and rewrite the momentum of the remaining line as $\sum\limits_{i=1}^{D-1}\al_i\ob{\mb p}_i+\wt{\mb P}_D$, where $\al_i$ are numbers and $\wt{\mb P}_D$ is a partial sum of external momenta. Now, since all these $D$ lines can become 0 simultaneously, it implies that $\wt{\mb P}_D=\mb 0$. This is consistent with our assumption of nonlocal soft limit (all partial sums but $\mb P$ remain finite) only if $\wt{\mb P}_D=\mb P=\sum\mb k^\mathrm{(L)}$ (or if, trivially, $\wt{\mb P}_D=-\mb P=\sum\mb k^\mathrm{(R)}$). Therefore, removing the first $D-1$ lines will result in a tree graph where the $D$'th line is a tree line connecting a left subgraph with external momenta $\{\mb k^\mathrm{(L)}\}$ and a right subgraph with external momenta $\{\mb k^\mathrm{(R)}\}$. Removing this remaining tree line thus makes the left subgraph and right subgraph disconnected.\footnote{When the degeneracy of the region is of order $r\geq 1$, one can follow the same argument here to show that the corresponding nonlocal cut will separate the graph into $r+1$ disconnected subgraphs. }
  \item We still need to show that, in the remaining graph after the removal of the $D$ lines, the $B_L$ left points are fully connected, and so are the $B_R$ right points. Suppose this is not true, and a genuine subset of $\{\mb k^\mathrm{(L)}\}$ is disconnected from others, then it implies that the sum of all momenta in this subset is zero, and this contradicts our assumption of the nonlocal soft limit. Therefore, the $B_L$ left points must be fully connected, so are $B_R$ the right points.
\end{enumerate}

Now we have established the equivalence between the degenerate region and the nonlocal cut, and therefore we have proved the second statement of Theorem \ref{thm_signal}.

\paragraph{The loop integral is analytic at finite $\mb P$.} It remains to prove the first statement of Theorem \ref{thm_signal}, namely, the loop integral is analytic at finite $\mb P$. By virtue of the lemma of degenerate singularity, we only need to show that a degenerate region never appears in the loop momentum space when all partial sums $\wt{\mb P}_i$, including $\mb P\equiv\sum\mb k^\mathrm{(L)}$, are held finite. Suppose it were otherwise, and we do have a degenerate region and the order of degeneracy is 1. (Again, degeneracies of higher order can be handled similarly.) Repeating the analysis of previous paragraphs, we see that, in this case, there must be at least one partial sum of external momenta approaching 0, which contradicts the original condition of all partial sums held finite. Therefore, there cannot be a degenerate region in this case and the loop integral must be analytic in $\mb P$ when all partial sums are held finite in the physical region. This completes the proof of the first statement of Theorem \ref{thm_signal}.

\subsection{Some simple lemmas}
\label{sec_lemma}

At the end of this section, we collect several simple yet useful lemmas. These results will be used in the analysis of subsequent sections.

\begin{lemma}[Perfect bipartition]\label{lem_perbip}
 An irreducible cut reduces the original graph into exactly two disconnected subgraphs; There exists no subgraph that is isolated from both the left and the right graphs. This is called a perfect bipartition. 
\end{lemma}
\emph{Proof:} If there exists a third connected component other than the left or the right graph, then we can always connect it to the left or to the right graph (but not to both), by putting back several cut lines. The resultant graph is still a valid nonlocal cut, showing that the original cut is not irreducible. 

This almost trivial observation has several useful consequences for an irreducible cut graph:
\begin{enumerate}
  \item All bulk vertices belong to either the left or the right graph; There can be no bulk vertex that is connected to neither the left nor the right graph.
  \item It is impossible to cut all internal lines ending at the same bulk vertex. Otherwise this bulk vertex will be isolated from the left and right subgraphs.
  \item A cut line in an irreducible cut must connect the left subgraph to the right subgraph. Proof: By the main lemma, the two endpoints of a cut line must belong to either the left or right subgraph. If they both belong to the left (or right) graph, then we can put the cut line back, and the resultant graph is again a valid cut, showing that the original cut is not irreducible.  
\end{enumerate}
Note that the property of perfect bipartition no longer holds for reducible cuts. In a reducible cut graph, we can have multiple disconnected components. In particular, there can be a bulk vertex such that all bulk lines attached to it are cut. In this case, we will have no control over the relative softness of these lines. This makes the resonance argument badly inapplicable, and also makes the computation of nonlocal signals difficult. Fortunately, this happens only to reducible cuts, which are also nonminimal cuts, meaning that they do not contribute to the leading nonlocal signal. 

\begin{lemma}[Maximal degree]\label{lem_maxdeg}
The degree $D$ of an irreducible cut must be smaller than $L+1$ where $L$ is the number of loops in the whole graph. 
\end{lemma}
\emph{Proof:} By Lemma \ref{lem_perbip}, in an irreducibly cut graph, a cut line must connect the left and the right subgraphs. Given that the left (right) subgraph is itself connected, we can always make use of a collection of bulk lines in the left subgraph to connect all endpoints of all cut lines that are attached to the left subgraph. We can do the same on the right side. With these additional lines, we have constructed a $D-1$ loop graph which is a subgraph of the original graph. Since the number of loops in a subgraph must not exceed the number of loops $L$ in the original graph, we conclude that $D\leq L+1$.\\[2mm]

Till now, we have been considering how to find the nonlocal signal for a given bipartition of the graph. However, not all bipartition leads to nonlocal signals. Let us call a bipartition that contains nonlocal signals \emph{signal-bearing}. It would be useful to know a priori whether a bipartition is signal-bearing or not. Here we provide two necessary conditions for a signal-bearing bipartition.

\begin{figure}
\centering
\includegraphics[width=0.5\textwidth]{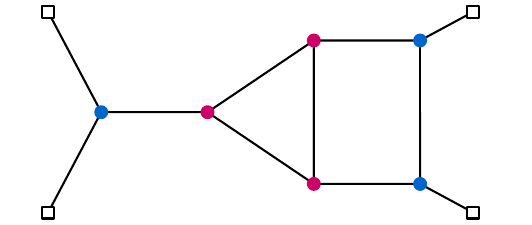}
\caption{Illustration of internal and external vertices. The magenta dots denote internal vertices, the blue dots denote external vertices, and the little white box denotes boundary points.} 
  \label{fig_IntVer}
\end{figure}

To state these conditions, we introduce the concept of the \emph{internal vertex} and the \emph{external vertex}, which will be useful in the following sections. An internal vertex is a vertex to which all lines attached are bulk propagators. An external vertex is a vertex to which there is at least one bulk-to-boundary propagator attached. We stress that both internal and external vertices are bulk vertices. We illustrate these definitions in Fig.\ \ref{fig_IntVer}.

With the above preparation, we can now ready to state two necessary conditions for a signal-bearing bipartition.

\begin{lemma}[Signal-bearing bipartition]\label{lem_signal}
A signal-bearing bipartition of a graph has the following two properties:
\begin{enumerate}
  \item A group of external points connected at a common vertex must be on the same side of the bipartition.
  \item For any two external vertices on the same side, there must exist a path of bulk propagators connecting them without passing through any external vertices on the other side. 
\end{enumerate}
\end{lemma}
The proof of this lemma is trivial. If either of the above two conditions is not met, a nonlocal cut would be impossible.

\begin{figure}
\centering
\includegraphics[width=\textwidth]{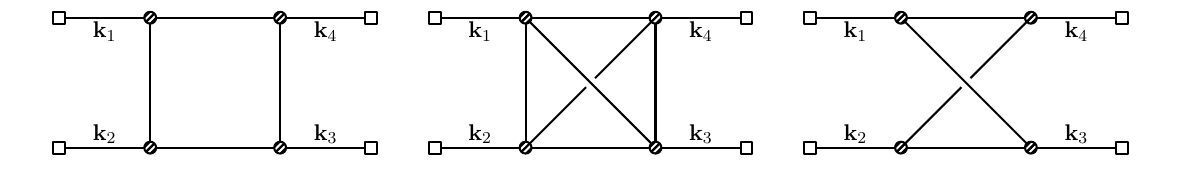}
\caption{Illustration of the second condition in Lemma \ref{lem_signal}. In the limit $\mb k_1+\mb k_2\to\mb 0$, the left and middle graphs could give rise to a nonlocal signal while the right graph cannot.} 
  \label{fig_boxes}
\end{figure}

Although trivial to prove, this lemma is useful in our analysis. The first condition shows that, not only the boundary points, but also all external vertices, are unambiguously separated into two groups by the bipartition. So we can meaningfully talk about left external vertices and right external vertices. The second condition shows that ``disconnected soft limits'' are signal-less. This is illustrated in Fig.\ \ref{fig_boxes}. 

\section{On-Shell Factorization of Bulk-Free Graphs}
\label{sec_bulkfree}

The signal detection algorithm from the last section gives us a practical way to search for nonlocal signals in a graph. By executing all possible nonlocal cuts associated with a given nonlocal soft limit, we are guaranteed to locate the origins of all possible nonlocal signals. However, the signal detection algorithm does not provide explicit expressions for the nonlocal signals. In the present and the next sections, we are going to work out explicit expressions for the leading nonlocal signals in a given soft limit for various types of graphs. These expressions will make the property of on-shell factorization explicit at any loop orders: The leading nonlocal signal from a minimal cut of degree $D$ naturally factorizes into three pieces: a left subgraph, a right subgraph, and a nonanalytic piece which we call the $(D-1)$-loop melon signal. Also, the cutting rule follows as a natural byproduct of our proof of the factorization theorem. That is, all cut propagators can be replaced by their real parts, and thereby the time orderings between any two endpoints of a cut propagator can be avoided.

In this section, we consider graphs of arbitrary loop order that contain no internal vertices. This amounts to saying that the fields corresponding to all internal lines have no self or mutual interactions, except the couplings to the external mode. Thus we can think of these graphs as the expectation values of a bunch of free fields in the bulk, with external legs acting as external sources for these bulk fields and their products. For this reason, we call such graphs bulk-free graphs. As we shall see, the analysis for bulk-free graphs turns out to be free from many subtleties in most general graphs. Thus, the factorization of nonlocal signals for such graphs holds under very general assumptions about the theory. 

For definiteness, we shall first present and prove the factorization theorem for arbitrary bulk-free graphs under a relatively restricted set of assumptions. After completing the proof, we shall explore the consequences of loosening some of the assumptions made in the theorem.

\subsection{Factorization theorem for bulk-free graphs}

\begin{theorem}[On-shell factorization of bulk-free graphs]\label{thm_bulkfree}
Let $\mathcal{G}(\{\mb k\})$ be a $B$-point and $L$-loop graph without any internal vertices, whose all bulk propagators can be massive with arbitrary (possibly dS-boost-breaking) dispersions. Then, for an arbitrary nonlocal bipartition of the graph $\{\mb k^\mathrm{(L)}\}\cup \{\mb k^\mathrm{(R)}\}$, the amplitude $\mathcal{G}(\{\mb k\})$ exhibits a nonlocal signal in the $\mb P\equiv \sum\{\mb k^\mathrm{(L)}\}\to \mb 0$ limit, if and only if there exists a nonlocal cut associated with the bipartition. When a nonlocal bipartition exists for this bipartition, the minimal cut is unique. The explicit expression for the nonlocal signal in this limit is factorized into three parts:
\begin{keyeqn}
\begin{align}
\label{eq_bulkfreefac}
  \lim_{P\to 0}\mathcal{G}(\{\mb k\})=\sum_{\substack{\cc_1,\cdots,\cc_D=\pm\\\mathrm{nonlocal}}} \mathcal{G}_{\cc_1\cdots\cc_D}^\mathrm{(L)}\Big(\{\mb k^\mathrm{(L)}\}\Big)\mathcal{G}_{\cc_1\cdots\cc_D}^\mathrm{(R)}\Big(\{\mb k^\mathrm{(R)}\}\Big)\mathfrak{M}_{\cc_1\cdots\cc_D}(P)+\text{terms analytic in $P$}.
\end{align}
\end{keyeqn}
\end{theorem}
Here the subscript ``nonlocal'' below the summation sign means to sum over all possible combinations of $\cc_1,\cdots,\cc_D\in\{\pm\}$ subject to the condition $\sum\limits_{i=1}^D\cc_i\wt\nu_i\neq 0$. All the nonanalyticity in $P$ at $P=0$ is encoded in $\mathfrak{M}_{\cc_1\cdots\cc_D}(P)$, which we call the ``melon signal,'' and its explicit expression is:
\begin{align}
\label{eq_frakM}
  \mathfrak{M}_{\cc_1\cdots\cc_D}(P)
  \equiv \FR{P^{3(D-1)}}{(4\pi)^{(5D-3)/2}}\Gamma\bgb -\sum\limits_{i=1}^D\cc_i\ii\wt\nu_i-\fr{3}2(D-1) \\ \fr{3}2 D+ \sum\limits_{i=1}^D \cc_i\ii\wt\nu_i\edb\prod_{\ell=1}^{D}\bigg\{\Gamma\Big[\fr32+ \cc_\ell\ii\wt\nu_\ell, - \cc_\ell\ii\wt\nu_\ell \Big]\Big(\FR{P}2\Big)^{2\ii\cc_\ell\wt\nu_\ell}\bigg\}.
\end{align}
In the above expression, the repeated indices are not summed unless explicitly stated. On the other hand, the left subgraph $\mathcal{G}^\mathrm{(L)}_{\cc_1\cdots\cc_D}$ and right subgraph $\mathcal{G}^\mathrm{(R)}_{\cc_1\cdots\cc_D}$ are given by the usual SK integrals, subject to the condition $\sum\mb k^\mathrm{(L)}=-\sum\mb k^\mathrm{(R)}=\mb 0$, and with additional insertions of time powers $(-\tau_i)^{3/2+\cc_i\ii\wt\nu_i}$ at each endpoint of a cut line. More explicitly: 
\begin{align}
\label{eq_TL}
  \mathcal{G}_{\cc_1\cdots\cc_D}^\mathrm{(L)}\Big(\{\mb k^\mathrm{(L)}\}\Big)
  =&\sum_{\aa_1,\cdots,\aa_{V_L}=\pm}\int\prod_{i=1}^{V_L}\Big[\ii\aa_i\,\di\tau_i\,(-\tau_i)^{p_i}\Big]\prod_{j=1}^{B_L}\Big[C_{\aa_j}(k_i;\tau_j)\Big]{\blue\prod_{\ell\in V_{CL}}(-\tau_{\ell})^{3/2+\cc_\ell\ii\wt\nu_\ell}}\n\\
  &\times \int\prod_{m=1}^{N_L}\bigg[\FR{\di^3\mb q_m}{(2\pi)^3}\bigg]\prod_{k=1}^{I_L}D_{\aa_{k1}\aa_{k2}}(p_k;\tau_{k1},\tau_{k2}).
\end{align}
\begin{align}
\label{eq_TR}
  \mathcal{G}_{\cc_1\cdots\cc_D}^\mathrm{(R)}\Big(\{\mb k^\mathrm{(R)}\}\Big)
  =&\sum_{\aa_1,\cdots,\aa_{V_R}=\pm}\int\prod_{i=1}^{V_R}\Big[\ii\aa_i\,\di\tau_i\,(-\tau_i)^{p_i}\Big]\prod_{j=1}^{B_R}\Big[C_{\aa_j}(k_i;\tau_j)\Big]{\blue\prod_{\ell\in V_{CR}}(-\tau_{\ell})^{3/2+\cc_\ell\ii\wt\nu_\ell}}\n\\
  &\times \int\prod_{m=1}^{N_R}\bigg[\FR{\di^3\mb q_m}{(2\pi)^3}\bigg]\prod_{k=1}^{I_R}D_{\aa_{k1}\aa_{k2}}(p_k;\tau_{k1},\tau_{k2}).
\end{align}
Here we are assuming the left subgraph $\mathcal{G}_{\cc_1\cdots\cc_D}^\mathrm{(L)}(\{\mb k^\mathrm{(L)}\})$ has $B_L$ external lines, $I_L$ internal lines, $V_L$ vertices, and $N_L$ independent loops. The conventions for writing (\ref{eq_TL}) is the same as (\ref{eq_Tk}). In particular, the time variables and SK indices in all propagators should be properly identified with the corresponding variables and indices at the vertices. The right subgraph $\mathcal{G}_{\cc_1\cdots\cc_D}^\mathrm{(R)}(\{\mb k^\mathrm{(R)}\})$ is specified similarly. In addition, we use $V_{CL}$ to denote the set of left endpoints of the $D$ cut lines, and use $V_{CR}$ to denote the set of right endpoints of the $D$ cut lines.

\subsection{Proof of the factorization theorem}

\paragraph{Outline of the proof.} The proof of Theorem \ref{thm_bulkfree} would also be long and technical. Thus we also outline the main idea of the proof before spelling out all the details. 
\begin{enumerate}
  \item We first show that, for a given nonlocal bipartition, if a nonlocal cut exists, then there will be a unique minimal cut associated with this nonlocal bipartition. So, we can focus on this minimal (and irreducible) cut for the subsequent analysis. Then, according to Lemma \ref{lem_perbip}, this minimal cut makes a perfect bipartition of the graph, and we get a left subgraph, a right subgraph, and $D$ cut lines. 
  
  \item We then take the partial MB representation for the graph, and rewrite the loop momentum integral to make the structure manifest that the graph is separated into the left and right subgraphs, together with $L$ cut lines. 
 
 \item Note that the left and right subgraphs are functions external momenta $\{\mb k\}$ as well as momenta of all cut lines $\{\mb q\}$. We then show that both subgraphs are analytic in all of $\{\mb q\}$ as all $\mb q\to \mb 0$.
 
 \item We can then decouple the left and right subgraphs from the loop momentum integral by restricting ourselves to the ``degenerate soft region'' where all $D$ cut lines are soft, in accordance with the analysis in the proof of Theorem \ref{thm_signal}. In this particular case, we can compute the leading nonanalytic term explicitly by finishing a ``melon integral.''
  
  \item We then analyze the pole structure of the whole graph in the Mellin space, and show that the nonlocal signal receives contribution only from the poles arising from the MB representation of the cut propagators. By collecting the residues at these poles, we finish the Mellin integral and get the final result (\ref{eq_bulkfreefac}), and thus complete the proof of the factorization theorem. 
  
  \item As a direct corollary of our proof, we show that, so long as the nonlocal signal is the only concern, we can replace all cut propagators by their real parts. As a result, the time integrals over all cut propagators are automatically factorized. This can be viewed as a cutting rule for computing nonlocal signals.
  
\end{enumerate}

Below we carry out these steps in detail.

\paragraph{Unique minimal cut.}
We first show that, if there exists a nonlocal cut for the given nonlocal bipartition $\{\mb k^\mathrm{(L)}\}\cup\{\mb k^\mathrm{(R)}\}$, then the irreducible cut is unique, and this irreducible cut is automatically a minimal cut. In fact, by our assumption of the bulk-free graph, there is no internal vertex. So, any cut line in the given nonlocal cut must connect two external vertices. Then, according to Lemma \ref{lem_signal}, in a signal-bearing bipartition, any external vertex is either a left vertex (L) or a right vertex (R). So, all cut lines can be classified as LL, RR, or LR, according to the sides of its two endpoints. We can restore all LL and RR lines, and the resultant graph is still a valid cut. However, we cannot restore any LR line, since this operation would reconnect the left and right subgraphs. So, we have found a unique irreducible cut, in which all LR lines are cut. 

Also, as will be made clear in the following analysis, the restored LL or RR lines in the above ``cut reduction'' procedure do not lead to independent new signals. So, we shall only consider the minimal cut in the rest of the proof.

\paragraph{Loop momentum integral in the partial MB representation.} We still work with the partial MB representation, and the graph $\mathcal{G}(\{\mb k\})$ still has the form of (\ref{eq_TkMellin}). To study the analytic properties of $\mathcal{G}(\{\mb k\})$ as a function of momentum partial sums, let us look at the loop momentum integral $\mathbb{L}(\{\mb k\})$. In this section, we suppress the dependence of $\mathbb{L}(\{\mb k\})$ on Mellin variables to avoid notational clutter.

 The whole graph $\mathcal{G}(\{\mb k\})$ has $L$ independent loop integral variables. With the minimal cut of the graph given, we can choose these $L$ independent loop momenta as follows. First, from Lemma \ref{lem_maxdeg}, we see that the $D$ cut lines belong to a $(D-1)$-loop subgraph of the whole graph. Furthermore, it is possible to choose the $D-1$ momenta flowing in these $D-1$ cut lines as independent integral variables. Second, after executing the nonlocal cut, the left subgraph is generally an $N_L$-loop graph with $N_L\geq 0$ and the right subgraph generally an $N_R$-loop graph with $N_R\geq 0$. Furthermore, we have $N_L+N_R+D-1=L$. Thus, we can choose the remaining $N_L+N_R$ integral variables from $N_L$ of loop lines in the left subgraph and $N_R$ of loop lines in the right subgraph. 

With the above choice made, we can rearrange the whole loop integral into the following form:
\begin{align}
\label{eq_MBloopInt}
  \mathbb{L}\big(\{\mb k\}\big)
  =&\int\prod_{\ell=1}^{D-1}\bigg[\FR{\di^3\mb q_\ell}{(2\pi)^3}\bigg]\big|\mb q_1\big|^{-2s_{1\bar1}} \cdots \big|\mb q_{D-1}\big|^{-2s_{(D-1)\ob{(D-1)}}}\Big|\mb P-\sum_{i=1}^{D-1}\mb q_i\Big|^{-2s_{D\ob{D}}}\n\\
  &\times\mathbb{L}^\mathrm{(L)}\Big(\{\mb k^\mathrm{(L)}\};\{\mb q\}\Big)\mathbb{L}^\mathrm{(R)}\Big(\{\mb k^\mathrm{(R)}\};\{\mb q\}\Big).
\end{align}
Here $\mathbb{L}^\mathrm{(L)}$ is the loop integral of the $N_L$-loop left subgraph. As made explicit, it is a function of all external momenta of the left subgraph $\{\mb k^\mathrm{(L)}\}$, as well as the momenta of $D$ cut lines, namely $\mb q_1,\cdots,\mb q_{D}$. Here, for convenience, we denote the momentum of the $D$'th cut line by $\mb q_D\equiv \mb P-\sum\limits_{\ell=1}^{D-1}\mb q_\ell$ with $\mb P\equiv \sum\mb k^\mathrm{(L)}$, although we stress that $\mb q_D$ is not an independent loop momentum variable. Then the left loop integral can be written as:
\begin{align}
\label{eq_leftLoopInt}
  \mathbb{L}^\mathrm{(L)}\Big(\{\mb k^\mathrm{(L)}\};\{\mb q\}\Big)
=\int\prod_{\ell=1}^{N_L}\bigg[\FR{\di^3\mb q_{\ell}^\mathrm{(L)}}{(2\pi)^3}\bigg]\prod_{i=1}^{I_L}\Big|\mb p_i^\text{(L)}\Big|^{-2s_{i\bar i}^\text{(L)}},
\end{align}
where $\mb p_i^\text{(L)}$ is the momentum of the $i$'th propagator.
Likewise, the momentum integral of the right subgraph can be expressed as:
\begin{align}
  \mathbb{L}^\mathrm{(R)}\Big(\{\mb k^\mathrm{(R)}\};\{\mb q\}\Big)
=\int\prod_{\ell=1}^{N_R}\bigg[\FR{\di^3\mb q_{\ell}^\mathrm{(R)}}{(2\pi)^3}\bigg]\prod_{i=1}^{I_R}\Big|\mb p_i^\text{(R)}\Big|^{-2s_{i\bar i}^\text{(R)}}.
\end{align}

Now, we have put the Mellin-space loop integral $\mathbb{L}\big(\{\mb k\}\big)$ into a proper form. The next task is to detect potential nonanalytic behavior in $\mathbb{L}\big(\{\mb k\}\big)$ in $P$ as $P\to 0$ while all external momenta $\mb k$ and partial sums $\wt{\mb P}\neq\mb P$ held fixed. In the following, we shall show that the loop integrals of the left and right subgraphs are both analytic in $P$ as $P\to 0$, and therefore, all possible nonanalytic behavior must be from the first line of (\ref{eq_MBloopInt}).

\paragraph{Analyticity of left and right subgraphs.} Now we show that the Mellin-space loop integrals for the left and right subgraphs are analytic in each of $q_i$ $(i=1,\cdots,D)$ when we send all $q_i\to 0$ simultaneously. We prove the above statement for the left subgraph and the treatment for the right subgraph is identical. 

Suppose the left subgraph has $N_L$ independent loops. If $N_L=0$ (tree graph), then it automatically allows a Taylor expansion around $q_\ell\to 0$ $(\ell=1,\cdots,D)$ in the Mellin representation, and the leading term is automatically finite. 

If $N_L\neq 0$ (loop graph), let there be $V_L$ vertices in the left subgraph. By our assumption of the main theorem, these $V_L$ vertices are all external vertices, to which all bulk-to-boundary propagators and the cut propagators attach. Therefore, it is clear that the loop momentum integral $\mathbb{L}^\mathrm{(L)}$ given in (\ref{eq_leftLoopInt}) depends on external momenta (including $\{\mb k^\mathrm{(L)}\}$, $\{\mb q\}$, and $\mb P$) only through certain combinations, namely, only through the total momenta injected into each of the $V_L$ vertices. Let the total momentum flowing into the $i$'th vertex be $\mb K_i$. Clearly, we can write
\bge
  \mb K_i=\sum_{j=1}^{B_L}A_{ij}\mb k_j^\mathrm{(L)}- \sum_{\ell=1}^{D} B_{i\ell}\mb q_\ell,~~~~(i=1,\cdots,V_L)
\ede
where the coefficients $A_{ij}$ and $B_{i\ell}$ take their values from $\{0,+1\}$. 

We can choose loop momentum variables $\mb q_\ell^\mathrm{(L)}$ $(\ell=1,\cdots,N_L)$ in such a way that the momentum $\mb p_i^\text{(L)}$ of the $i$'th loop propagator $(i=1,\cdots,I_L)$ takes the following form:
\begin{align}
  \mb p_i^\text{(L)}=
  \begin{cases} 
  \mb q_i^\mathrm{(L)}, &(i=1,\cdots, N_L) \\[1mm]
  \sum\limits_{\ell=1}^{N_L}\al_{i\ell}\mb q_\ell^\mathrm{(L)}+\sum\limits_{j=1}^{V_L}\be_{ij}\mb K_j, &(i=N_L+1,\cdots, I_L)
  \end{cases}
\end{align}
where the coefficients $\al_{i\ell}$ take values from $\{0,\pm 1\}$, and $\be_{ij}$ take values from $\{0,+1\}$. (See Footnote \ref{fn_beta} for the proof.) 

By our construction of the nonlocal soft limit, the first term of the above expression, $\sum A_{ij}\mb k_j^\mathrm{(L)}$, remains finite for all $i$ when taking the soft limit. Therefore, taking the soft limit is equivalent to evaluating the left subgraph at the finite values of external momenta $\mb K_i=\sum A_{ij}\mb k_j^\mathrm{(L)}$. According to the first part of Theorem \ref{thm_signal}, the loop integral is analytic in the external momenta for generic finite values in the physical region. Therefore, we conclude that the left subgraph is analytic in $\mb q_\ell$ and $\mb P$ in the simultaneous soft limit $q_\ell\to 0$ ($\ell=1,\cdots,D$). In this limit, the left subgraph allows a Taylor expansion, whose zeroth order term is 
\begin{align}
    \mathbb{L}^\mathrm{(L)}\Big(\{\mb k^\mathrm{(L)}\}\Big)
    \equiv&~\mathbb{L}^\mathrm{(L)}\Big(\{\mb k^\mathrm{(L)}\};\{\mb 0\}\Big)\n\\
=&\int\prod_{\ell=1}^{N_L}\bigg[\FR{\di^3\mb q_{\ell}^\mathrm{(L)}}{(2\pi)^3}\bigg]\prod_{i=1}^{I_L}\bigg|\sum_{\ell=1}^{N_L}\al_{i\ell}\mb q_\ell^\mathrm{(L)}+\sum_{j=1}^{V_L}\sum_{k=1}^{B_L}\be_{ij}A_{jk}\mb k_k^\mathrm{(L)}\bigg|^{-2s_{i\bar i}^\text{(L)}}.
\end{align}

In exactly the same way, we can show that the right subgraph is also analytic in $q_\ell$  in the same soft limit. 

\paragraph{Factorization of loop momentum integral.}

The above analysis shows that we can Taylor expand both the left and right subgraphs around $\mb q_\ell=\mb 0$ and $\mb P=\mb 0$. When we insert such expansions back into the full loop integral $\mathbb{L}$ in (\ref{eq_MBloopInt}), the higher order terms in $\mb q_\ell$ and $\mb P$ would eventually contribute to higher powers of $P$ when we finish the $\mb q_\ell$ integral, which decreases faster in the $P\to 0$ limit. Therefore, to single out the leading term, we should keep the leading term only, which simply means that we can directly set $\mb q_\ell=\mb P=\mb 0$ in both $\mathbb{L}^\mathrm{(L)}$ and $\mathbb{L}^\mathrm{(R)}$. In turn, it means that we can move $\mathbb{L}^\mathrm{(L)}$ and $\mathbb{L}^\mathrm{(R)}$ out of the $\mb q_\ell$-integral. The result is:
\begin{align}
\label{eq_Lkfactor}
  \mathbb{L}\big(\{\mb k\}\big)
  =&~ \mathbb{L}^\mathrm{(L)}\Big(\{\mb k^\mathrm{(L)}\}\Big)\mathbb{L}^\mathrm{(R)}\Big(\{\mb k^\mathrm{(R)}\}\Big) \mathbb{M}_{D-1}(P)\Big[1+\order{P}\Big],
\end{align}
where $\mathbb{M}_{D-1}(P)$ is the $(D-1)$-loop melon integral in the Mellin representation:
\begin{align}
\label{eq_melonInt}
\mathbb{M}_{D-1}(P)
  =&\int\prod_{\ell=1}^{D-1}\bigg[\FR{\di^3\mb q_\ell}{(2\pi)^3}\big|\mb q_\ell\big|^{-2s_{\ell\bar\ell}}\bigg]  \Big|\mb P-\sum_{i=1}^{D-1}\mb q_i\Big|^{-2s_{D \ob{D}}} . 
\end{align}
This integral can be directly done with details shown in App.\ \ref{app_int}, and the result is:
\begin{align}
\label{eq_melonIntResult}
\mathbb{M}_{D-1}(P)=&~\FR{P^{3(D-1)-2s_{1\bar1\cdots D\ob{D}}}}{(4\pi)^{3(D-1)/2}}\Gamma\bgb s_{1\bar1\cdots D\ob{D}}-\fr{3(D-1)}2 \\ \fr{3D}2-s_{1\bar1\cdots D\ob{D}}\edb\prod_{\ell=1}^{D}\Gamma\bgb\fr32-s_{\ell\bar\ell}\\ s_{\ell\bar\ell}\edb.
\end{align}
It is already clear that the result of the loop momentum integral at the leading order of $P$ factorizes into three pieces, the left subgraph $\mathbb{L}^\mathrm{(L)}$, the right subgraph $\mathbb{L}^\mathrm{(R)}$, and the melon integral $\mathbb M_{D-1}(P)$. In the $P\to 0$ limit, both the left and right subgraphs become independent of $P$, and all the $P$ dependences, and thus all potential nonanalyticities in $P$, are from the melon integral $\mathbb M_{D-1}(P)$. From (\ref{eq_melonIntResult}) we see that the (non)analyticity of $\mathbb M_{D-1}(P)$ in $P$ is fully controlled by the value of the power $-2s_{1\bar1\cdots D\ob{D}}$. After we perform the Mellin integrals, this power will take its value from the location of poles in all Mellin variables. Therefore, the (non)analyticity of the graph would be controlled by the locations of these poles. A nonanalytic piece appears whenever we take poles such that $-2s_{1\bar1\cdots D\ob{D}}\notin\mathbb{N}$. So the next step is to analyze the pole structure of the partial MB representation of the graph $\mathcal{G}$ as a function of all Mellin variables.

\paragraph{Pole structure.} 
A Mellin integral contour runs along a path from $-\ii\infty$ to $+\ii\infty$, and in all cases of our interest, the Mellin integrand is meromorphic. So, we should close the contour properly and pick up the residues of appropriate poles. In our case, the power $P^{3(D-1)-2s_{1\bar1\cdots D\ob{D}}}$ in the melon integral in (\ref{eq_melonIntResult}) together with the limit $P\to 0$ shows that we should close the MB contour from the left half plane. 

Now, let us look at the pole structure of the Mellin integrand. There are three types of poles:
\begin{enumerate}
  \item \emph{Poles from time integrals.} The Mellin-space time integral $\mathbb{T}(\{\mb k\})$ in (\ref{eq_TIntMellin}) only contains right poles of all Mellin variables. The simplest way to see this is to consider a single layer of integral:
  \bge
  \label{eq_SLTI}
    \int_{-\infty}^0\di\tau_i\,(-\tau_i)^{P_i-2S_i}e^{\ii\aa_i E_i\tau_i}  
  \ede 
The integral is converged when $\tau_i\to-\infty$ as is guaranteed by the Bunch-Davies initial condition. Any singularity in $S_i$ must be from the IR limit $\tau_i\to 0$. In this limit, we can expand the exponential factor $e^{\ii\aa_iE_i\tau_i}=\sum(\ii\aa_iE_i\tau_i)^n/n!$, and we see that the integral (\ref{eq_SLTI}) is divergent if $1+P_i-2S_i+n=0$ for all $n=0,1,2,\cdots$. They correspond to right poles of $S_i$. In fact, we can do this integral explicitly, and get:
\bge
  \int_{-\infty}^0\di\tau_i\,(-\tau_i)^{P_i-2S_i}e^{\ii\aa_i E_i\tau_i} =\FR{1}{(\ii\aa_i E_i)^{1+P_i-2S_i}}\Gamma(1+P_i-2S_i).
\ede
The aforementioned poles are encoded in the Euler $\Gamma$-function. The above argument can be generalized to arbitrarily nested time integrals to show that the full time integral, including all possible nesting functions, only produces right poles in the Mellin variables. We refer the readers to \cite{Qin:2023bjk} for details. Since we only need to pick up left poles, the time-integral poles are irrelevant to us. 
  
  \item \emph{Poles from the loop momentum integral.}
Second, we have a set of poles from an Euler $\Gamma$ factor in the melon integral (\ref{eq_melonIntResult}). The poles from $\Gamma(\fr32-s_{\ell\bar\ell})$ are right poles and can be discarded. On the other hand, there is a series of left poles from the factor $\Gamma(s_{1\bar1\cdots D\ob{D}}-\fr{3(D-1)}2)$:
\bge
  \sum_{\ell=1}^D(s_\ell+\bar s_\ell)=-n+\FR{3(D-1)}2.~~~~(n=0,1,2,\cdots) 
\ede
This set of poles arise due to the divergence of the melon loop integral in the UV limit where all loop momenta goes to infinity. For this reason we call it the UV pole. Clearly, the UV pole forces the combination $-2s_{1\bar 1\cdots D\ob{D}}$ to be an integer and thus its contribution to the melon integral is analytic in $P$ at $P=0$. So, to evaluate the nonlocal signal, we can discard this set of poles as well.

  \item \emph{Infrared poles from Hankel functions.}
Third, we have $2D$ sets of poles from the Euler $\Gamma$ factors in the MB representation of the $D$ cut propagators. Assuming distinct masses $\wt\nu_i~(i=1,\cdots,D)$ for all lines, these $\Gamma$ factors are:
\begin{align}
  \prod_{i=1}^D\Gamma\Big[s_i+\FR{\ii\wt\nu_i}2,s_i-\FR{\ii\wt\nu_i}2,\bar s_i+\FR{\ii\wt\nu_i}2,\bar s_i-\FR{\ii\wt\nu_i}2\Big].
\end{align}
So the corresponding poles are:
\begin{align}
\label{eq_IRpoles}
   s_i=-n_i-\FR{\cc_i\ii\wt\nu_i}2,
   ~~~~~~\bar s_i=-\bar n_i-\FR{\dd_i\ii\wt\nu_i}2,
   ~~~~~~(i=1,\cdots,D)
\end{align}
where $\cc_i,\dd_i=\pm$. Note that these poles are from the MB representation of the bulk propagators, which correspond to a late-time expansion of the bulk propagators. The positions of these poles encode the information of the scaling dimensions of the late-time modes, and for this reason we shall call them infrared (IR) poles. As we shall show below, these poles are the only sources of nonlocal signals.

\end{enumerate}

For principal scalars we have $\wt\nu_i\in\mathbb R_+$ and thus the IR poles can generate noninteger powers for $K$ in the melon integral (\ref{eq_melonIntResult}). So we should consider all possible combinations of IR poles to get the full nonanalytic contributions to the graph. That is, we should consider all possible choices of $\cc_i$ and $\dd_i$ in (\ref{eq_IRpoles}).

However, some combinations of $\cc_i$ and $\dd_i$ do not yield nonanalytic results. There are two cases where this can happen. 

First, we have a string of $\Gamma$ factors $1/\Gamma[s_{1\bar 1},\cdots,s_{D\ob D}]$ in the melon loop integral (\ref{eq_melonIntResult}) which give a set of zeros of degree 1 when $s_{\ell\bar\ell}$ are nonpositive integers. These zeros show that we should discard all pole combinations with $\cc_i=-\dd_i$, and only keep the poles with $\cc_i=\dd_i$ for all $i=1,\cdots, D$. So, after this step, the possible signal poles reduce to: 
\begin{align}
\label{eq_NLpoles}
   s_i=-n_i-\FR{\cc_i\ii\wt\nu_i}2,
   ~~~~~~\bar s_i=-\bar n_i-\FR{\cc_i\ii\wt\nu_i}2,
   ~~~~~~(i=1,\cdots,D)
\end{align}

Second, it may happen that, for a particular choice of $\cc_i$, the combination $\sum\limits_{i=1}^D \cc_i\wt\nu_i$ vanishes. For instance, when the degree of cut $D$ is an even integer and all $D$ cut lines share the identical mass, we can take $D/2$ of $\cc_i$ to be $+1$ and the other $D/2$ of $\cc_i$ to be $-1$. The combination $\sum\cc_i\wt\nu_i$ is 0. Similar things can also happen when all masses are distinct. For instance, we can have $D=3$ and $\wt\nu_1=1$, $\wt\nu_2=2$, and $\wt\nu_3=3$. Then the choice $\cc_1=\cc_2=-\cc_3$ clearly makes the combination $\sum\cc_i\wt\nu_i$ vanish. In such situations, the melon integral (\ref{eq_melonIntResult}) would again give a positive integer power of $P$ and thus are analytic at $P=0$. So we should discard such combinations as well.

\paragraph{Nonlocal signal cutting rule.}
A direct consequence of choosing the nonlocal poles in (\ref{eq_NLpoles}) is that we can replace the propagators of all $D$ cut lines by their real parts, known as the nonlocal signal cutting rule \cite{Tong:2021wai,Qin:2022lva,Qin:2022fbv,Qin:2023bjk}. This is a direct generalization of the nonlocal signal cutting rule for one-loop graph proposed in \cite{Qin:2023bjk}. To see this, we note that the MB representation for a bulk propagator (\ref{eq_DScalarMB1}) contains a factor $e^{\mp\ii\pi(s_i-\bar s_i)}$. When evaluating this factor with the nonlocal poles (\ref{eq_NLpoles}), we get $(-1)^{n_i-\bar n_i}$ which is real. The whole propagator (\ref{eq_DScalarMB1}) then becomes real as well. This is why we can replace the cut propagators by their real parts. Furthermore, after this replacement, all four propagators $D_{\aa\bb}$ with $\aa,\bb=\pm$ become identical, and the time-ordering $\theta$-functions in $D_{\pm\pm}$ automatically disappear. So, the time integral factorizes into a left part and a right part.

Let us emphasize again that the cutting rule presented here is conceptually different from the nonlocal cut introduced earlier. As we showed above, the cutting rule $D_{\aa\bb}\to \text{Re}\,D_{\aa\bb}$ is a consequence of taking nonlocal signal poles (\ref{eq_NLpoles}) for all cut propagators. On the other hand, a nonlocal cut only indicates which propagators should be made soft in order to generate a nonlocal signal. It is not a priori clear that we can take nonlocal signal poles to integrate out Mellin variables of all the cut propagators. While the procedure of ``taking signal poles for all cut lines'' is always possible for bulk-free graphs, we will see in the next section that this is not the case for more general graphs.

\paragraph{Completing the Mellin integral.} 
Now we are ready to finish all the Mellin integrals in (\ref{eq_TkMellin}) for a bulk-free graph and complete the proof of Theorem \ref{thm_bulkfree}. Since we are only concerned with the leading nonlocal signal, we only need to collect residues of leading nonlocal poles for the Mellin variables of all cut lines, namely $n_i=\bar n_i=0$ for all $i=1,\cdots, D$ in (\ref{eq_NLpoles}). Then, the melon integral (\ref{eq_melonIntResult}), together with the factors from MB representations of cut lines (collectively denoted as $\mathbb{D}(s_i,\bar s_i)$ in (\ref{eq_TkMellin})), gives exactly the melon signal $\mathfrak{M}_{\cc_1\cdots\cc_D}(P)$ in (\ref{eq_frakM}). Since the time integrals have been fully factorized into a left part and a right part, finishing the rest of Mellin integrals simply generates the standard SK integrals for the left subgraph and the right subgraph, as shown in (\ref{eq_TL}) and (\ref{eq_TR}). In particular, the additional time-power insertion $(-\tau_i)^{3/2+\ii\cc_i\wt\nu_i}$ comes from the cut propagator, and is nothing but the leading mode of a bulk propagator in the late-time limit. Thus we see that, despite of all the complications of loops, the cut lines can still be expanded at the late-time limit so far as the nonlocal signal is the only concern.

\subsection{Discussions}

From the proof of Theorem \ref{thm_bulkfree}, we see that the assumption of bulk-free graphs brings two great simplifications. First, whenever nonlocal cuts exist, the minimal cut is guaranteed to be unique. Second, after taking the minimal cut, the resulting left and right subgraphs are both analytic and finite when $P\to 0$. Importantly, the arguments leading to these simplifications do not rely on the type of couplings and the dispersions of bulk fields. Thus, we can immediately loosen some of the conditions of Theorem \ref{thm_bulkfree} as follows.
\begin{enumerate}
  \item As a trivial extension, we can change the external lines to be (nearly) massless scalars (such as the inflaton fluctuations), massless spin-2 gravitons, or arbitrary mixtures of them. These are the most relevant cases for CC applications.
  \item We can generalize the type of internal fields to include nonzero spins and more exotic (and possibly dS boost breaking) dispersions. The tensor structure introduced by the spinning fields does not affect the analytical properties, and the factorization of the leading nonlocal signal into three pieces still holds in this case. Of course, the explicit expression such as the melon signal (\ref{eq_frakM}) will be changed, but it is still calculable so long as the MB representation of bulk propagators can be analytically found. As an example, we can consider a vector field of mass $m$ with dS-boost-breaking chemical potential $\mu$. In terms of the vector mass parameter $\wt\nu=\sqrt{m^2-1/4}$ and the dimensionless chemical potential $\wt\mu=\mu/H$, its propagator can be found to be \cite{Qin:2022fbv}:
  \bge
    D_{-+}^{(h)}=\FR{e^{-\pi h\wt\mu}}{2k}\mathrm{W}_{ \ii h\wt\mu,\ii\wt\nu}(2\ii k\tau_1)\mathrm{W}_{-\ii h\wt\mu,\ii\wt\nu}(-2\ii k\tau_2),
  \ede
where $\mathrm{W}_{\ka,\mu}(z)$ is the Whittaker W function.
The MB representation of this propagator is:
  \begin{align}
  \label{eq_WhitMB}
  D_{-+}^{(h)}=&~e^{-\pi h\wt\mu}\int_{-\ii\infty}^{\ii\infty}\FR{\di s}{2\pi\ii}\FR{\di \bar s}{2\pi\ii}\, e^{+\ii\pi (s-\bar s)/2}(2k)^{-(s+\bar s)}
    (-\tau_1)^{-s+1/2}(-\tau_2)^{-\bar s+1/2}\n\\
    &\times {}_2 {\mathcal{F}}_1\left[\bgm s-\ii\wt\nu,s+\ii\wt\nu\\s-\ii h\wt\mu+\fr12\edm\middle|\,\fr12\right]
    {}_2 {\mathcal{F}}_1\left[\bgm \bar s-\ii\wt\nu,\bar s+\ii\wt\nu\\ \bar s+\ii h\wt\mu+\fr12\edm\middle|\,\fr12\right],
  \end{align}
where ${}_2\mathcal{F}_1$ is the dressed hypergeometric function, defined in App.\ \ref{app_notation}. Thus we see that the pole structure of this propagator is very similar to the case of a massive scalar. The integrand also possesses two pairs of IR poles at $s=-n\mp\ii \wt\nu$ and $\bar s=-\bar n\mp\ii \wt\nu$.\footnote{The cutting rule, on the other hand, is obscured for the propagators with nonzero helical chemical potential if we use the representation (\ref{eq_WhitMB}). As shown in \cite{Qin:2022lva}, the MB representation is not unique, and in this case, one can use the so-called partially resolved MB representation for the bulk propagator to make the cutting rule manifest.}

  \item We can include arbitrary dS covariant or dS breaking couplings, such as uncontracted time derivatives or fully contracted space derivatives. A time derivative on external lines is inconsequential, while a time derivative on an internal line only shifts the position of right poles produced by the Mellin-space time integral, which is again irrelevant for the computation of nonlocal signals. On the other hand, the spatial derivatives all become factors of momenta, which do not affect the analytic property. The only caveat is that a spatial derivative on a cut line would make the signal decrease faster than nonderivative couplings as $P\to 0$. 
\end{enumerate}

\section{On-Shell Factorization of Arbitrary Graphs}
\label{sec_arbitrarygraph}

In the previous section, we studied nonlocal signals in arbitrary bulk-free graphs, and derived an explicit and factorized expression for the nonlocal signal in the given nonlocal soft limit. The restriction to bulk-free graphs implies that there are no internal vertices, and this significantly simplifies the analysis. As a result, we are able to extend the factorization theorem to rather general couplings and field species, as discussed above. 

However, we are ultimately interested in nonlocal signals in more general graphs, including arbitrary loop graphs with internal vertices. In this case, the analysis can be significantly involved. Thus, in the following, we shall first discuss several complications with loop graphs containing internal vertices, and then focus on a particular case where the minimal cut is unique with respect to a given bipartition to which we can directly generalize the on-shell factorization theorem proved in the previous section for bulk-free graphs. For graphs with multiple minimal cuts, we will see that there can be a peculiar type of nonlocal signals, which we call hybrid signals. This hybrid signal further complicates the analysis. Fortunately, as we shall see, such hybrid signals are absent for graphs with complete dS covariance. Thus, we can still formulate an on-shell factorization theorem for dS covariant graphs with most general loop topology. Below we spell out these points in turn.

\subsection{Complications with internal vertices}
\label{sec_complication}

\paragraph{Leading signal.}
The first complication for graphs with internal vertices is that there could be multiple distinct irreducible cuts associated with a given nonlocal bipartition. See Fig.\ \ref{fig_tribox} for an example. Then, according to Theorem \ref{thm_signal}, each of these cuts could potentially contribute a nonlocal signal and we need to collect all of them to get the full nonanalytic piece in the corresponding soft limit.

This complication is partially relieved if we only consider the leading signal. As shown in (\ref{eq_NLSignalDef}), each of the nonlocal signals in a given soft limit $P\to 0$ scale as $P^{\al_\ell\pm\ii\omega_\ell}$, where both $\al_\ell$ and $\omega_\ell$ are positive real numbers. Thus, the leading signal, namely the signal with the smallest $\al_\ell$, dominates over all nonlocal signals in the soft limit.    

It then remains to understand which cut gives the leading signal. Intuitively, we expect that a nonlocal signal would be more subleading if we cut more lines. The reason is that a nonlocal signal is generated from on-shell resonance of heavy particles, whose number density is diluted by the expanding physical volume $V$ according to $1/V\sim 1/a^3\sim (-\tau)^3$. From the analysis around (\ref{eq_resonance}), we learn that the late-time limit $\tau\to 0$ is translated to a squeezed limit $P\to 0$ in the correlation function. So, by asking one more internal line to contribute the nonlocal signal, we pay the price of an additional suppression factor of $P^3$. Indeed, this intuition is very well confirmed by the explicit expression of the melon signal (\ref{eq_frakM}), in which we can read the real part of the power of $P$ as $P^{3(D-1)}$, where $D$ is the degree of the cut.

The above analysis seems to suggest that we only need to consider minimal cuts in order to get the leading signal. However, there are further complications, as we shall discuss below.

\paragraph{Derivative couplings.}
An immediate complication comes from derivative couplings. In a general setup where the dS boosts are explicitly broken (such as in the EFT of inflation \cite{Cheung:2007st}), we can have spatial-derivative couplings without temporal-derivative counterparts. Suppose we have spatial derivatives $-\pd_i^2$ acting on an endpoint of a soft bulk propagator, then, in the momentum space, this gives rise to a factor of $P^2$ where $P$ is the momentum of the soft propagator. As a result, it might be possible that the signal from a minimal cut decreases even faster than the signal from a nonminimal cut, countering our earlier intuition. 

We note that only spatial derivatives change the scaling behavior of the signal in the soft limit. Neither the time derivatives nor the non-scale-invariant time powers at each interaction vertex generate similar complications. This can be very easily seen using the partial MB representation, in which the time and momentum variables are fully separated, and thus any change in the time integral $\mathbb{T}(\{k\};\{s,\bar s\})$ cannot alter the power counting of momentum partial sums in the momentum sector $\mathbb{L}(\{\mb k\};\{s,\bar s\})$.

The complication of derivative couplings can be addressed by defining the \emph{drop} $\Delta$ of a nonlocal cut, which measures how fast the would-be signal decreases in the late-time/squeezed limit. In the late-time limit, a massive propagator behaves like $k^{3}\times $nonanalytic factor. Then, when there are spatial derivatives, we can define the total drop of a cut of Degree $D$ as:
\begin{align}
  \Delta = 3D+\sum_{\ell=1}^D \de_i,
\end{align}
where $\de_i\geq 0$ denotes the total number of spatial derivatives acting on both sides of the $i$'th cut propagator. Then, the leading signal would be given by the cut of minimal drop $\Delta$ rather than the minimal degree $D$. This is indeed the case if there is a unique cut of minimal drop $\Delta$ in a given bipartition. Due to another complication we shall discuss below, when there are multiple cuts sharing the same minimal drop, the situation can be more complicated.

On the other hand, we note that the complication of spatial derivatives is always irrelevant in dS covariant graphs. The reason is simple: Whenever there is a derivative coupling in a dS covariant graph, the spatial and temporal derivatives must appear together. So, in the above example of $\ld\supset-(\pd_i^2\si)\cdots$ we must have terms like $\ld\supset [\si''-\pd_i^2\si]\cdots$. While the spatial derivatives make the signal decrease faster in the soft limit, the temporal derivatives do not. Thus, the temporal derivatives give the leading signal, whose scalings are identical to our naive counts of number of cuts. Thus, we conclude that, in dS covariant graphs, the leading nonlocal signal is given by the minimal cut, at least when the minimal cut is unique. 

Thus, in the following analysis, when we mention a minimal cut, we always mean a cut of minimal degree $D$ for dS covariant graphs, or a cut of minimal drop $\Delta$ for boost breaking graphs. 

There are additional complications with graphs containing internal vertices when the minimal cuts are not unique in a given nonlocal soft limit, to be discussed below. For now, the analysis of the derivative couplings is sufficient for us to formulate the on-shell factorization theorem for arbitrary loop diagrams with unique minimal cut. Thus, we shall first formulate this theorem in Sec.\ \ref{sec_uniquecut} before considering the most general case in Sec.\ \ref{sec_multiplecuts}.

\subsection{On-shell factorization with unique minimal cut}
\label{sec_uniquecut}

The above analysis shows that the complication from internal vertices can be largely circumvented if we only consider the leading signal, and if there exists a unique minimal cut associated with a given nonlocal bipartition. Equipped with this knowledge, we are ready to generalize the on-shell factorization to arbitrary loop diagrams with unique minimal cut, as summarized in the following theorem. As we shall see, this is the most general situation where we can formulate a theorem with certain mathematical rigor.

\begin{theorem}[On-shell factorization with unique minimal cut]\label{thm_UMC}
Let $\mathcal{G}(\{\mb k\})$ be an arbitrary $B$-point and $L$-loop graph with dS covariant propagators and couplings, and let $\{\mb k^\mathrm{(L)}\}\cup\{\mb k^\mathrm{(R)}\}$ be a nonlocal bipartition of $\mathcal{G}(\{\mb k\})$. Then, if there exists a unique minimal cut of degree $D$ associated with this nonlocal bipartition, then the nonlocal signal in the limit $\mb P\equiv\sum\mb k^\mathrm{(L)}\to\mb 0$ factorizes into the following three pieces at the leading order in $P$:
\begin{keyeqn}
\begin{align}
\label{eq_gfacUMC}
  \lim_{P\to 0}\mathcal{G}(\{\mb k\})
  = &\sum_{\substack{\cc_1,\cdots,\cc_D=\pm\\\text{nonlocal}}} \mathcal{G}_{\cc_1\cdots\cc_D}^\mathrm{(L)}\Big(\{\mb k^\mathrm{(L)}\}\Big)\mathcal{G}_{\cc_1\cdots\cc_D}^\mathrm{(R)}\Big(\{\mb k^\mathrm{(R)}\}\Big)\mathfrak{M}_{\cc_1\cdots\cc_D}(P)\n\\
  +&~\text{(subleading nonlocal signals)}+\text{(terms analytic in $P$)}.
\end{align}
\end{keyeqn}
\end{theorem}
Here the melon signal $\mathfrak{M}_{\cc_1\cdots\cc_D}(K)$, the left subgraph $\mathcal{G}_{\cc_1\cdots\cc_D}^\mathrm{(L)}(\{\mb k^\mathrm{(L)}\})$, and the right subgraph $\mathcal{G}_{\cc_1\cdots\cc_D}^\mathrm{(R)}(\{\mb k^\mathrm{(R)}\})$ are identical to the case of Theorem \ref{thm_bulkfree}, and are respectively given by (\ref{eq_TL}), (\ref{eq_TR}), and (\ref{eq_frakM}).\\[1mm]

\noindent\emph{Proof.} We focus on the nonlocal signal generated from the minimal cut with degree $D$. Since a minimal cut is necessarily irreducible, by Lemma \ref{lem_perbip}, it generates a perfect bipartition of the graph. As a result, the graph is fully resolved into a left subgraph, a right subgraph, and $D$ cut lines. Then, the proof goes in the same way as we did for Theorem \ref{thm_bulkfree} in the last section. The difference is that the left and right subgraphs are not guaranteed to be analytic in $q_\ell$ as $q_\ell \to 0$ where $q_\ell$ is the momentum of the $\ell$'th cut line ($\ell =1,\cdots, D$). Due to the existence of additional cuts, the left and right subgraphs can exhibit additional nonanalytic behaviors in $q_\ell \to 0$. However, thanks to the assumption of unique minimal cut, the nonlocal signals arising from those additional nonanalytic terms would be subleading in the soft limit $P\to 0$. 

Taking account of additional nonanalytic terms in the left and right subgraphs, we can write the soft limit $q_\ell\to 0$ of the Mellin-space loop integral of the left subgraph, as,
\begin{align}
\label{eq_facLC}
  \lim_{q_\ell\to 0}\mathbb{L}^\mathrm{(L)}\Big(\{\mb k^\mathrm{(L)}\},\{\mb q\}\Big)=\sum_{n=0}^\infty\FR{1}{n !}q^n\bigg[\pd^{(n)}\mathbb{L}^\mathrm{(L)}\Big(\{\mb k^\mathrm{(L)}\},\{\mb q\}\Big)\bigg]_{q=0}+\text{nonanalytic term}.
\end{align}
That is, the analytic part of the left loop integral can be Taylor expanded around $q_\ell \to 0$, and the leading order ($n=0$) term is generically nonzero. The nonanalytic term, on the other hand, typically decreases to zero in the $q_\ell\to 0$ limit, so long as the interactions are IR finite. Indeed, from the expression of the melon signal in (\ref{eq_frakM}), we see that a cut of degree $D$ yield a nonanalytic term (namely, a signal) that scales like $P^{3(D-1)+\ii\sum\cc_\ell\wt\nu_\ell}$. 

So, the leading contribution from the left subgraph in the $q_\ell\to 0$ limit is from the $q_\ell$-independent term of its analytic part. Then, the rest of the proof goes in exactly the same way as the last section. Now, so long as the couplings are dS covariant, the decreasing speed of a nonlocal signal is exactly given by the degree of the corresponding nonlocal cut. Therefore, it is guaranteed that the nonanalytic terms of left or right subgraphs are subdominant relative to the nonlocal signal from the minimal cut. So, the leading nonlocal signal must be from the term explicitly shown in (\ref{eq_facLC}). Then, going through all similar steps as in the proof of Theorem \ref{thm_bulkfree}, we end up with the factorized results in (\ref{eq_gfacUMC}). This completes the proof of Theorem \ref{thm_UMC}.

Clearly, the above proof can be readily generalized to the case of boost breaking interactions and dispersions. To address the complication of spatial derivatives, we only need to change the condition of unique minimal cut to the unique cut of minimal drop $\Delta$. It is also straightforward to see that the cut of minimal drop is necessarily an irreducible cut, so the perfect bipartition (left subgraph $+$ right subgraph $+$ cut lines) still applies by Lemma \ref{lem_perbip}.

\subsection{On-shell factorization with multiple minimal cuts}
\label{sec_multiplecuts}

Now let us consider the most general case where there exist more than one minimal cut in a given nonlocal bipartition. As alluded to several times, things get even more complicated if the minimal cut is not unique for a given bipartition. To appreciate the difficulty in this case, we need to introduce another subtlety associated with internal vertices, namely the $\de$-constraint in the Mellin space.

\paragraph{$\bm{\de}$-constraint from internal vertices.} The subtlety is this:
 While it may be possible to cut all the bulk lines attached to a given internal vertex, it is impossible to get nonlocal signals from all these lines, due to the $\delta$-constraint of an internal vertex in the Mellin space. 

To see this point more explicitly, we consider an internal vertex with the time variable $\tau$ and let there be $N$ bulk propagators attached to this vertex. Then, if we cut all these $N$ bulk propagators, we may want to naively apply the cutting rule $D_{\aa\bb}\to \text{Re}\,D_{\aa\bb}$, which would allow us to remove all the time orderings between $\tau$ and any other time variables. In this case, the $\tau$-integral can be isolated:
\begin{align}
  \int_{-\infty}^0\di \tau(-\tau)^{p}\prod_{i=1}^N\,\text{Re}\,D^{(\wt\nu_i)}(p_i,\tau,\tau_i).
\end{align}
Now, in the Mellin space, each propagator $D^{(\wt\nu_i)}$ gives rise to a factor of $(-\tau)^{-2s_{i\bar i}}$. Thus, the time integral yields a $\de$-function:
\bge
\label{eq_IVdelta}
  \int_{-\infty}^0\di \tau(-\tau)^{p-2s_{1\bar1\cdots N\ob{N}}}=2\pi\de\big[ \ii(1+p-2s_{1\bar 1\cdots N\ob{N}})\big].
\ede
On the other hand, as we showed in the last section, for all $N$ lines to contribute nonlocal signals, it is important to take the (IR) signal poles. In the current situation, it means that we should choose:
\begin{align}
  &s_1=-n_1+\FR{\ii\cc_1\wt\nu_1}2,\qquad \bar s_1=-\bar n_1+\FR{\ii\cc_1\wt\nu_1}2\cdots,\n\\
  &s_N= -n_N+\FR{\ii\cc_N\wt\nu_N}2,\qquad \bar s_N=-\bar n_N+\FR{\ii\cc_N\wt\nu_N}2,
\end{align}
where $\cc_1,\cdots,\cc_N=\pm1$. So, the possibility that all these lines together contribute a nonlocal signal with $\sum\ii\cc_i\wt\nu_i\notin \mathbb{R}$ is obviously inconsistent with the $\de$-function constraint (\ref{eq_IVdelta}). This gives us a sharp example showing that, while we can take a nonlocal cut by making certain lines soft, we cannot get nonlocal signals from all these soft lines. As a result, the cutting rule does not apply. 

The $\de$-constraints add another complication to our analysis. In the partial MB representation, we need to finish all Mellin integrals to recover the ordinary expressions for the graph. As shown in the proof of Theorem \ref{thm_bulkfree}, we only need to pick up left poles of all Mellin variables, and the only left poles are UV poles from the loop momentum integrals, and the IR poles from the MB representation of the bulk massive propagators. Now, the $\de$-constraint introduces another possibility: we can carry out integration over some Mellin variables by the $\de$-constraint, and the resulting expression would flip the side of poles of other Mellin variables. While we can postpone the use of $\de$-constraint to deal with the side-flipping problem, we cannot avoid the $\de$-constraint altogether. As we shall see below, the $\de$-constraint may lead to peculiar signals when there are multiple minimal cuts in a graph.

\paragraph{Hybrid signal from multiple minimal cuts.} Now we come back to the discussion of multiple minimal cuts. In this case, our first guess may be that the leading nonlocal signal in the soft limit is the sum of the nonlocal signals corresponding to each of the nonlocal cuts. However, our experience with explicit examples suggests that there can be additional contributions to the leading nonlocal signal, which can have the same fall-off as those signals from a minimal cut, but can not be obtained by adding nonlocal signals from all minimal cuts alone. This is closely related to the $\de$-constraint discussed above. 

At this point, we suggest the readers have a look at the two-point mixing example in Sec.\ \ref{sec_2ptmix} before reading on. To see the peculiarity of this example, we note that, in all examples hitherto considered, the nonlocal signal was obtained by taking signal poles for both of Mellin variables of all cut lines: 
\begin{align}
  &s= -n\mp \ii\wt\nu/2,
  &&\bar s= -\bar n\mp \ii\wt\nu/2.
\end{align}
 Using our intuition of on-shell resonance, we may say that the signal comes from the resonance of both endpoints of a cut propagator. Now, we see that, in the example of 2-point mixing graph in Fig.\ \ref{fig_2ptmix}, one can generate a nonlocal signal by asking the two modes attached to the two cubic vertices to resonate, while the two modes at the two-point mixing vertex are ``locked'' by the $\de$-constraint. This yields a nonlocal signal that scales with the same real power as the signal from a single soft propagator, namely $P^0$, but with its imaginary power being $P^{\pm\ii\wt\nu_1\pm\ii\wt\nu_2}$, which we call a \emph{hybrid signal}. So, in this two-point mixing example, we would expect that the leading signals have three terms:
\begin{align}
  &P^{\pm 2\ii\wt\nu_1},
  &&P^{\pm 2\ii\wt\nu_2},
  &&P^{\pm\ii\wt\nu_1\pm\ii\wt\nu_2}.
\end{align}
The first two are ordinary signals, corresponding to cutting only the left internal line or only the right internal line in Fig.\ \ref{fig_2ptmix}, and requiring both modes of a single line to resonate. The last one, namely the hybrid signal, corresponds the peculiar situation where we cut two lines, but the resulting signal scales as if it is from cutting one line.
 
We expect that this feature also appears in more general loop graphs whenever there are multiple minimal cuts: The full leading signal is not only from the ordinary signal of each individual minimal cut, but also include hybrid signals from several one-side resonances. It is actually not clear if we would get more peculiar signals in more complicated loop graphs, and this is the major reason that we cannot get a simple and neat factorization theorem for the most general loop diagram.
 
One possible way to see the appearance of the hybrid signal is to think of a loop graph with internal vertices as a soft limit of a bulk-free graph without any internal vertices: We can artificially insert auxiliary external legs with energies $E_i$ to all internal vertices and thus turn them into external vertices. Then, we get a bulk-free graph which appears much simpler. The original graph can then be recovered by taking $E_i\to 0$ for all the auxiliary external legs. Then, one may attempt to think of the hybrid signal as a combination of ordinary signals in the bulk-free graph in the corresponding soft limit. However, the real situation is not as trivial. The reason is that, when we talk about ordinary signals, we always assume that the momenta $P_j$ of the signal-generating bulk propagators are much softer than any other external momenta, where the resonance picture is well applicable. That is, we need $P_j/E_i\ll 1$ for all $i$ and $j$. However, when we turn off the momenta of all auxiliary external legs, the above assumption no longer holds. As a consequence, we need to know how to take analytical continuation of the bulk-free graph from the region $P_j/E_i\ll1$ to $P_j/E_i\to \infty$. This is a nontrivial task which we leave for a future study.

\paragraph{Absence of hybrid signals in dS covariant graphs.} Although the hybrid signals make the analysis of nonlocal signals rather intractable, they are likely absent in dS covariant graphs. In the example of two-point mixing graph in Sec.\ \ref{sec_2ptmix}, the absence of the hybrid signal in the dS covariant limit is clearly observed. There, we have a simple explanation: a covariant mixing (such as a mass mixing) can always be rotated away by a linear field redefinition. Of course this explanation does not apply directly to arbitrary loop graphs, in which case we have to resort to other arguments. 

\begin{figure}
\centering
\includegraphics[width=\textwidth]{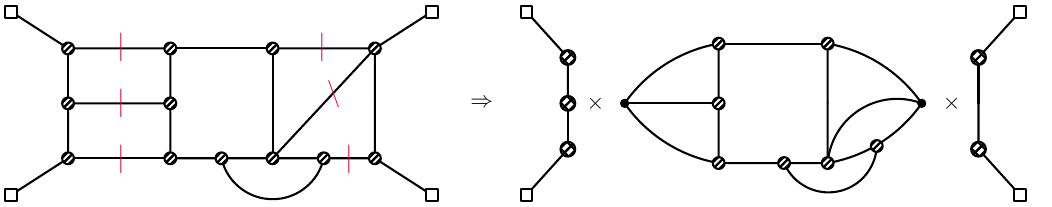}
\caption{Illustration of the absence of hybrid signals in dS covariant graphs. The leading nonlocal signal in any loop diagram can be understood as a two-point function on the late-time boundary. In a dS covariant graph, the resulting two-point function respects the boundary conformal symmetry, and thus must be diagonal in the scaling dimensions at its two endpoints. This precludes all hybrid signals.} 
  \label{fig_doublecut}
\end{figure}

Fortunately, we can find such an argument from CFT, using our interpretation of nonlocal signals as two-point function at the future boundary. (See the end of Sec.\ \ref{sec_scatamp}, and also \cite{Qin:2023bjk} for more detailed expositions.) From our proof of Theorem \ref{thm_bulkfree}, we see that the leading nonlocal signal is obtained by cutting open all soft lines while pitching all hard lines. Here the ``pitching'' procedure means that we can take all hard lines out of the loop integral as we did around (\ref{eq_Lkfactor}). As a result, nonlocal signals in arbitrary loop graphs can also be viewed as the late-time limit of a two-point function. We illustrate this ``cut and pinch'' procedure in Fig.\ \ref{fig_doublecut}. Then, so long as the bulk graph is dS covariant, the corresponding boundary two-point correlator will be conformal covariant, and thus diagonal in the scaling dimensions:
\bge
  \lim_{\tau_1,\tau_2\to 0}G(P;\tau_1,\tau_2)=\sum_{\Delta_1,\Delta_2} G_{\Delta_1\Delta_2}(P)(-\tau_1)^{\Delta_1}(-\tau_2)^{\Delta_2},
\ede
where $\Delta_{1,2}$ denote the scaling dimension of the boundary two-point function $G_{\Delta_1\Delta_2}(P)\propto P^{\Delta_1+\Delta_2-3}$. The crucial point is that bulk dS isometry (or equivalently, the boundary conformal symmetry) demands that $\Delta_1=\Delta_2$ or $\Delta_1=\ob{\Delta}_2$ where $\ob{\Delta}_2=3-\Delta_2$ is the shadow counterpart of $\Delta_2$. The latter case with $\Delta_1=\ob{\Delta}_2$ is irrelevant to us since it makes $G(P)$ independent of $P$ and only produces a contact term $\propto\delta^{(3)}(\mb x)$ in position space.\footnote{However, in this part, there is still a factor $(-\tau_1)^{\Delta_1}(-\tau_2)^{\ob{\Delta}_2}$ which could be nonanalytic in the time variables. They would produce a local signal instead of nonlocal signal.} Thus the only possible nonanalytic behavior must have $\Delta_1=\Delta_2\equiv\Delta$. Therefore the hybrid signal with $\wt\nu_1\neq\wt\nu_2$ is precluded since it corresponds to $G(P;\tau_1,\tau_2)\propto (-P\tau_1)^{\pm\ii\wt\nu_1}(-P\tau_2)^{\pm\ii\wt\nu_2}$ in the late-time limit, inconsistent with the conformal constraint. 

\paragraph{On-shell factorization of most general dS covariant graphs.} Above we have shown that the main complication from internal vertices with non-unique minimal cuts is the generation of hybrid signals, however, this complication is absent if the graph respects the full dS isometry. Thus, if we restrict ourselves to dS covariant processes, it would be possible to formulate the following theorem for leading nonlocal signals in arbitrary loop graphs: 

\begin{theorem}[Nonlocal signals in general graphs]\label{thm_generalgraph}
Let $\mathcal{G}(\{\mb k\})$ be an arbitrary $B$-point ($B\geq 4$) and $L$-loop ($L\geq 0$) graph with dS covariant propagators and couplings, and let $\{\mb k^\mathrm{(L)}\}\cup\{\mb k^\mathrm{(R)}\}$ be a nonlocal bipartition of $\mathcal{G}(\{\mb k\})$. Then, the leading nonlocal signal in the limit $\mb P\equiv\sum\mb k^\mathrm{(L)}\to\mb 0$ is contributed (probably exclusively) by the sum of signals coming from all minimal cuts:
\begin{align}
\label{eq_gfacMMC}
  \lim_{P\to 0}\mathcal{G}(\{\mb k\})
  = &\sum_\text{C}\sum_{\substack{\cc_1,\cdots,\cc_{D}=\pm\\\text{nonlocal}}} \mathcal{G}_{\cc_1\cdots\cc_{D}}^{\mathrm{L}(C)}\Big(\{\mb k^\mathrm{(L)}\}\Big)\mathcal{G}_{\cc_1\cdots\cc_{D}}^{\mathrm{R}(C)}\Big(\{\mb k^\mathrm{(R)}\}\Big)\mathfrak{M}^{(C)}_{\cc_1\cdots\cc_{D}}(K) \n\\
  +&~\text{(other nonlocal signals)}+\text{(terms analytic in $K$)}.
\end{align}
\end{theorem}
Here $C$ labels all minimal cuts with respect to the given bipartition, and $D$ denotes the degree of the minimal cuts. The three factors $\mathcal{G}_{\cc_1\cdots\cc_{D}}^{\mathrm{L}(C)}(\{\mb k^\mathrm{(L)}\})$, $\mathcal{G}_{\cc_1\cdots\cc_{D}}^{\mathrm{R}(C)}(\{\mb k^\mathrm{(R)}\})$, $\mathfrak{M}^{(C)}_{\cc_1\cdots\cc_{D}}(K)$ denote the left subgraph, the right subgraph, and the melon signal generated by the cut $C$. The term denoted by ``other nonlocal signals'' is probably subleading in the $P\to 0$ limit.

\subsection{Discussions}
Theorem \ref{thm_generalgraph} shows that the leading nonlocal signal of an arbitrary (dS covariant) graph is always generated by one (or several if the minimal cut is not unique) $(D-1)$-loop melon subgraph with $D\geq 1$, which is formed by all the cut lines in the corresponding minimal cut. This can be intuitively understood as from a ``cut and pinch" process: We first identify a minimal cut that could generate a melon signal, and then pinch all the other lines and loops. There is also a boundary OPE perspective: The cut lines are soft and we can push them to the future boundary of dS spacetime, where the massive modes become boundary operators with scaling dimensions $3/2\pm\ii\wt\nu_\ell$. Then we can do the OPE to the endpoints of the cut lines, such that they are pinched together and the soft lines become a melon graph. See \cite{Qin:2023bjk} for more details.

It is interesting to see how we can pinch a loop. In fact, this is related to the UV pole of the Mellin-space loop momentum integral $\mathbb{L}(\{\mb k\};\{s,\bar s\})$ in \eqref{eq_MellinLoopInt}. The simplest example is a bubble integral (namely the 1-loop melon integral):
\begin{align}
\mathbb{M}_1(P)= \int \FR{\di^3\mb q}{(2\pi)^3}|\mb q|^{-2s_{1\bar1}}|\mb P-\mb q|^{-2s_{2\bar2}} = \FR{P^{3-2s_{1\bar12\bar2}}}{(4\pi)^{3/2}}\Gamma\bgb s_{1\bar12\bar2}-\fr32,\fr32-s_{1\bar1},\fr32-s_{2\bar2}\\3-s_{1\bar12\bar2},s_{1\bar1},s_{2\bar2}\edb.
\end{align}
Clearly, the integral $\mathbb{M}_1(P)$ possesses a set of UV poles at  $s_{1\bar12\bar2}=-n+3/2$ with $n=0,1,\cdots$. If we take the residue of $\mathbb{M}_1(P)$ at the leading UV pole $s_{1\bar12\bar2}=3/2$, we get:
\bge
\underset{s_{1\bar12\bar2}=3/2}{\text{Res}}\Big[\mathbb{M}_1(P)\Big] = \FR{1}{4\pi^2}.
\ede
This shows that, at the leading UV pole, the residue of the bubble integral reduces to a constant number. In this sense, the bubble integral is ``pinched" to a point.

This is in fact a more general statement. Consider a particular layer of arbitrary loop momentum integral in the Mellin space, which corresponds to a particular sub-loop structure in the original loop graph:
\bge
\mathbb L = \int \FR{\di^3\mb q}{(2\pi)^3} \prod_{i=1}^N |\mb q + \mb p_i|^{-2s_{i\bar i}},
\ede
where $\mb p_i$ is an arbitrary combination of external momenta and loop momenta other than $\mb q$.
In general, $\mathbb L$ is a function of $\{s_i,\bar s_i\}$ with some poles: When $\{s_i,\bar s_i\}$ take certain values, this integral could diverge  in the hard region of $\mb q$, and the divergent part can be obtained by:
\bge
\mathbb L|_{\text{divergent part}} = \int_{\Lambda}^\infty \FR{\di^3\mb q}{(2\pi)^3} \prod_{i=1}^N |\mb q + \mb p_i|^{-2s_{i\bar i}}.
\ede
Here $\Lambda$ is a hard scale: $\Lambda \gg |\mb p_i|$. Thus we can then neglect all the $\mb p_i$, and therefore:
\bge
\mathbb L|_{\text{divergent part}} = \int_{\Lambda}^\infty \FR{\di^3\mb q}{(2\pi)^3} \prod_{i=1}^N |\mb q |^{-2s_{i\bar i}} = \int_{\Lambda}^\infty \FR{\di^3\mb q}{(2\pi)^3}  |\mb q |^{-2\sum s_{i\bar i}} .
\ede
Now notice that:
\bge
 \int \FR{\di^3\mb q}{(2\pi)^3}  |\mb q |^{-2\sum s_{i\bar i}} = \FR{1}{\pi} \de\big[\ii(3-2\sum s_{i\bar i}) \big],
\ede
we can then extract the possible UV divergence of $\mathbb L$:
\begin{align}
\mathbb L|_{\text{divergent part}} =& \int \FR{\di^3\mb q}{(2\pi)^3}  |\mb q |^{-2\sum s_{i\bar i}}-\int_0^{\Lambda} \FR{\di^3\mb q}{(2\pi)^3}  |\mb q |^{-2\sum s_{i\bar i}}\n\\
=&~\FR{1}{\pi} \de\big[\ii(3-2\sum s_{i\bar i}) \big]-\FR{1}{2\pi^2}\FR{\Lambda^{3-2\sum s_{i\bar i}}}{3-2\sum s_{i\bar i}}.\end{align}
Therefore, $\mathbb L$ indeed has a UV pole $\sum s_{i\bar i}=3/2$, and the residue is:
\bge
\underset{\sum s_{i\bar i}=3/2}{\text{Res}} \, \mathbb L  = \FR{1}{4\pi^2}.
\ede
That is, when we calculate the Mellin integral, we are able to take the UV pole of any loop integral and pinch the loop to a constant.

Now we can understand our theorem \eqref{eq_gfacMMC} with this pinch idea. All the soft lines will be diluted due to the exponential expansion of the spacetime. Obviously we cannot pinch all the loops, otherwise we will not obtain a signal. So we should pinch all the lines and loops but leave one melon subgraph with minimal degree, which corresponds to a minimal cut. If there are multiple minimal cuts, we should sum the signals together since we should sum residues at all the allowable pole configurations.

\section{Examples}
\label{sec_example}

Below we will present some concrete examples of four-point inflation correlators, and explicitly show how our theorems work. For definiteness, we set all the external lines to be the conformal scalar $\varphi$ with $m^2=2$, all the bulk lines to be massive scalars $\si_\ell$ in the principal series with different mass parameters $\wt\nu_\ell>0$. It is assumed that all fields are coupled directly without any spatial or temporal derivatives, but with arbitrary power dependences $(-\tau_i)^{p_i}$ in the coupling. For simplicity we also assume that $\wt\nu_\ell$ take generic values such that $\sum\limits_\ell \cc_\ell\wt\nu_\ell\neq 0$ for any $\cc_\ell\in\{0,\pm1\}$ with at least one $\cc_\ell\neq 0$. We will compute the Mellin-space loop integral \eqref{eq_MellinLoopInt} and analyze its possible pole structures which could generate a nonlocal signal.

\paragraph{Notations.}
In this section, we use solid lines to denote massless inflaton, and the external momenta $\mb k_1,\cdots,\mb k_4$ are flowing inwards. We specify the nonlocal soft limit to be the s-channel squeezed limit, namely $\mb k_s\equiv \mb k_1+\mb k_2=-\mb k_3-\mb k_4$ and $\mb k_s \to 0$. 
Each internal line $\ell$ is labeled by its mass $\wt\nu_\ell$. In the Mellin space, we introduce two Mellin variables $s_\ell, \bar s_\ell$ to the two endpoints of the bulk line $\ell$, respectively. We use an arrow to clarify our choice: The arrow always points from $s_\ell$ to $\bar s_\ell$. The independent loop momenta $\mb q_i$ will be explicitly labeled in graphs, flowing in the arrowed direction.

\subsection{Melon graph}
\begin{figure}
\centering
\includegraphics[width=0.53\textwidth]{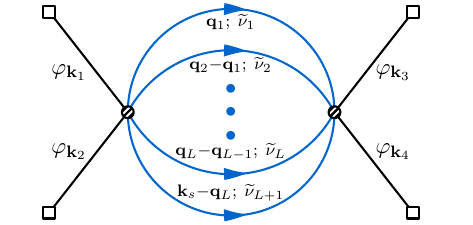}
\caption{The 4-point $L$-loop melon graph.} 
  \label{fig_melon}
\end{figure}

As a first example, let us consider the $L$-loop melon graph as shown in Fig.\ \ref{fig_melon}.  This graph is bulk-free, and there is only one nonlocal cut, namely cutting all the $(L+1)$ massive propagators simultaneously.

Let us write down the expression for the melon graph. We neglect the constant factor $-\tau_f/(2k)$ in the bulk-to-boundary propagator \eqref{eq_CSProp}, and absorb the factor $-\tau$ into the interaction $(-\tau_i)^{p_i}$, then the melon graph is expressed as the following:
\begin{align}
  \mathcal{G}=-\sum_{\aa,\bb=\pm}\aa\bb\int\di\tau_1\di\tau_2\,(-\tau_1)^{p_1}(-\tau_2)^{p_2}e^{+\ii\aa k_{12}\tau_1+\ii\bb k_{34}\tau_2}\mathcal{M}^{(L)}(k_s;\tau_1,\tau_2).
\end{align}
Here $\mathcal{M}^{(L)}(k_s;\tau_1,\tau_2)$ is the $L$-loop integral of the $L+1$ massive propagators:
\begin{align}
  \mathcal{M}^{(L)}(k_s;\tau_1,\tau_2)=&\int\prod_{i=1}^{L}\bigg[\FR{\di^3\mb q_i}{(2\pi)^3}\bigg]D_{\aa\bb}^{(\wt\nu_1)}\Big(|\mb q_1|;\tau_1,\tau_2\Big)D_{\aa\bb}^{(\wt\nu_2)}\Big(|\mb q_2-\mb q_1|;\tau_1,\tau_2\Big)\n\\
  &\times\cdots D_{\aa\bb}^{(\wt\nu_{L})}\Big(|\mb q_L-\mb q_{L-1}|;\tau_1,\tau_2\Big)D_{\aa\bb}^{(\wt\nu_{L+1})}\Big(|\mb k-\mb q_L|;\tau_1,\tau_2\Big).
\end{align}
Then we apply partial MB representation to extract the nonlocal signal from the Mellin-space loop integral \eqref{eq_MellinLoopInt}, and the procedure is totally parallel with the proof of Theorem \ref{thm_bulkfree}.
The result is given by \eqref{eq_bulkfreefac}, and the nonlocal signal is:
\begin{align}
\label{eq_GNLmelon}
\mathcal{G}_{\text{NL}}=\sum_{\substack{\cc_1,\cdots,\cc_{L+1}=\pm\\\mathrm{nonlocal}}} \mathcal{G}_{\cc_1\cdots\cc_{L+1}}^\mathrm{(L)}(k_{12})\mathcal{G}_{\cc_1\cdots\cc_{L+1}}^\mathrm{(R)}(k_{34})\mathfrak{M}_{\cc_1\cdots\cc_{L+1}}(k_s).
\end{align}
As we can see, the nonanalyticity of $k_s$ is fully encoded in the melon signal $\mathfrak{M}_{\cc_1\cdots\cc_{L+1}}(k_s)$, whose explicit form of is given by \eqref{eq_frakM}. We also find that the signal contains $2^L$ (generally different) frequencies with respect to $\log k_s$, given by $|\omega_{\cc_1\cdots\cc_{L+1}}|$:
\bge
\omega_{\cc_1\cdots\cc_{L+1}} \equiv 2\sum_{\ell=1}^{L+1} \cc_\ell\wt\nu_\ell,
\ede
where $\cc_1,\cdots,\cc_L$ can take values from $\pm$. Each choice gives rise to a value of the frequency, modulo the overall sign-flip of $\cc_\ell$. 

Furthermore, the left and the right subgraphs in (\ref{eq_GNLmelon}) can be directly calculated by \eqref{eq_TL} and \eqref{eq_TR}, and the results are:
\begin{align}
  \mathcal{G}_{\cc_1\cdots\cc_{L+1}}^\mathrm{(L)}(k_{12})
  =&\sum_{\aa=\pm}\int(\ii\aa)\,\di\tau_1\,(-\tau_1)^{p_1}e^{\aa\ii k_{12}\tau_1}{\prod_{\ell=1}^{L+1}(-\tau_1)^{3/2+\cc_\ell\ii\wt\nu_\ell}}\n\\
 =&~ 2k_{12}^{-1-P_1}\cos\Big(\FR\pi2 P_1\Big)\Gamma(1+P_1),\\
  \mathcal{G}_{\cc_1\cdots\cc_{L+1}}^\mathrm{(R)}(k_{34})
  =&\sum_{\bb=\pm}\int(\ii\bb)\,\di\tau_2\,(-\tau_2)^{p_2}e^{\bb\ii k_{34}\tau_2}{\prod_{\ell=1}^{L+1}(-\tau_2)^{3/2+\cc_\ell\ii\wt\nu_\ell}}\n\\
 =&~ 2k_{34}^{-1-P_2}\cos\Big(\FR\pi2 P_2\Big)\Gamma(1+P_2),
\end{align}
where $P_{1,2}\equiv p_{1,2}+3(L+1)/2+\ii\omega_{\cc_1\cdots\cc_{L+1}}/2$.

\subsection{Two-point mixing}
\label{sec_2ptmix}

\begin{figure}
\centering
\includegraphics[width=0.55\textwidth]{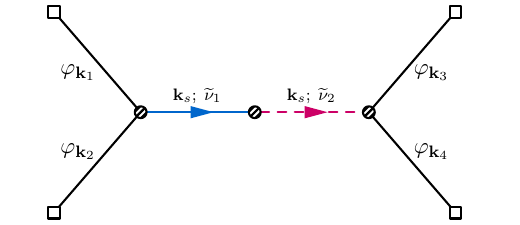}
\caption{The 4-point correlator with tree-level $s$-channel exchange of two mixed massive modes. The mixing is generally not dS covariant. } 
  \label{fig_2ptmix}
\end{figure}

From now on, let us consider graphs that contain internal vertices. The simplest case is a tree-level two-point mixing, shown in Fig.\ \ref{fig_2ptmix}. Given the subtleties associated with internal vertices, we will investigate this example more closely than the previous one. The purpose of this example is to explicitly show the appearance of a hybrid signal for non-dS covariant couplings and the absence of this signal for dS covariant couplings.

Since we are looking for nonlocal signals, let us assume that we can cut the graph as usual. The expression for this graph is the following:
\begin{align}
 \mathcal{G}_{\aa\bb\cc}=-\ii \aa\bb\cc \int\prod_{i=1}^3\Big[\di\tau_i(-\tau_i)^{p_i}
\Big]e^{\ii\aa k_{12}\tau_1+\ii\bb k_{34}\tau_2}D_{\aa\cc}^{(\wt\nu_1)}(k_s;\tau_1,\tau_3)D_{\cc\bb}^{(\wt\nu_2)}(k_s;\tau_3,\tau_2).
\end{align}
As always, we neglected unimportant prefactors from the bulk-to-boundary propagators.
We first cut the $\wt\nu_1$ propagator. It is not a priori clear that we can take a symmetric cut $D\To \text{Re}\,D$. So, we take a safer cut:
\bge
  D_{++}^{(\wt\nu_1)}(k_s;\tau_1,\tau_3)
  \To D_{-+}^{(\wt\nu_1)}(k_s;\tau_1,\tau_3)+{\blue\Big[D_{+-}^{(\wt\nu_1)}(k_s;\tau_1,\tau_3)-D_{-+}^{(\wt\nu_1)}(k_s;\tau_1,\tau_3)\Big]\theta(\tau_3-\tau_1)}.
\ede 
By a cut we mean to remove the blue term. Similarly,
\bge
  D_{++}^{(\wt\nu_2)}(k_s;\tau_3,\tau_2)
  \To D_{+-}^{(\wt\nu_2)}(k_s;\tau_3,\tau_2)+\Big[D_{-+}^{(\wt\nu_2)}(k_s;\tau_3,\tau_2)-D_{+-}^{(\wt\nu_2)}(k_s;\tau_3,\tau_2)\Big]\theta(\tau_3-\tau_2),
\ede 
but here we do not remove any term.

 Under this cut, the integral of $\tau_1$ and $\tau_2$ is factorized. The original integral now becomes two parts: A (fully) factorized (F) part, and a (partially) nested (N) part which involves integral of $\theta(\tau_3-\tau_2)$. For example, we can write $\mathcal{G}_{+\pm\pm}=\mathcal{G}_{+\pm\pm}^\text{(F)}+\mathcal{G}_{+\pm\pm}^\text{(N)}$. 

Let us first look at the nested part:
\begin{align}
 &\mathcal{G}_{+++}^\text{(N)}=-\ii\int_\tau D_{-+}^{(\wt\nu_1)}(D_{+-}^{(\wt\nu_2)}-D_{-+}^{(\wt\nu_2)})\theta(\tau_3-\tau_2)e^{+\ii k_{12}\tau_1+\ii k_{34}\tau_2},
\end{align}
where $\int_\tau\cdots\equiv\int\prod\limits_{i=1}^3[\di\tau_i(-\tau_i)^{p_i}]\cdots$. We use the MB representation for the massive propagators:
\begin{align}
\label{eq_G+++N}
\mathcal{G}_{+++}^\text{(N)}=&~\FR{-\ii}{(4\pi)^2}\int_s \Big(\FR{k_s}2\Big)^{-2s_{1\bar12\bar2}}
e^{+\ii\pi(s_1-\bar s_1)}(e^{-\ii\pi(s_2-\bar s_2)}-e^{+\ii\pi(s_2-\bar s_2)})\Gamma\Big[\text{IR poles}\Big]\n\\
&\times \int_\tau \theta(\tau_3-\tau_2)e^{+\ii k_{12}\tau_1+\ii k_{34}\tau_2}.
\end{align}
Here and below, we use $\int_s\cdots\equiv\int\prod\limits_{i=1}^2\big(\fr{\di s_i}{2\pi\ii}\fr{\di \bar s_i}{2\pi\ii}\big)$, and, 
\bge
\Gamma\Big[\text{IR poles}\Big]\equiv\prod_{i=1}^2\Gamma\Big[s_i+\FR{\ii\wt\nu_i}2,s_i-\FR{\ii\wt\nu_i}2,\bar s_i+\FR{\ii\wt\nu_i}2,\bar s_i-\FR{\ii\wt\nu_i}2\Big].
\ede
The other four branches can be calculated similarly. It is easy to check that the time integral will only give right poles of the Mellin variables. The next step is to integrate out these Mellin variables. Since the integrand of \eqref{eq_G+++N} is proportional to $k_s^{-2s_{1\bar12\bar2}}$ and $k_s\to0$ in the squeezed limit, we should sum residues at left poles, and the only left poles are the IR poles. The leading poles are the following:
\bge
s_1 = - \cc_1\ii\wt\nu_1/2,\qquad \bar s_1 =  -\cc_2\ii \wt\nu_1,\qquad s_2 = - \cc_3\ii\wt\nu_2,\qquad \bar s_2 = - \cc_4\ii\wt\nu_2.
\ede
However, the factor $e^{-\ii\pi(s_2-\bar s_2)}-e^{+\ii\pi(s_2-\bar s_2)}$ in the integrand of \eqref{eq_G+++N} requires that $\cc_3=-\cc_4$, otherwise $\mathcal G^{(\text{N})}_{+++}$ would vanish. Furthermore, to generate a signal with nonzero frequency, we should take $\cc_1=\cc_2$ such that $\mathcal G^{(\text{N})}_{+++} \propto k_s^{\pm 2\ii\wt\nu_1}$. This is the case that corresponds to a minimal cut on Line 1. Of course there is a similar signal $\propto k_s^{\pm2\ii \wt\nu_2}$ corresponding to a minimal cut on Line 2. These signals are ordinary signals of Fig.\ \ref{fig_2ptmix}.

Now let us look at the factorized part. In this part, we have:
\begin{align}
  &\mathcal{G}_{+++}^\text{(F)}=-\ii\int_\tau D_{-+}^{(\wt\nu_1)}D_{+-}^{(\wt\nu_2)}e^{+\ii k_{12}\tau_1+\ii k_{34}\tau_2} ,
  &&\mathcal{G}_{++-}=+\ii\int_\tau D_{-+}^{(\wt\nu_1)}D_{+-}^{(\wt\nu_2)}e^{+\ii k_{12}\tau_1-\ii k_{34}\tau_2}, \\
  &\mathcal{G}_{+-+}^\text{(F)}=+\ii\int_\tau D_{+-}^{(\wt\nu_1)}D_{-+}^{(\wt\nu_2)}e^{+\ii k_{12}\tau_1+\ii k_{34}\tau_2} ,
  &&\mathcal{G}_{+--}^\text{(F)}=-\ii\int_\tau D_{+-}^{(\wt\nu_1)}D_{-+}^{(\wt\nu_2)}e^{+\ii k_{12}\tau_1-\ii k_{34}\tau_2}.
\end{align}
The other four branches can be directly obtained by taking the complex conjugate. In total,
\begin{align}
  \mathcal{G}^\text{(F)}
  \equiv&\sum_{\aa,\bb,\cc=\pm}\mathcal{G}_{\aa\bb\cc}^\text{(F)} 
  = 8\int_\tau\text{Im}\Big[D_{+-}^{(\wt\nu_1)}(k_3;\tau_1,\tau_3)D_{-+}^{(\wt\nu_2)}(k_3;\tau_3,\tau_2)\Big]\sin(k_{12}\tau_1)\sin(k_{34}\tau_2).
\end{align}
Now, using the MB representation for the massive propagators, we get:
\begin{align}
\label{eq_GF}
  \mathcal{G}^\text{(F)} 
  =\FR{8}{(4\pi)^2}\int_{s}\Big(\FR{k_s}2\Big)^{-2s_{1\bar12\bar2}}\text{Im}\Big[e^{-\ii\pi(s_{12}-s_{\bar1\bar2})}\Big]\Gamma\Big[\text{IR poles}\Big]\int_\tau\sin(k_{12}\tau_1)\sin(k_{34}\tau_2).
\end{align}
The three time integrals give:
\begin{align}
  &\int\di\tau_1(-\tau_1)^{p_1+3/2-2s_1}\sin( k_{12}\tau_1)=-\FR{\Gamma[p_1+\fr52-2s_1]\sin [\fr{\pi}2(p_1+\fr52-2s_1) ]}{k_{12}^{p_1+5/2-2s_1}},\\
  &\int\di\tau_2(-\tau_2)^{p_2+3/2-2s_2}\sin( k_{34}\tau_2)=-\FR{\Gamma[p_2+\fr52-2s_2]\sin [\fr{\pi}2(p_2+\fr52-2s_2) ]}{k_{34}^{p_2+5/2-2s_2}},\\
  &\int\di\tau_3(-\tau_3)^{p_3+3-2s_{\bar1\bar2}}=2\pi \de\big[ \ii (p_3+4-2s_{\bar1\bar2})\big].
\end{align}
We can first use the $\de$-function to remove $\bar s_2$, then we find the integrand of \eqref{eq_GF} is proportional to $k_s^{-2s_{12}}$. For the other three Mellin variables, we take the following left poles to collect leading terms proportional to $k_s^{\ii(\wt\nu_1+\wt\nu_2)}$:
\begin{align}
  &s_1=-\ii\wt\nu_1/2, \qquad \bar s_2=-\ii\wt\nu_2/2, \qquad s_{\bar 1}=-\ii\aa\wt\nu_1/2-n.~~~(n=0,1,\cdots)
\end{align}
Together with the complex conjugate, the final result is:
\begin{align}
  \mathcal{G}^\text{(F)} 
  =&~\FR{8}{(4\pi)^2}\Big(\FR{k_s}2\Big)^{+\ii\wt\nu_1+\ii\wt\nu_2-p_3-4}\sin\Big[\FR{\pi}2(\ii\wt\nu_1+\ii\wt\nu_2+p_3+4)\Big]\sum\text{Res}\Big[\text{IR poles}\Big]\n\\
  &\times\FR{\Gamma[p_1+\fr52+\ii\wt\nu_1,p_2+\fr52+\ii\wt\nu_2]\sin[\fr{\pi}2(p_1+\fr52+\ii\wt\nu_1)]\sin[\fr{\pi}2(p_2+\fr52+\ii\wt\nu_2)]}{k_{12}^{p_1+5/2+\ii\wt\nu_1}k_{34}^{p_2+5/2+\ii\wt\nu_2}}.
\end{align}
This is a hybrid signal as discussed in Sec.\ \ref{sec_multiplecuts}. 
The summation of residues at IR poles is given by
\begin{align}
  &\sum\text{Res}\Big[\text{IR poles}\Big]\n\\
  =&\sum_{n=0}^\infty\sum_{\aa=\pm}\FR{(-1)^n}{n!}\Gamma\Big[-\ii\wt\nu_1,-\ii\wt\nu_2,-\ii\aa\wt\nu_1-n,\fr{\ii(\aa\wt\nu_1+\wt\nu_2)+p_3+4}{2}+n,\fr{\ii(\aa\wt\nu_1-\wt\nu_2)+p_3+4}{2}+n\Big]\n\\
  =&-\FR{2\pi\Gamma[-\ii\wt\nu_1,-\ii\wt\nu_2]}{\sinh\pi\wt\nu_1}\,\text{Im}\,{}_2\mathcal{F}_1\left[\bgm\fr{\ii(\wt\nu_1-\wt\nu_2)+4+p_3}{2},\fr{\ii(\wt\nu_1+\wt\nu_2)+4+p_3}{2}\\ 1+\ii\wt\nu_1\edm\middle|1\right].
\end{align}
When $p_3=-4$ (covariant mass mixing), the above dressed hypergeometric function is simplified to ${}_2\mathcal{F}_1[...|1]=-4/(\wt\nu_1^2-\wt\nu_2^2)$ and is manifestly real. It can be checked that the above $\text{Im}\,{}_2\mathcal{F}_1=0$ for all integer $p_3\leq-4$. Therefore, we conclude that there can be a hybrid signal for non-covariant mixing but such a signal does not exist for covariant mass mixing, in agreement with the general analysis in Sec.\ \ref{sec_multiplecuts}. 

\subsection{Double-bubble graph}
\begin{figure}
\centering
\includegraphics[width=0.5\textwidth]{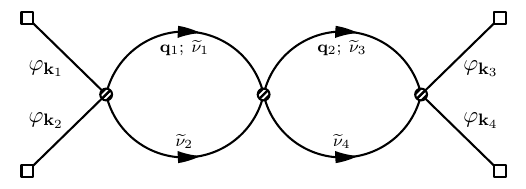}
\caption{The 4-point double-bubble graph.} 
  \label{fd_doublebubble}
\end{figure}
The third example is the double-bubble graph in Fig.\ \ref{fd_doublebubble}, which is like a loop version of the tree-level 2-point mixing graph. Below we mainly focus on the loop integral under the partial MB representation, but we also keep in mind that the time integral at the internal vertices could give extra $\de$-functions. The loop integral in Mellin space is:
\bge
\mathbb L = \int \FR{\di^3 \mb q_1}{(2\pi)^3}\FR{\di^3 \mb q_2}{(2\pi)^3} |\mb q_1|^{-2s_{1\bar1}}|\mb k_s-\mb q_1|^{-2s_{2\bar2}}|\mb q_2|^{-2s_{3\bar3}}|\mb k_s-\mb q_2|^{-2s_{4\bar4}}.
\ede
The two loop momentum integrals are factorized, and the integral can be easily computed:
\bge
\mathbb L = \FR{k_s^{6-2s_{1\bar12\bar23\bar34\bar4}}}{(4\pi)^3}\Gamma\bgb
{\blue s_{1\bar12\bar2}-\fr32},\fr32-s_{1\bar1},\fr32-s_{2\bar2},{\blue s_{3\bar34\bar4}-\fr32},\fr32-s_{3\bar3},\fr32-s_{4\bar4}\\3-s_{1\bar12\bar2},s_{1\bar1},s_{2\bar2},3-s_{3\bar34\bar4},s_{3\bar3},s_{4\bar4}
\edb.
\ede
There are two sets of left poles from this loop integral, coming from the factor $\Gamma(s_{1\bar12\bar2}-3/2)$ and $\Gamma(s_{3\bar34\bar4}-3/2)$, respectively. In fact, they correspond to the UV poles of the left and right loops. If we take one of the UV poles, say $s_{3\bar34\bar4}=3/2$,  then the leading contribution becomes $k_s^{3-2s_{1\bar12\bar2}}$, and we can then set $s_{1,\bar1,2,\bar2}$ again at the IR poles:
\bge
s_1=\bar s_1=-\FR{\ii\cc_1\wt\nu_1}2,\qquad s_2=\bar s_2=-\FR{\ii\cc_2\wt\nu_2}2,
\ede
and the loop integral $\mathbb L$ is proportional to $k_s^{3+2\ii\cc_1\wt\nu_1+2\ii\cc_2\wt\nu_2}$.
This corresponds to the minimal cut of the left loop, namely cutting Line 1 and Line 2.
Similarly, we can cut the right loop, namely Line 3 and Line 4, and take the UV pole: $s_{1\bar12\bar2}=3/2$, then we will get a signal proportional to $k_s^{3+2\ii\cc_1\wt\nu_3+2\ii\cc_2\wt\nu_4}$. These are all ordinary signals. Of course,  we cannot take both UV poles for the purpose of generating a signal, since the result will simply be a constant.

The time integral of the internal vertex could give another type of signal. For simplicity, we take a direct coupling for the internal vertex with an arbitrary power $p_3$, and then the time integral gives:
\bge
\int_{-\infty}^0 \di\tau_3\, (-\tau_3)^{p_3+4\times 3/2-2s_{\bar1\bar234}} =2\pi  \de\big[ \ii(p_3+7-2s_{\bar1\bar234})\big].
\ede
We can use this $\de$-function to remove one Mellin variable. Then, the integrand of $\mathbb L$ becomes proportional to $k_s^{-1-p_3-2s_{12\bar3\bar4}}$. To integrate out the remaining Mellin variables, we take residues of the integrand at the leading IR poles:
\bge
s_1=-\FR{\ii\cc_1\wt\nu_1}2,\qquad s_2=-\FR{\ii\cc_2\wt\nu_2}2,\qquad \bar s_3=-\FR{\ii\cc_3\wt\nu_3}2,\qquad \bar s_4=-\FR{\ii\cc_4\wt\nu_4}2.
\ede
we find that there is a signal of frequency $|\cc_1\wt\nu_1+\cc_2\wt\nu_2+\cc_3\wt\nu_3+\cc_4\wt\nu_4|$, which is generated by the resonances of the four modes at the two external vertices. This is exactly a hybrid signal discussed in Sec.\ \ref{sec_multiplecuts}. However, for dS-covariant coupling, namely $p_3=-4$, this signal should vanish, just like the case of a covariant 2-point mass mixing.\footnote{One can see this point more clearly by taking spectral decompositions of the two loops, as was done in \cite{Xianyu:2022jwk}.}
\subsection{Double-triangle graph}
\begin{figure}
\centering
\includegraphics[width=0.5\textwidth]{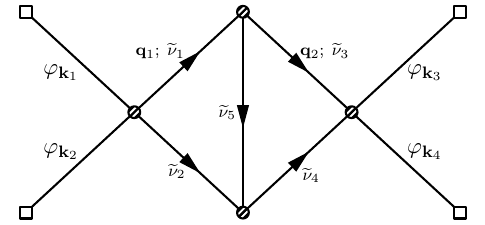}
\caption{The 4-point double-triangle graph.} 
  \label{fig_2triangle}
\end{figure}
The next example is the double-triangle graph Fig.\ \ref{fig_2triangle}.
We can write down the loop integral in the Mellin space:
\begin{align}
  \mathbb{L}=&\int\FR{\di^3\mb q_1}{(2\pi)^3}\FR{\di^3\mb q_2}{(2\pi)^3}|\mb q_1|^{-2s_{1\bar 1}}|\mb k_s-\mb q_1|^{-2s_{2\bar 2}}|\mb q_2|^{-2s_{3\bar 3}}|\mb k_s-\mb q_2|^{-2s_{4\bar 4}}|\mb q_1-\mb q_2|^{-2s_{5\bar 5}}
\end{align}
First consider the $\mb q_2$ integral:
\bge
\wt{\mathbb L}= \int \FR{\di^3\mb q_2}{(2\pi)^3}|\mb q_2|^{-2s_{3\bar3}}|\mb k_s-\mb q_2|^{-2s_{4\bar4}}|\mb q_1-\mb q_2|^{-2s_{5\bar5}}.
\ede
The integral allows various different series expansions, depending on the shape of the triangle formed by vectors $\mb k_s$ and $\mb q_1$; See App. \ref{app_int}. Especially,
\begin{align}
\label{eq_wtLSoft}
\wt{\mathbb L}|_{q_1\ll k_s} =&~
\FR{k_s^{3-2s_{3\bar34\bar45\bar5}}}{(4\pi)^{3/2}} \Gamma\bgb {\blue s_{3\bar34\bar45\bar5}-\fr{3}2},\fr{3}2-s_{3\bar35\bar5},\fr{3}2-s_{4\bar4} \\
  3-s_{3\bar34\bar45\bar5},s_{3\bar34\bar4},s_{5\bar5} \edb \n\\
  &+\FR{q_1^{3-2s_{3\bar35\bar5}}k_s^{-2s_{4\bar4}}}{(4\pi)^{3/2}}
\Gamma\bgb {\red s_{3\bar35\bar5}-\fr32},\fr32-s_{3\bar3},\fr32-s_{5\bar5}\\
3-s_{3\bar35\bar5},s_{3\bar3},s_{5\bar5} \edb,\\
\label{eq_wtLHard}
\wt{\mathbb L}|_{k_s\ll q_1} =&~
\FR{q_1^{3-2s_{3\bar 34\bar 45\bar 5}}}{(4\pi)^{3/2}}
  \Gamma\bgb {\blue s_{3\bar34\bar45\bar5}-\fr{3}2},\fr{3}2-s_{3\bar34\bar4},\fr{3}2-s_{5\bar5} \\
  3-s_{3\bar34\bar45\bar5},s_{3\bar34\bar4},s_{5\bar5} \edb \n\\
  &+\FR{k_s^{3-2s_{3\bar34\bar4}}q_1^{-2s_{5\bar5}}}{(4\pi)^{3/2}}\Gamma\bgb
{\red s_{3\bar34\bar4}-\fr32},\fr32-s_{3\bar3},\fr32-s_{4\bar4}\\3-s_{3\bar34\bar4},s_{3\bar3},s_{4\bar4}
\edb.
\end{align}
As we can see, the UV pole of the right loop $s_{3\bar34\bar45\bar5}=3/2$ appears in both regions of $\mb q_1$. Once we take this UV pole, we find the residue is $\wt{\mathbb L} \to 1/(4\pi^2)$, so the loop integral $\mathbb L$ is simplified to a bubble integral:
\begin{align}
\mathbb L \to&~ \FR{1}{(4\pi)^2} \int \FR{\di^3\mb q_1}{(2\pi)^3}|\mb q_1|^{-2s_{1\bar1}}|\mb k_s-\mb q_1|^{-2s_{2\bar2}}\n\\
=&~ \FR{k_s^{3-2s_{1\bar12\bar2}}}{(4\pi)^{7/2}} \Gamma\bgb
s_{1\bar12\bar2}-\fr32,\fr32-s_{1\bar1},\fr32-s_{2\bar2}\\3-s_{1\bar12\bar2},s_{1\bar1},s_{2\bar2}
\edb,
\end{align}
then we take leading IR poles:
\bge
s_1=\bar s_1 = \FR{\ii\cc_1\wt\nu_1}2, \qquad s_2=\bar s_2 =  \FR{\ii\cc_2\wt\nu_2}2,
\ede
and we obtain an ordinary signal proportional to $k_s^{3+2\ii\cc_1\wt\nu_1+2\ii\cc_2\wt\nu_2}$, which corresponds to cutting Line 1 and Line 2.

We can also take the UV pole coming from the integral of $\mb q_1$. Notice that this UV pole comes from the hard region $q_1\gg k_s$, so we can substitute $\wt{\mathbb L}$ by the expansion \eqref{eq_wtLHard} into the loop integral $\mathbb L$. The first line of \eqref{eq_wtLHard} gives:
\bge
\FR{k_s^{6-2s_{1\bar12\bar23\bar34\bar45\bar5}}}{(4\pi)^3}\Gamma\bgb
{\red s_{1\bar12\bar23\bar34\bar45\bar5}-3},3-s_{1\bar13\bar34\bar45\bar5},\fr32-s_{2\bar2}, {\blue s_{3\bar34\bar45\bar5}-\fr{3}2},\fr{3}2-s_{3\bar34\bar4},\fr{3}2-s_{5\bar5} \\
\fr92-s_{1\bar12\bar23\bar34\bar45\bar5}, s_{1\bar13\bar34\bar45\bar5}-\fr32, s_{2\bar2} , 3-s_{3\bar34\bar45\bar5},s_{3\bar34\bar4},s_{5\bar5}
\edb.
\ede
There are two left poles from the loop integral: a UV pole for the whole graph $s_{1\bar12\bar23\bar34\bar45\bar5}=3$ which gives no signal, and the UV pole for the right loop which we have discussed before. So there is no new leading signal. The second line of \eqref{eq_wtLHard} gives:
\bge
\FR{k_s^{6-2s_{1\bar12\bar23\bar34\bar45\bar5}}}{(4\pi)^3}
  \Gamma\bgb {\blue s_{1\bar12\bar25\bar5}-\fr{3}2},\fr{3}2-s_{1\bar15\bar5},\fr{3}2-s_{2\bar2},{\red s_{3\bar34\bar4}-\fr{3}2},\fr{3}2-s_{3\bar3},\fr{3}2-s_{4\bar4}\\
  3-s_{1\bar12\bar25\bar5},s_{1\bar15\bar5},s_{2\bar2},3-s_{3\bar34\bar4},s_{3\bar3},s_{4\bar4}\edb .
\ede
As we can see, we recover the left loop UV pole $s_{1\bar12\bar25\bar5}=3/2$.
When taking residue at this pole, we find $\mathbb L \propto k_s^{3-2s_{3\bar34\bar4}}$, so we can take residues at the following leading IR poles:
\bge
s_3 = \bar s_3 = \mp \FR{\ii\cc_3\wt\nu_3}2, \qquad s_4=\bar s_4 = \mp \FR{\ii\cc_4\wt\nu_4}2,
\ede
and we get another ordinary signal proportional to $k_s^{3+2\ii\cc_3\wt\nu_3+2\ii\cc_4\wt\nu_4}$, corresponding to cutting Line 3 and Line 4.

The other two left poles $s_{3\bar34\bar4}=3/2$ and $s_{3\bar35\bar5}=3/2$ are more subtle. For example, we can take $s_{3\bar34\bar4}=3/2$, then the residue of $\mathbb L$ is propoertional to $k_s^{3-2s_{1\bar12\bar25\bar5}}$. Naively, one may expect that we can take the corresponding IR poles, corresponding to cutting Line 1, Line 2, and Line 5 simultaneously, and we would obtain a signal with frequency $|2\cc_1\wt\nu_1+2\cc_2\wt\nu_2+2\cc_3\wt\nu_5|$. However, this is not the case, since when Lines 1,2,5 become soft together, the integral of $\tau_3$ and $\tau_4$ will be factorized, giving additional $\de$-constraints. 

In fact, assuming direct couplings, the time integrals at the two internal vertices give:
\begin{align}
 &\int_{-\infty}^0\di\tau_3\,(-\tau_3)^{p_3+3\times 3/2-2s_{\bar1\bar25}} = 2\pi  \de\big[\ii(p_3+11/2-2s_{\bar1\bar25})\big],\\
 &\int_{-\infty}^0\di\tau_4\,(-\tau_4)^{p_4+3\times 3/2-2s_{\bar3\bar4\bar5}} = 2\pi \de\big[\ii(p_4+11/2-2s_{34\bar5})\big].
\end{align}
Then we see that the $k_s$ power becomes $k_s^{-5-p_3-p_4-2s_{12\bar3\bar4}}$. This suggests that there is a hybrid signal from the four modes attached to the ``outer'' two vertices, with frequencies $|\cc_1\wt\nu_1+\cc_2\wt\nu_2+\cc_3\wt\nu_3+\cc_4\wt\nu_4|$. Similar to the previous cases, this signal should vanish for dS covariant couplings, namely when $p_3=p_4=-4$.
 
\subsection{Double-box graph} 
\begin{figure}
\centering
\includegraphics[width=0.5\textwidth]{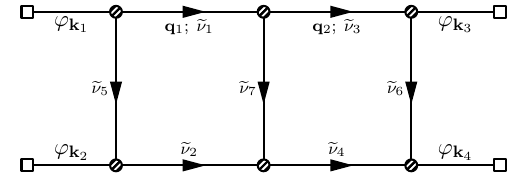}
\caption{The 4-point (planar) double-box graph.} 
  \label{fig_doublebox}
\end{figure}
The case of the double-box graph Fig.\ \ref{fig_doublebox} is very similar to the double-triangle graph. The loop integral becomes:
\begin{align}
  \mathbb{L}=&\int\FR{\di^3\mb q_1}{(2\pi)^3}|\mb q_1|^{-2s_{1\bar 1}}|\mb k_s-\mb q_1|^{-2s_{2\bar 2}}|\mb k_1-\mb q_1|^{-2s_{5\bar5}}\n\\
  &\times \int \FR{\di^3\mb q_2}{(2\pi)^3} |\mb q_2|^{-2s_{3\bar 3}}|\mb k_s-\mb q_2|^{-2s_{4\bar 4}}|\mb k_3+\mb q_2|^{-2s_{6\bar6}}|\mb q_1-\mb q_2|^{-2s_{7\bar 7}}.
\end{align}
Again we first calculate the integral of $\mb q_2$:
\bge
\wt{\mathbb L} = \int \FR{\di^3\mb q_2}{(2\pi)^3} |\mb q_2|^{-2s_{3\bar 3}}|\mb k_s-\mb q_2|^{-2s_{4\bar 4}}|\mb k_3+\mb q_2|^{-2s_{6\bar6}}|\mb q_1-\mb q_2|^{-2s_{7\bar 7}}.
\ede
Depending on the shapes of vectors $\mb k_s$, $\mb q_1$ and $\mb k_3$, $\wt{\mathbb L}$ can be expanded as different series. Here we focus on the hard region:
\begin{align}
\wt{\mathbb L}|_{q_1\gg k_3 \gg k_s} = &~ \FR{q_1^{3-2s_{3\bar34\bar46\bar67\bar7}}}{(4\pi)^{3/2}} \Gamma\bgb
{\blue s_{3\bar34\bar46\bar67\bar7}-\fr32},\fr32-s_{3\bar34\bar46\bar6},\fr32-s_{7\bar7}\\3-s_{3\bar34\bar46\bar67\bar7},s_{3\bar34\bar46\bar6},s_{7\bar7}
\edb  \n\\
&+ \FR{k_3^{3-2s_{3\bar34\bar46\bar6}}q_1^{-2s_{7\bar7}}}{(4\pi)^{3/2}} \Gamma\bgb
{\red s_{3\bar34\bar46\bar6}-\fr32},\fr32-s_{3\bar34\bar4},\fr32-s_{6\bar6}\\
3-s_{3\bar34\bar46\bar6},s_{3\bar34\bar4},s_{6\bar6}
\edb\n\\
&+\FR{k_s^{3-2s_{3\bar34\bar4}}k_3^{-2s_{6\bar6}}q_1^{-2s_{7\bar7}}}{(4\pi)^{3/2}}\Gamma\bgb
{\red s_{3\bar34\bar4}-\fr32},\fr32-s_{3\bar3},\fr32-s_{4\bar4}\\3-s_{3\bar34\bar4},s_{3\bar3},s_{4\bar4}
\edb.
\end{align}
Plug the hard region expression into the original integral and keeping terms nonanalytic in $k_s$, we obtain:
\begin{align}
\label{eq_verylongL}
&~\mathbb L \To \FR{k_s^{6-2s_{1\bar12\bar23\bar34\bar46\bar67\bar7}}k_1^{-2s_{5\bar5}}}{(4\pi)^3}\n\\
&\times \Gamma\bgb
{\red s_{1\bar12\bar23\bar34\bar46\bar67\bar7}-3},3-s_{1\bar13\bar34\bar46\bar67\bar7},\fr32-s_{2\bar2},
{\blue s_{3\bar34\bar46\bar67\bar7}-\fr32},\fr32-s_{3\bar34\bar46\bar6},\fr32-s_{7\bar7}\\
\fr92-s_{1\bar12\bar23\bar34\bar46\bar67\bar7},s_{1\bar13\bar34\bar46\bar67\bar7}-\fr32,s_{2\bar2},3-s_{3\bar34\bar46\bar67\bar7},s_{3\bar34\bar46\bar6},s_{7\bar7}
\edb\n\\
&+\FR{k_s^{3-2s_{1\bar12\bar27\bar7}}k_1^{-2s_{5\bar5}}k_3^{3-2s_{3\bar34\bar46\bar6}}}{(4\pi)^3} \Gamma\bgb
{\red s_{1\bar12\bar27\bar7}-\fr32},\fr32-s_{1\bar17\bar7},\fr32-s_{2\bar2},{\red s_{3\bar34\bar46\bar6}-\fr32},\fr32-s_{3\bar34\bar4},\fr32-s_{6\bar6}\\
3-s_{1\bar12\bar27\bar7},s_{1\bar17\bar7},s_{2\bar2},3-s_{3\bar34\bar46\bar6},s_{3\bar34\bar4},s_{6\bar6}
\edb\n\\
&+\FR{k_s^{3-2s_{3\bar34\bar4}}k_1^{3-2s_{1\bar12\bar25\bar57\bar7}}k_3^{-2s_{6\bar6}}}{(4\pi)^3}\Gamma\bgb
{\red s_{1\bar12\bar25\bar57\bar7}-\fr32},\fr32-s_{1\bar12\bar27\bar7},\fr32-s_{5\bar5},
{\red s_{3\bar34\bar4}-\fr32},\fr32-s_{3\bar3},\fr32-s_{4\bar4}\\
3-s_{1\bar12\bar25\bar57\bar7},s_{1\bar12\bar27\bar7},s_{5\bar5},
3-s_{3\bar34\bar4},s_{3\bar3},s_{4\bar4}
\edb\n\\
&+\FR{k_s^{6-2s_{1\bar12\bar23\bar34\bar47\bar7}}k_1^{-2s_{5\bar5}}k_3^{-2s_{6\bar6}}}{(4\pi)^3}\Gamma\bgb
{\blue s_{1\bar12\bar27\bar7}-\fr32},\fr32-s_{1\bar17\bar7},\fr32-s_{2\bar2},{\red s_{3\bar34\bar4}-\fr32},\fr32-s_{3\bar3},\fr32-s_{4\bar4}\\3-s_{1\bar12\bar2},s_{1\bar1},s_{2\bar2},3-s_{3\bar34\bar4},s_{3\bar3},s_{4\bar4}
\edb.
\end{align}
In the first term on the right hand side, there is a UV pole $s_{3\bar34\bar46\bar67\bar7}=3/2$ for the right loop. We can take this pole, which amounts to pinching the right loop. Then, the integrand is proportional to $k_s^{3-2s_{1\bar12\bar2}}$, and then we can take the following IR poles:
\bge
s_1=\bar s_1 = -\FR{\ii\cc_1\wt\nu_1}2,\qquad s_2=\bar s_2=-\FR{\ii\cc_2\wt\nu_2}2,
\ede
    and obtain a signal of frequency $2|\wt\nu_1\pm\wt\nu_2|$, corresponding to cutting Line 1 and Line 2. Notice that we cannot take poles $s_1 = -\bar s_1 = -\ii\cc_1\wt\nu_1/2$ to construct a signal of frequency $2\wt\nu_2$, because the factor $\Gamma(s_{1\bar13\bar34\bar46\bar67\bar7}-3/2)$ in the denominator will becomes $\Gamma(s_{1\bar1})$ at the UV pole $s_{3\bar34\bar46\bar67\bar7}=3/2$.

In the last term on the right hand side of \eqref{eq_verylongL}, there is another UV pole $s_{1\bar12\bar27\bar7}=3/2$, at which the integrand is proportional to $k_s^{3-2s_{3\bar34\bar4}}$. Then we take IR poles:
\bge
s_3=\bar s_3 = -\FR{\ii\cc_3\wt\nu_3}2,\qquad s_4=\bar s_4=-\FR{\ii\cc_4\wt\nu_4}2,
\ede
which indicates another ordinary signal of frequency $2|\wt\nu_3\pm \wt\nu_4|$, corresponding to cutting Line 3 and Line 4.

Other possible left poles lead to either no new signals, or signals constrained by the $\de$-functions from time integral. For instance, we can take another left pole $s_{3\bar34\bar4}=3/2$ in the fourth line, then the integrand is proportional to $k_s^{3-2s_{1\bar12\bar27\bar7}}$. However, if we try to cut Line 1, 2, 7 together, we will encounter the two $\de$-functions from the integral of $\tau_5$ and $\tau_6$:
\begin{align}
&\int_{-\infty}^0 \di\tau_5\,(-\tau_5)^{p_5+3\times 3/2-2s_{\bar137}} = 2\pi \de\big[\ii(p_5+11/2-2s_{\bar137})\big],\\
&\int_{-\infty}^0 \di\tau_6\,(-\tau_6)^{p_6+3\times 3/2-2s_{\bar24\bar7}} =2\pi  \de\big[\ii(p_6+11/2-2s_{\bar24\bar7})\big],
\end{align}
which make the integrand proportional to $k_s^{-5-p_5-p_6-2s_{12\bar3\bar4}}$. This could generate a hybrid signal of frequency $|\cc_1\wt\nu_1+\cc_2\wt\nu_2+\cc_3\wt\nu_3+\cc_4\wt\nu_4|$ in the non-dS-covariant case.

\subsection{Non-planar double-box graph} 
\begin{figure}
\centering
\includegraphics[width=0.5\textwidth]{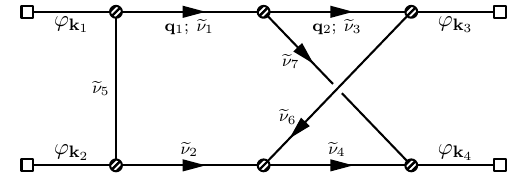}
\caption{The 4-point non-planar double-bubble graph.} 
  \label{fig_NPdoublebox}
\end{figure}

Now let us consider the non-planar double box graph Fig.\ \ref{fig_NPdoublebox}. The loop integral is:
\begin{align}
  \mathbb{L}=&\int\FR{\di^3\mb q_1}{(2\pi)^3}|\mb q_1|^{-2s_{1\bar 1}}|\mb k_s-\mb q_1|^{-2s_{2\bar 2}}|\mb k_1-\mb q_1|^{-2s_{5\bar5}}\n\\
  &\times \int \FR{\di^3\mb q_2}{(2\pi)^3} |\mb q_2|^{-2s_{3\bar 3}}|\mb k_s+\mb k_3-\mb q_1+\mb q_2|^{-2s_{4\bar 4}}|\mb k_3+\mb q_2|^{-2s_{6\bar6}}|\mb q_1-\mb q_2|^{-2s_{7\bar 7}}.
\end{align}
We still consider the $\mb q_2$ integral $\wt{\mathbb L}$ first, namely the second lien of the above expression. In particular, in the region $q_1\gg k_3 \gg k_s$, we have:
\begin{align}
\wt{\mathbb L}|_{q_1\gg k_3\gg k_s}=&~
\FR{q_1^{3-2s_{3\bar34\bar46\bar67\bar7}}}{(4\pi)^{3/2}}\Gamma\bgb
{\blue s_{3\bar34\bar46\bar67\bar7}-\fr32},\fr32-s_{3\bar36\bar6},\fr32-s_{4\bar47\bar7}\\
3-s_{3\bar34\bar46\bar67\bar7},s_{3\bar36\bar6},s_{4\bar47\bar7}
\edb\n\\
&+\FR{k_3^{3-2s_{3\bar36\bar6}}q_1^{-2s_{4\bar47\bar7}}}{(4\pi)^{3/2}}\Gamma\bgb
{\red s_{3\bar36\bar6}-\fr32},\fr32-s_{3\bar3},\fr32-s_{6\bar6}\\
3-s_{3\bar36\bar6},s_{3\bar3},s_{6\bar6}
\edb.
\end{align}
Inserting the above expansion back into the original loop integral, and keeping the nonanalytic terms in $k_s$, we obtain:
\begin{align}
\label{eq_nonplanar}
\mathbb L \To&~ \FR{k_s^{6-2s_{1\bar12\bar23\bar34\bar46\bar67\bar7}}k_1^{-2s_{5\bar5}}}{(4\pi)^3}\n\\
&\times \Gamma\bgb
{\red s_{1\bar12\bar23\bar34\bar46\bar67\bar7}-3},3-s_{1\bar13\bar34\bar46\bar67\bar7},\fr32-s_{2\bar2},{\blue s_{3\bar34\bar46\bar67\bar7}-\fr32},\fr32-s_{3\bar36\bar6},\fr32-s_{4\bar47\bar7}\\
\fr92-s_{1\bar12\bar23\bar34\bar46\bar67\bar7},s_{1\bar13\bar34\bar46\bar67\bar7}-\fr32,s_{2\bar2},3-s_{3\bar34\bar46\bar67\bar7},s_{3\bar36\bar6},s_{4\bar47\bar7}
\edb\n\\
&+\FR{k_s^{3-2s_{1\bar12\bar24\bar47\bar7}}k_1^{-2s_{5\bar5}}k_3^{-3-2s_{3\bar36\bar6}}}{(4\pi)^3}\n\\
&\times\Gamma\bgb
{\red s_{1\bar12\bar24\bar47\bar7}-\fr32},\fr32-s_{1\bar14\bar47\bar7},\fr32-s_{2\bar2},{\red s_{3\bar36\bar6}-\fr32},\fr32-s_{3\bar3},\fr32-s_{6\bar6}\\
3-s_{1\bar12\bar24\bar47\bar7},s_{1\bar14\bar47\bar7},s_{2\bar2},3-s_{3\bar36\bar6},s_{3\bar3},s_{6\bar6}
\edb.
\end{align}
We can take the UV pole $s_{3\bar34\bar46\bar67\bar7}$ from the right (nonplanar) loop and thus the integrand is proportional to $k_s^{3-2s_{1\bar12\bar2}}$, and then, we can cut Line 1 and Line 2 and take IR poles:
\bge
s_1=\bar s_1 = -\FR{\ii\cc_1\wt\nu_1}2,\qquad s_2=\bar s_2 = -\FR{\ii\cc_2\wt\nu_2}2,
\ede
which generates a signal of frequency $2|\wt\nu_1\pm \wt\nu_2|$. On the other hand, we cannot take the UV pole from the $\mb q_1$ integral, which will not give us any signal. Indeed, the minimal cut for this graph is unique, which is exactly cutting Line 1 and Line 2.
\subsection{Bubble-on-bubble graph}

\begin{figure}
\centering
\includegraphics[width=0.48\textwidth]{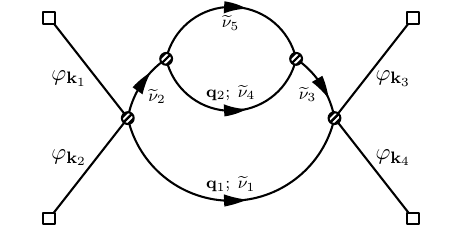}
\caption{The 4-point bubble-on-bubble graph.} 
  \label{fig_bubbleonbubble}
\end{figure}
Our final example is the bubble-on-bubble graph Fig.\ \ref{fig_bubbleonbubble}. The loop integral can be calculated explicitly:
\begin{align}
\mathbb L =& \int \FR{\di^3\mb q_1}{(2\pi)^3}\FR{\di^3\mb q_2}{(2\pi)^3}|\mb q_1|^{-2s_{1\bar1}}|\mb k_s-\mb q_1|^{-2s_{2\bar23\bar3}} |\mb q_2|^{-2s_{4\bar4}}|\mb k_s-\mb q_1-\mb q_2|^{-2s_{5\bar5}}\n\\
=& \int \FR{\di^3\mb q_1}{(2\pi)^3}|\mb q_1|^{-2s_{1\bar1}}|\mb k_s-\mb q_1|^{-2s_{2\bar23\bar3}} \times \FR{|\mb k_s-\mb q_1|^{3-2s_{4\bar45\bar5}}}{(4\pi)^{3/2}}\Gamma\bgb
{\blue s_{4\bar45\bar5}-\fr32},\fr32-s_{4\bar4},\fr32-s_{5\bar5}\\3-s_{4\bar45\bar5},s_{4\bar4},s_{5\bar5}
\edb\n\\
=&~\FR{k_s^{6-2s_{1\bar12\bar23\bar34\bar45\bar5}}}{(4\pi)^3}\Gamma\bgb
{\red s_{1\bar12\bar23\bar34\bar45\bar5}-3},\fr32-s_{1\bar1},3-s_{2\bar23\bar34\bar45\bar5},{\blue s_{4\bar45\bar5}-\fr32},\fr32-s_{4\bar4},\fr32-s_{5\bar5}\\
3-s_{4\bar45\bar5},\fr92-s_{1\bar12\bar23\bar34\bar45\bar5},s_{1\bar1},s_{4\bar4},s_{5\bar5},s_{2\bar23\bar34\bar45\bar5}-\fr32
\edb.
\end{align}
We can first take the UV pole of the small loop at $s_{4\bar45\bar5}=3/2$, then the integrand of $\mathbb L$ is proportional to $k_s^{3-2s_{1\bar12\bar23\bar3}}$. Next we should specify appropriate poles for $s_{1,\bar1,2,\bar2,3,\bar3}$. As before, the factor $\Gamma(s_{1\bar1})$ in the denominator requires that we should select leading IR poles $s_1=\bar s_1 = \mp \ii\cc_1 \wt\nu_1$. Then, we can choose leading IR poles of $s_{2,\bar2,3,\bar3}$. For example, we can take leading poles $s_2=\bar s_2 = \mp \ii \cc_2\wt\nu_2$ and $s_3=-\bar s_3=\mp\ii\cc_3\wt\nu_3$ (notice that there is no factor $\Gamma(s_{3\bar3})$ in the denominator), which gives a signal proportional to $k_s^{3+2\ii\cc_1\wt\nu_1+2\ii\cc_2\wt\nu_2}$. This corresponds to cutting Line 1 and Line 2. Similarly, we can cut Line 1 and Line 3 and obtain a signal proportional to $k_s^{3+2\ii\cc_1\wt\nu_1+2\ii\cc_3\wt\nu_3}$.

However, if we try to cut Line 2, and Line 3 simultaneously, there would be an additional $\de$-constraint coming from the integral of $\tau_3$ and $\tau_4$. For example, if we consider the $\mathcal G_{+-+-}$ channel, then the time integrals give:
\begin{align}
&\int \di\tau_3\,(-\tau_3)^{p_3+3\times3/2-2s_{\bar245}} = 2\pi\de\big[\ii(p_3+11/2-2s_{\bar245})\big],\\
&\int \di\tau_4\,(-\tau_4)^{p_4+3\times3/2-2s_{3\bar4\bar5}} =2\pi \de\big[\ii(p_4+11/2-2s_{3\bar4\bar5})\big],
\end{align}
and thus the integrand is now proportional to $k_s^{-5-p_3-p_4-2s_{1\bar12\bar3}}$, which indicates signals of frequencies $|2\cc_1\wt\nu_1+\cc_2\wt\nu_2+\cc_3\wt\nu_3|$ in the non-dS-covariant case, corresponding to the oscillations of the four outer modes.

\section{Conclusion and Outlook}
\label{sec_conclusion}
Inflation correlators are central observables in Cosmological Collider physics, playing a similar role as the scattering amplitudes and $S$-matrix for scattering experiments in Minkowski spacetime. It is thus crucial to have a good understanding of the analytical structure of inflation correlators. Among all kinds of singularities, the nonlocal signal produced by massive exchanges is special: On the one hand, it gives rise to a characteristic oscillatory signal in the physical region and thus is the main observable in CC physics. On the other hand, it produces a branch cut starting from the origin in the complex plane of the momentum transfer, and this feature has no flat spacetime correspondence. For these reasons, studying the analytical structure of the nonlocal signal is useful and important for both CC phenomenology and general dS QFTs.

In this work, we generalize our previous result in \cite{Qin:2023bjk} to all loop orders: With the help of partial Mellin-Barnes representation, we state and prove a factorization theorem \eqref{eq_gfacMMC}, which can be used to detect and identify all possible nonlocal signal in an arbitrary graph. 
At 1-loop order, a nonlocal signal could appear when two internal lines become soft simultaneously, and we can cut these two soft lines, pinch their endpoints and get a bubble signal.
Similarly, the nonlocal signal of an arbitrary graph is associated with a nonlocal cut, and the leading signal in the nonlocal soft limit $\mb P\to 0$ corresponds to the minimal cut. The signal appears when all these lines become soft simultaneously, and we can pinch their endpoints and get a melon subgraph, and the nonanalyticity of $P$ is manifest in its melon signal.

Partial Mellin-Barnes representation has been proven suitable and useful when studying inflation correlators, and there are still many interesting targets to explore with this tool. Under this representation, the graph breaks into different pieces in \eqref{eq_TkMellin}, and both the time and loop momentum integrals are much simplified. Moreover, the external momenta only appear in the loop integral $\mathbb L$, while the external energies only appear in the time integral $\mathbb T$ (if there is no degeneracy between momenta and energies, of course). Therefore, it seems that considering the loop integral $\mathbb L$ is enough to analyze the nonlocal signals. However, this is not exactly the case, since time integral $\mathbb T$ could give some $\de$-functions of Mellin variables, which can result in the hybrid signals briefly discussed in Sec.\ \ref{sec_multiplecuts} and Sec.\ \ref{sec_2ptmix}.
Although such signals are absent in bulk-free graphs and forbidden by the conformal symmetry in graphs with dS-covariant couplings, it will be interesting to have a detailed study of such hybrid signals.

Also, as mentioned in the Introduction, there are other kinds of nonanalyticity, including the local signal associated with the partial external energy sum. Since the energy only appears in time integrals, we can derive similar detection rules for such local signals using partial MB representation. This requires a careful study of the Mellin-space time integral $\mathbb T$ in \eqref{eq_TIntMellin}, and we will pursue this topic in another work.

Branch cuts associated with CC signals are in a sense unique to dS amplitudes, which reflect the particle production and resonances in an inflationary universe. So they are absent in the corresponding flat-space amplitudes. It is well known that we have full control of a meromorphic function if we know its behavior at all the singularities (for example, residues at simple poles). The case of functions with branch cuts is similar: once we know the discontinuity along every branch cut, we can restore the function using the dispersion relation. We have seen that the massive inflation correlators have a branch cut starting from the origin of the complex plane of certain momentum transfers, and we are able to calculate the nonanalyticity using our factorization theorem. It is thus natural to ask whether we can reconstruct the full correlator using dispersion relation. This is not as trivial as it sounds, since our result is only valid around the squeezed limit. However, to do a dispersion integral along the full branch cut, we need a more comprehensive understanding of the nonanalyticity away from the origin point. Meanwhile, we should also identify all the other singular structures on the complex plane for the dispersion integrals to work. We also leave this question to a future work.

\paragraph{Acknowledgments.} We thank Clifford Cheung, Bing-Chu Fan, Xi Tong, Yi Wang, and Yuhang Zhu for useful discussions. ZX thanks the HKUST Jockey Club Institute for Advanced Study for hospitality during the completion of this work. This work is supported by the National Key R\&D Program of China (2021YFC2203100), NSFC under Grant No.\ 12275146, an Open Research Fund of the Key Laboratory of Particle Astrophysics and Cosmology, Ministry of Education of China, and a Tsinghua University Initiative Scientific Research Program. 

\newpage
\begin{appendix}

\section{Notations}
\label{app_notation}

In this work, we draw heavily use of following shorthands for products and fractions of Euler $\Gamma$ functions:
\begin{align}
  \Gamma\left[ z_1,\cdots,z_m \right]
  \equiv&~ \Gamma(z_1)\cdots \Gamma(z_m) ,\\
  \Gamma\left[\bgm z_1,\cdots,z_m \\w_1,\cdots, w_n\edm\right]
  \equiv&~\FR{\Gamma(z_1)\cdots \Gamma(z_m)}{\Gamma(w_1)\cdots \Gamma(w_n)}.
\end{align}
We also use the dressed version of generalized hypergeometric functions, defined in the following way when the series converges: 
\begin{align}
\label{eq_dressedF}
  {}_p\mathcal{F}_q\left[\bgm a_1,\cdots,a_p \\ b_1,\cdots,b_q \edm  \middle| z \right]
  =\sum_{n=0}^\infty\Gamma\left[\begin{matrix}
        a_1+n, \cdots, a_p+n \\
        b_1+n, \cdots, b_q+n
    \end{matrix}\right]\FR{z^n}{n!}.
\end{align}
The definition can be extended beyond the radius of convergences by analytical continuation. 

A central formula for our partial MB representation for massive inflation correlators is the following MB representation of the Hankel functions: 
 \bge
  \text{H}_{\nu}^{(j)}(az)=\int_{-\ii\infty}^{\ii\infty}\FR{\di s}{2\pi\ii}\FR{(az/2)^{-2s}}{\pi}e^{(-1)^{j+1}(2s-\nu-1)\pi\ii/2}\Gamma\Big[s-\FR{\nu}{2},s+\FR{\nu}{2}\Big].~~~~(j=1,2)
\ede

Finally, we collect frequently used symbols in Table.\ \ref{tab_notations}.

\begin{table}[tbph] 
 \centering
  \caption{List of symbols}
  \vspace{2mm}
 \begin{tabular}{lll}
  \toprule[1.5pt]
    Symbol 
   &\multicolumn{1}{c}{} & Equation \\ \hline 
   $\mb k_i$ $(i=1,\cdots,B)$ & Momentum of an external line  & \\
   $\mb k_i^\mathrm{(L)}$ $(i=1,\cdots,B_L)$ & Momentum of an external line in the left sugraph & \\
   $\mb k_i^\mathrm{(R)}$ $(i=1,\cdots,B_R)$ & Momentum of an external line in the right sugraph & \\
   $\mb p_i$ $(i=1,\cdots,I)$ & Momentum of an internal line  & \\
   $\mb q_i$ $(i=1,\cdots,L)$ & Loop momentum variable & \\
   $\mb q_i^\mathrm{(L)}$ $(i=1,\cdots,N_L)$ & Loop momentum variable of the left subgraph& \\
   $\mb q_i^\mathrm{(R)}$ $(i=1,\cdots,N_R)$ & Loop momentum variable of the right subgraph & \\
   $\mb K_i$ $(i=,1\cdots,V)$ & Total external momentum flowing into the $i$'th vertex & \\
   $\wt{\mb P}_i$  & Partial sum of external momenta & \\
   $\mb P\equiv\sum\mb k^\mathrm{(L)}$ & Total external momentum of the left subgraph & \\
   $\wt\nu\equiv\sqrt{m^2-9/4}$ & Mass parameter of a principal scalar of mass $m$ & \\ 
   $\mathcal{G}(\{\mb k\})$ & (Amplitude of) a graph & (\ref{eq_Tk})\\ 
   $C_\aa(k;\tau)$ & Bulk-to-boundary propagator of a conformal scalar & \eqref{eq_CSProp}\\
   $D_{\aa\bb}(k;\tau_1,\tau_2)$ & Bulk propagator of a massive scalar & \eqref{eq_Dmp}-\eqref{eq_Dpmpm}\\
   $\aa,\bb,\cc,\cdots$ & SK indices or other indices taking values from $\pm 1$ & \\
   $s_i,\bar s_i$ & Mellin variable & \\
   $\mathbb{T}(\{\mb k\};\{s,\bar s\})$ & Mellin-space time integral & \eqref{eq_TIntMellin}\\ 
   $\mathbb{L}(\{\mb k\};\{s,\bar s\})$ & Mellin-space loop momentum integral & \eqref{eq_MellinLoopInt} \\
   $\mathbb{M}_L(P)$ & $L$-loop melon integral in Mellin space & \eqref{eq_melonIntResult}\\
   $\mathfrak{M}_{\cc_1\cdots\cc_D}(P)$ & Melon signal & \eqref{eq_frakM} \\
   \bottomrule[1.5pt] 
 \end{tabular}
 \label{tab_notations}
 \end{table}

\section{Useful Integrals}
\label{app_int}
\paragraph{Melon integral.}  
To derive the melon signal $\mathfrak{M}_{\cc_1\cdots\cc_D}(P)$ (\ref{eq_frakM}), we need to compute the Mellin-space melon integral \eqref{eq_melonInt}, and the result is given in \eqref{eq_melonIntResult}. Below we will show the calculation explicitly. The basic tool is the bubble integral:
\bge
\int \FR{\di^3\mb q}{(2\pi)^3}|\mb q|^{-2s_{1\bar1}}|\mb P-\mb q|^{-2s_{2\bar2}} = \FR{P^{3-2s_{1\bar12\bar2}}}{(4\pi)^{3/2}}
\Gamma\bgb
s_{1\bar12\bar2}-\fr32,\fr32-s_{1\bar1},\fr32-s_{2\bar2}\\
3-s_{1\bar12\bar2},s_{1\bar1},s_{2\bar2}
\edb.
\ede

We start with the definition:
\begin{align}
\mathbb{M}_{D-1}(P)
  =&\int\prod_{\ell=1}^{D-1}\bigg[\FR{\di^3\mb q_\ell}{(2\pi)^3}\big|\mb q_\ell\big|^{-2s_{\ell\bar\ell}}\bigg]  \Big|\mb P-\sum_{i=1}^{D-1}\mb q_i\Big|^{-2s_{D \ob{D}}} . 
\end{align}
We first consider the integral of $\mb q_1$, which is a bubble integral:
\bge
\label{eq_q1int}
\int \FR{\di^3\mb q_1}{(2\pi)^3} \big|\mb q_1\big|^{-2s_{1\bar1}} \Big|\mb P-\sum_{i=1}^{D-1}\mb q_i\Big|^{-2s_{D\ob D}}
=
\FR{|\mb P-\sum_{i=2}^{D-1}\mb q_i|^{3-2s_{1\bar1D\ob D}}}{(4\pi)^{3/2}}\Gamma\bgb
{\red s_{1\bar1D\ob D}-\fr32},\fr32-s_{1\bar1},\fr32-s_{D\ob D}\\
{\blue 3-s_{1\bar1D\ob D}},s_{1\bar1},s_{D\ob D}
\edb.
\ede
Then we can finish the integral of $\mb q_2$, which is also a bubble integral:
\bge
\label{eq_q2int}
\int \FR{\di^3\mb q_2}{(2\pi)^3} \big|\mb q_2\big|^{-2s_{2\bar2}} \Big|\mb P-\sum_{i=2}^{D-1}\mb q_i\Big|^{3-2s_{1\bar1D\ob D}}
=
\FR{|\mb P-\sum_{i=3}^{D-1}\mb q_i|^{6-2s_{1\bar12\bar2D\ob D}}}{(4\pi)^{3/2}}\Gamma\bgb
s_{1\bar12\bar2D\ob D}-3,\fr32-s_{2\bar2},{\blue 3-s_{1\bar1D\ob D}}\\
\fr92-s_{1\bar12\bar2D\ob D},s_{2\bar2},{\red s_{1\bar1D\ob D}-\fr32}
\edb.
\ede
We see that the two $\Gamma$ functions in red are canceled between \eqref{eq_q1int} and \eqref{eq_q2int}, and so are those in blue. We can then repeat the procedure and finish the full integral, and the result is exactly \eqref{eq_melonIntResult}:
\bge
\mathbb{M}_{D-1}(P)=\FR{P^{3(D-1)-2s_{1\bar1\cdots D\ob{D}}}}{(4\pi)^{3(D-1)/2}}\Gamma\bgb s_{1\bar1\cdots D\ob{D}}-\fr{3(D-1)}2 \\ \fr{3D}2-s_{1\bar1\cdots D\ob{D}}\edb\prod_{\ell=1}^{D}\Gamma\bgb\fr32-s_{\ell\bar\ell}\\ s_{\ell\bar\ell}\edb.
\ede
\paragraph{Expansion of loop integrals}
For more complicated Mellin-space loop momentum integrals, it is very difficult to compute the closed-form result. However, we can calculate the leading contribution in special hierachical limits of the momentum configuration. For instance, we consider the triangle integral:
\bge
\label{eq_triangleint}
\mathbb L = \int \FR{\di^3\mb q}{(2\pi)^3}|\mb q|^{-2s_{1\bar1}}|\mb k_s-\mb q|^{-2s_{2\bar2}}|\mb k_1-\mb q|^{-2s_{3\bar3}}.
\ede
This is the loop integral of a triangle with external momenta $\mb k_s$, $\mb k_1$ and $\mb k_2 = -\mb k_1-\mb k_s$. Furthermore, we can consider the squeezed limit $k_1\gg k_s$. Then the leading result is:
\begin{align}
\label{eq_triangleintResult}
\mathbb L|_{k_1\gg k_s} = &~\FR{k_1^{3-2s_{1\bar12\bar23\bar3}}}{(4\pi)^{3/2}}\Gamma\bgb
s_{1\bar12\bar23\bar3}-\fr32,\fr32-s_{1\bar1},\fr32-s_{1\bar1},\fr32-s_{2\bar23\bar3}\\3-s_{1\bar12\bar23\bar3},s_{1\bar1},s_{2\bar23\bar3}
\edb\n\\
&+ \FR{k_s^{3-2s_{1\bar12\bar2}}k_1^{-2s_{3\bar3}}}{(4\pi)^{3/2}}\Gamma\bgb
s_{1\bar12\bar2}-\fr32,\fr32-s_{1\bar1},\fr32-s_{2\bar2}\\
3-s_{1\bar12\bar2},s_{1\bar1},s_{2\bar2}
\edb + \mathcal O\Big( \FR{k_s}{k_1}\Big).
\end{align}
The detailed derivation can be found in the Appendix B of \cite{Qin:2023bjk}. Here we provide a much simpler proof which could be applied to general situations, including the 2-loop examples in Sec.\ \ref{sec_example}.

In the limit $k_s\to 0$, there could be two contributions. One piece is analytic in $k_s$, then we can set $k_s\to 0 $ in the integrand of \eqref{eq_triangleint}. Equivalently, we can expand  the integrand and take the leading contribution in the region $q \gg k_1$:
\bge
\mathbb L|_{q \gg k_1} = \int \FR{\di^3\mb q}{(2\pi)^3}|\mb q|^{-2s_{1\bar12\bar2}}|\mb k_1-\mb q|^{-2s_{3\bar3}},
\ede 
which becomes a bubble integral and gives the first line of \eqref{eq_triangleintResult}. Another piece is nonanalytic in $k_s$. This piece appears in the region $q \ll k_1$, so the factor $|k_1-q|^{-2s_{3\bar3}}$ becomes $|k_1|^{-2s_{3\bar3}}$, and we obtain:
\bge
\mathbb L|_{q \ll k_1} = |k_1|^{-2s_{3\bar3}}\int \FR{\di^3\mb q}{(2\pi)^3}|\mb q|^{-2s_{1\bar1}}|\mb k_s-\mb q|^{-2s_{2\bar2}},
\ede 
which is again a bubble integral and gives the second line of \eqref{eq_triangleintResult}.
\end{appendix}
 
\newpage
\bibliography{CosmoCollider} 
\bibliographystyle{utphys}

\end{document}